\documentclass{article}
\usepackage{amsmath}
\usepackage{indentfirst}
\usepackage{bm}
\usepackage{braket}
\usepackage[pdftex]{graphicx}
\usepackage{float}
\usepackage{titling}
\title{{\bf Cosmological correlation functions including a massive scalar field and\\ an arbitrary number of soft-gravitons}}
\author{Ryo Saito\\\\{\footnotesize {\sl Department of Physics, Osaka University, Toyonaka, Osaka 560-0043, Japan}}}
\date{ }
\def\vector#1{\mbox{\boldmath $#1$}}
\def\vk{{\bm k}}
\def\vx{{\bm x}}
\def\vy{{\bm y}}
\def\vz{{\bm z}}
\def\vq{{\bm q}}
\def\vl{{\bm l}}

\def\zetan{\zeta_{n}}
\def\gammat{\tilde{\gamma}}
\def\gaij{\gamma_{ij}}
\def\gaijt{\gammat_{ij}}
\def\gadot{\dot{\gamma}}
\def\gadij{\gadot_{ij}}
\def\binashi{\frac{H^2}{4\epsilon (k_2)^3}}
\def\biari{-\frac{H^3}{4\epsilon k_2}}
\def\binashii{\frac{H^2}{4\epsilon (k_1)^3}}
\def\biarii{-\frac{H^3}{4\epsilon k_1}}
\def\binashiq{\frac{H^2}{q^3}}
\def\biariq{-\frac{H^3}{q}}
\def\binashik{\frac{H^2}{4\epsilon k^3}}
\def\pidot{\dot{\pi}}
\def\gtil{\tilde{g}}
\def\parti{\partial_{i}}
\def\partj{\partial_{j}}
\def\partl{\partial_{l}}

\def\ttil{\tilde{t}}
\def\sigdot{\dot{\sigma}}
\def\zetadot{\dot{\zeta}}
\def\hankel{H_{\nu}^{(1)}}
\def\hankelc{H_{\nu}^{(1)*}}
\def\hankelm{H_{i\mu}^{(1)}}
\def\hankelmc{H_{i\mu}^{(1)*}}
\def\zetzet{\langle\zeta\zeta\rangle}
\def\gamzetzet{\langle\gamma\zeta\zeta\rangle}
\def\gamgamzetzet{\langle\gamma\gamma\zeta\zeta\rangle}
\def\sigmatip{e^{-\pi\textrm{Im}(\nu)}\frac{\pi}{4}H^2}
\def\csch{\textrm{csch}}
\def\coth{\textrm{coth}}
\def\cosh{\textrm{cosh}}
\def\sinh{\textrm{sinh}}

\def\pfq{{}_2F_2}
\def\bessel{J_{i\mu}}
\def\besselc{J_{-i\mu}}
\def\cnp{c_{n}^{+}}
\def\cnm{c_{n}^{-}}
  \makeatletter
    
    \@addtoreset{equation}{section}
  \makeatother

\begin{document}
\begin{titlingpage}
    \maketitle
    \begin{abstract}
    We study the imprint of a massive scalar particle on cosmological correlation functions, and suggest the way to determine the mass of the newly introduced particle, which is expected to be around $10^{14}$ GeV.
    After reviewing the basic theory by Maldacena and the effective field theory (EFT) of inflation by Cheung {\sl et al.}, we apply these two theories to construct new couplings of a massive scalar field with primordial fluctuations including an arbitrary number of gravitons.
    We compute some correlation functions including these couplings in the soft-graviton limit.
    We show that when the number of soft-gravitons is getting larger, the peak of the correlation function is shifted to larger mass of the scalar particle.
    In addition we derive a relation, which relates correlation functions with $N+1$ to $N$ soft-gravitons when the mass of the scalar particle becomes much higher than $10^{14}$ GeV, and confirm the relation by numerical analysis.
    \end{abstract}
\end{titlingpage}
\tableofcontents
\clearpage
\section{Introduction}
The methods to obtain the information during inflation have been developed in various ways. One of the most innovative works was done by Maldacena\cite{Mal}, who applied the quantum field theory to cosmology and computed the three point functions of primordial fluctuations, $\zeta$ (scalar fluctuation) and $\gaij$ (tensor fluctuation or `graviton'). Especially the three point function of $\zeta$ is important because it tells us the deviation from the Gaussian features of the cosmic state, which is called `non-Gaussianity'\cite{Komatsu}.\vspace{1mm}

Another astonishing work was done by Cheung {\sl et al.}, who constructed the `Effective Field Theory of Inflation'\cite{Effective}(The similar approaches are also shown in \cite{Senatore} or \cite{Weinberg}). In this theory, the scalar fluctuation $\zeta$ is interpreted as a Nambu-Goldstone boson $\pi$, which is associated with a spontaneous breaking of time diffeomorphism invariance. The broken symmetry allows us to consider more terms in a Lagrangian than Maldacena's, so we can discuss a more general theory using the EFT method.\vspace{1mm}

In addition recently, there have been many attempts to introduce other fields into the inflationary theory, such as `quasi-single field inflation' by Chen and Wang\cite{Quasi}.
The mass of the newly introduced particles can be expected to be around $10^{14}$ GeV, which can be estimated as the energy scale during inflation.
Therefore, inflation is expected to become a tool to seek for unknown particles, which cannot be detected in terrestrial accelerators\cite{Noumi}\cite{Cosmo}\cite{Baumann}\cite{SMcosmo}.\vspace{1mm}

In this paper, we apply the EFT method to introduce another scalar field $\sigma$ into the Maldacena's theory.
Such a way was already studied by Noumi {\sl et al.}\cite{Noumi}; they constructed some couplings with $\zeta$ and $\sigma$ using the EFT method.
Expanding their work, we study also a coupling with a graviton: `{\sl $\gamma\zeta\sigma$ coupling}'.
Then, we compute some correlation functions including this coupling in the soft-graviton limit:
\begin{eqnarray*}
\zetzet&=&f_1(m,c_1),\\
\gamzetzet&=&f_2(m,c_1,c_4),\\
\gamgamzetzet&=&f_3(m,c_4),
\end{eqnarray*}
where $c_1$ and $c_4$ are constants associated with the new couplings, $m$ is the mass of $\sigma$, and $f_1$, $f_2$ and $f_3$ are functions of them (the explicit forms are shown in (\ref{f1ori}) to (\ref{f3muzui})). 
If the observation tells the values of these correlation functions in the future, then it follows that there are three equations including three variables $c_1$, $c_4$ and $m$.
Therefore, solving the equations, we will be able to determine the mass of the unknown particle, which may be around $10^{14}$ GeV.\vspace{1mm}

Then, we generalize the theory above; we construct couplings of $\zeta$, $\sigma$ and $N$ soft-gravitons and compute correlation functions $\langle\gamma^{s_1}\cdots\gamma^{s_{N}}\zeta\zeta\rangle$ including these couplings.
By plotting this result as a function of $m$ for several $N$'s, we examine how the number of soft-gravitons affects the correlation function.
Finally, we derive a relation, which relates $\langle\gamma^{s_1}\cdots\gamma^{s_{N+1}}\zeta\zeta\rangle$ to $\langle\gamma^{s_1}\cdots\gamma^{s_{N}}\zeta\zeta\rangle$ in the $m\to\infty$ limit in this model, and confirm that our results are consistent with the relation numerically.\vspace{1mm}

This paper is organized as follows. In Section 2, we review the Maldacena's theory, especially the computation of $\gamzetzet$. In Section 3, we also review the EFT method and confirm that it actually derives the Maldacena's theory. In Section 4, we apply the EFT method to introduce $\sigma$, compute the correlation functions and examine the several features as mentioned above. Section 5 is devoted to the summary.
Note that in the following, we set $c=\hbar=8\pi G=1$.

\section{Review of the Maldacena's theory}
In this section, we review the basic theory constructed by Maldacena\cite{Mal}, which computed the three point functions of primordial fluctuations.
\subsection{Set-up}
The action of gravity and a scalar field $\phi$ (called `inflaton') with a flat FRW metric is
\begin{equation}\label{action0}
S=\frac{1}{2}\int d^{4}x\sqrt{-g}\left[R-g^{\mu\nu}\partial_{\mu}\phi\partial_{\nu}\phi-2V(\phi)\right],
 \end{equation}
where $R$ is the Ricci scalar, $V(\phi)$ is a potential of $\phi$ and the metric is
\begin{eqnarray} 
ds^2&=&-dt^2+a^2(t) d\vector{x}^2\nonumber\\
&=&a^2(t)(-d\eta^2+d\vector{x}^2).
 \end{eqnarray}
Note that $a(t)$ is the scale factor and $\eta$ is the conformal time. In addition, the slow-roll parameters are defined as
 \begin{eqnarray}
 \epsilon\equiv-\frac{\dot{H}}{H^2}=\frac{1}{2}\frac{\dot{\phi}^2}{H^2},\\
 \delta\equiv\frac{1}{2}\frac{\ddot{H}}{H\dot{H}}=\frac{1}{H}\frac{\ddot{\phi}}{\dot{\phi}},
 \end{eqnarray}
 where $H(t)$ is the Hubble parameter. In the equations above, we used the relation
 \begin{equation}
 \dot{H}=-\frac{1}{2}\dot{\phi}^2,
 \end{equation}
which can be obtained by solving the Einstein's equations.\vspace{3mm}

From now on, we will introduce the ADM formalism\cite{ADMpaper} in which the metric is written as
\begin{equation}\label{ADMmetric}
ds^2=-N^2 dt^2+h_{ij}(dx^{i}+N^{i}dt)(dx^{j}+N^{j}dt).
\end{equation}
Note that $h_{ij}$ is a spatial metric and $N$ and $N^{i}$ are undetermined coefficients called `lapse' and `shift' respectively. Using this metric, the background action (\ref{action0}) becomes
\begin{equation}
\begin{split}
S=\frac{1}{2}\int d^{4}x\sqrt{h}\biggl[NR^{(3)}-2NV &+N^{-1}(E_{ij}E^{ij}-E^2)\\
&+N^{-1}(\dot{\phi}-N^{i}\partial_{i}\phi)^2-Nh^{ij}\partial_{i}\phi\partial_{j}\phi\biggl],
\end{split}\label{ADMaction}
\end{equation}
where $R^{(3)}$ is a spatial scalar curvature and
\begin{eqnarray}
E_{ij}&\equiv&\frac{1}{2}(\dot{h}_{ij}-\nabla_{i}N_{j}-\nabla_{j}N_{i}),\\
E&\equiv&E^{i}_{i}.
\end{eqnarray}\vspace{3mm}

In this formalism, Maldacena used two gauges in order to describe the primordial fluctuations.\vspace{3mm}

\leftline{\underline{\textbf{Comoving gauge}}}\vspace{1mm}
In this gauge, the fluctuations appear in the metric while the inflaton field is homogeneous:
\begin{equation}\label{comoving}
\phi=\phi(t),\quad h_{ij}=a^2(t) e^{2\zeta}\left[\delta_{ij}+\gamma_{ij}+\frac{1}{2}\gamma_{il}\gamma_{lj}+\cdots\right],
\end{equation}
where $\zeta$ and $\gamma_{ij}$ are scalar and tensor fluctuations respectively. Note that we usually call $\gamma_{ij}$ as `graviton' field.\vspace{3mm}

\leftline{\underline{\textbf{Spatially-flat gauge}}}\vspace{1mm}
On the other hand in this gauge, the scalar fluctuation is imposed not on the spatial metric but on the inflaton field:
\begin{equation}\label{spatially}
\phi=\phi(\tilde{t})+\varphi(\tilde{t},\vector{x}),\quad \tilde{h}_{ij}=a^2(\tilde{t}) \left[\delta_{ij}+\tilde{\gamma}_{ij}+\frac{1}{2}\tilde{\gamma}_{il}\tilde{\gamma}_{lj}+\cdots\right].
\end{equation}

Note that in both of the gauges, we assume the transverse and traceless conditions about the graviton:
\begin{equation}
\partial_{i}\gamma_{ij}=0,\quad\gamma_{ii}=0.
\end{equation}
Also note that these two gauges are equivalent because we can go from one gauge to the other by the time reparametrization $\tilde{t}=t+T$. To the first order,
\begin{eqnarray}
T&=&-\frac{\varphi}{\dot{\phi}}\ ,\\
\zeta&=&HT\ =\ -H\frac{\varphi}{\dot{\phi}}\ .\label{varphizeta}
\end{eqnarray}\vspace{3mm}

From now on basically, we will use the comoving gauge (\ref{comoving}). In order to obtain the action written by $\zeta$ and $\gamma_{ij}$, we have to determine $N$ and $N^{i}$ appeared in our metric (\ref{ADMmetric}) at first. Since these quantities $N$ and $N^{i}$ are not dynamical variables, there are constraint equations:
\begin{equation}\label{constraints}
\frac{\delta S}{\delta N}=0,\quad\frac{\delta S}{\delta N^{j}}=0.
\end{equation}
Substituting the action (\ref{ADMaction}) into these equations, we obtain
\begin{eqnarray}
R^{(3)}-2V-N^{-2}(E_{ij}E^{ij}-E^2)-N^{-2}\dot{\phi}^2&=&0\ ,\\
\nabla_{i}\left[N^{-1}(E^{i}_{j}-\delta^{i}_{j}E)\right]&=&0\ .
\end{eqnarray}
Setting $N=:1+N_{1}$, in which $N_1$ is a first-order quantity, we can solve these equations to the first order:
\begin{eqnarray}
N_{1}&=&\frac{\dot{\zeta}}{H},\label{Quasirefone}\\
\partial_{i}N^{i}&=&-a^{-2}\frac{1}{H}\partial^{2}\zeta+\epsilon\,\dot{\zeta}.\label{Quasireftwo}
\end{eqnarray}
Notice that in this paper, $\partial^{2}\zeta$ and $(\partial\zeta)^2$ mean $\partial_{i}\partial^{i}\zeta$ and $\partial_{i}\zeta\partial^{i}\zeta$ respectively ($i = 1, 2, 3)$. 
Finally, substituting these solutions into the action (\ref{ADMaction}), we can obtain the action written by $\zeta$ and $\gamma_{ij}$.\vspace{3mm}

Even when using the spatially-flat gauge (\ref{spatially}), we can similarly solve the constraint equations (\ref{constraints}):
\begin{eqnarray}
N_{1}&=&\frac{\dot{\phi}}{2H}\varphi,\label{N1sp}\\
\partial_{i}N^{i}&=&\epsilon\frac{d}{dt}\left(-\frac{H}{\dot{\phi}}\varphi\right).\label{Nisp}
\end{eqnarray}
Recalling the relation between $\varphi$ and $\zeta$ (\ref{varphizeta}), let us redefine $\varphi$ as
\begin{equation}\label{zetandef}
\zeta_{n}\equiv-H\frac{\varphi}{\dot{\phi}},
\end{equation}
so that the solutions (\ref{N1sp}) and (\ref{Nisp}) become
\begin{eqnarray}
N_{1}&=&-\epsilon\,\zeta_{n}\label{n1mal},\\
\partial_{i}N^{i}&=&\epsilon\,\dot{\zeta_{n}}.\label{nimal}
\end{eqnarray}
We will use them in Section 3.

\subsection{Quantization of the fluctuations}
After substituting the solutions $N$ and $N^{i}$ into the action, we can obtain the quadradic action of $\zeta$:
\begin{equation}
S_{\zeta\zeta}=\int d^{4}x \ \epsilon\left[a^{3}(t)\dot{\zeta}^{2}-a(t)(\partial\zeta)^{2}\right].\label{quadaction}
\end{equation}
In order to quantize $\zeta$, it is useful to introduce the conformal time $\eta$, which is defined as $d\eta=dt/a$. If we assume that $\eta$ moves from $-\infty$ to $0$ when $t$ moves from $-\infty$ to $\infty$, it follows that
\begin{equation}
\eta=-\frac{1}{aH}.  
\end{equation}
Using this $\eta$, we can obtain the equation of motion from the action (\ref{quadaction}),
\begin{equation}
\zeta''-\frac{2}{\eta}\zeta'-\partial^{2}\zeta=0,\label{zetaeom}
\end{equation}
where $'$ means $d/d\eta$. Note that we are taking the de Sitter limit, in which $\epsilon$ or $H$ is constant and $\epsilon\ll 1$\vspace{3mm}.

The way to quantize $\zeta$ is the same as the usual QFT method\cite{Birrel}. Firstly, we represent $\zeta$ using the Fourier transform,
\begin{equation}\label{fourzeta}
\zeta(\eta,\vx)=\int\frac{d^3{k}}{(2\pi)^{3}}\left[u_{k}(\eta)a_{\scriptsize{\vk}}e^{i\vk\cdot\vx}+u_{k}^{*}(\eta)a_{\scriptsize{\vk}}^{\dagger}e^{-i\vk\cdot\vx}\right],
\end{equation}
where $u_{k}$ and $u_{k}^{*}$ are the solutions of E.O.M (\ref{zetaeom}) and $a_{\vk}$ and $a_{\vk}^{\dagger}$ are the annihilation and creation operators respectively.
They satisfy the commutation relation:
\begin{equation}\label{aacomm}
\left[a_{\vk}, a_{\vk'}^{\dagger}\right]=(2\pi)^{3}\delta^{3}(\vk-\vk').
\end{equation}
This equation and the canonical commutation relation,
\begin{equation}
\left[\zeta(\eta,\vx), \frac{\delta L}{\delta\zeta'}(\eta,\vy)\right]=i\delta^{3}(\vx-\vy),
\end{equation}
impose the normalization condition on $u_{k}$ and $u_{k}^{*}$, so that
\begin{eqnarray}
u_{k}(\eta)&=&\frac{H}{\sqrt{4\epsilon k^3}}(1+i k\eta)e^{-i k\eta},\label{uksol}\\
u_{k}^{*}(\eta)&=&\frac{H}{\sqrt{4\epsilon k^3}}(1-i k\eta)e^{i k\eta}.\label{ukcsol}
\end{eqnarray}
It follows that when $\eta\to 0$, the solutions $u_{k}$ and $u_{k}^{*}$ become constant ($H/\sqrt{4\epsilon k^{3}}$), which means that the scalar fluctuation $\zeta$ freezes after long time have passed. Usually, we assume that this freeze occurs when the wavelength of $\zeta$ crosses the Hubble horizon:
\begin{equation}\label{crosses}
\frac{a}{k}\sim H^{-1}.
\end{equation}\vspace{3mm}

Now that we have finished the quantization of $\zeta$, we are ready to compute the correlation functions. Under the assumption that the initial state of the universe is $\ket{0}$ (called `Bunch-Davies vacuum'), which is annihilated by $a_{\vk}$, the two point function can be easily computed just by using (\ref{fourzeta}), (\ref{aacomm}), (\ref{uksol}) and (\ref{ukcsol}). In the momentum space,
\begin{eqnarray}
\langle\zeta(\vk_{1})\zeta(\vk_{2})\rangle&=&\bra{0}\zeta(\vk_{1})\zeta(\vk_{2})\ket{0}\nonumber\\
&=&(2\pi)^{3}\delta^{3}(\vk_{1}+\vk_{2})\frac{H_{*}^{2}}{4\epsilon_{*}(k_{2})^{3}},\label{twopoint}
\end{eqnarray}
where the subscript $*$ means `evaluated when the wavelength crosses the Hubble horizon'.
This means from (\ref{crosses}),
\begin{equation}
\frac{a_{*}}{k}\sim H_{*}^{-1}.
\end{equation}
Note that in (\ref{twopoint}) we don't have to think about the evolution of the state because such terms are higher order. The formula appeared in (\ref{twopoint}) is called {\bf `power spectrum'}:
\begin{equation}\label{powerszeta}
P_{\zeta}(k)=\frac{H^{2}}{4\epsilon k^{3}}.
\end{equation}\vspace{3mm}

Similarly, we can quantize the graviton $\gamma_{ij}$. The quadratic action is
\begin{equation}\label{quadactiongamma}
S_{\gamma\gamma}=\frac{1}{8}\int d^{4}x\left[a^{3}(t)\dot{\gamma}_{ij}\dot{\gamma}_{ij}-a(t)\partial_{l}\gamma_{ij}\partial_{l}\gamma_{ij}\right].
\end{equation}
Then we have to represent $\gamma_{ij}$ using the Fourier transform:
\begin{equation}\label{fourgamma}
\gamma_{ij}(\eta,\vx)=\sum_{s=\pm}\int\frac{d^3 k}{(2\pi)^3}\epsilon_{ij}^{s}(k)\left[U_{k}^{s}(\eta)a_{\vk}e^{i\vk\cdot\vx}+U_{k}^{s*}(\eta)a_{\vk}^{\dagger}e^{-i\vk\cdot\vx}\right],
\end{equation}
where $U_{k}$ and $U_{k}^{*}$ are the solutions of E.O.M and $\epsilon_{ij}^{s}$ is a polarization tensor which satisfies the transverse and traceless condition, $k^{i}\epsilon_{ij}=\epsilon_{ii}=0$, and the orthonormality condition, $\epsilon_{ij}^{s}(k)\epsilon_{ij}^{s'}(k)=2\delta_{ss'}$. After the quantization in the same way as of $\zeta$, we can finally obtain
\begin{equation}
\langle\gamma^{s}(\vq_1)\gamma^{s'}(\vq_2)\rangle=(2\pi)^3 \delta^3 (\vq_1+\vq_2)\delta_{ss'}\frac{H_{*}^2}{(q_{2})^{3}},
\end{equation}
so that the power spectrum of $\gaij$ is
\begin{equation}\label{powersgamma}
P_{\gamma}(q)=\frac{H^2}{q^3}.
\end{equation}
Note that this result is different from $\zeta$'s (\ref{powerszeta}) just by a factor $4\epsilon$.

\subsection{Energy scale during inflation}
We can define the dimensionless power spectrum for (\ref{powerszeta}) and (\ref{powersgamma})\cite{TASI}:
\begin{equation}\label{dimensionless}
\Delta_{\zeta}^2\sim\frac{H^2}{\epsilon}\times\frac{1}{M_{\textrm{Pl}}^2},\qquad\Delta_{\gamma}^2\sim H^2\times\frac{1}{M_{\textrm{Pl}}^2},
\end{equation}
where $M_{\textrm{Pl}}\equiv(8\pi G)^{-1/2}(\equiv 1)$ is the reduced Planck mass; only in this subsection, we write this quantity visibly.
Therefore, {\sl the tensor-to-scalar ratio}, which is defined as
\begin{equation}\label{tensortoscalar}
r\equiv\frac{\Delta_{\gamma}^2}{\Delta_{\zeta}^2},
\end{equation}
has the order of $\epsilon$.\footnote{Choosing the appropriate coefficients for (\ref{dimensionless}), it follows that $r=16\epsilon_{*}$\cite{TASI}.}
Then, note that $M_{\textrm{Pl}}^2 H^2\sim V$; this is obtained by one of the Einstein's equations,
\begin{equation}
3M_{\textrm{Pl}}^2 H^2=\frac{1}{2}\dot{\phi}^2+V(\phi),
\end{equation}
under the slow-roll assumption that the potential energy dominates over the kinetic energy, $\dot{\phi}^2\ll V(\phi)$.
In addition, the observation tells $\Delta_{\zeta}^2\sim10^{-9}$, so (\ref{tensortoscalar}) becomes
\begin{equation}\label{energyscaleone}
r\sim\frac{V}{10^{-9}}\times\frac{1}{M_{\textrm{Pl}}^4}.
\end{equation}
Finally, considering $M_{\textrm{Pl}}\approx2.4\times10^{18}$ GeV, (\ref{energyscaleone}) suggests
\begin{equation}
V^{1/4}\sim r^{1/4}\times10^{16}\,\textrm{[GeV]}.
\end{equation}
In addition, using $M_{\textrm{Pl}}^2 H^2\sim V$,
\begin{equation}
H\sim r^{1/2}\times10^{14}\,\textrm{[GeV]}.
\end{equation}
Although this estimation is not completely accurate, we can consider $10^{14}$ GeV, which comes close to the GUT scale, as the energy scale during inflation.

\subsection{Three point functions}
In order to compute the three point functions, we have to derive the cubic action at first. Actually Maldacena derived it completely and computed all of the three point functions, $\langle\zeta\zeta\zeta\rangle$, $\langle\gamma\zeta\zeta\rangle$, $\langle\gamma\gamma\zeta\rangle$ and $\langle\gamma\gamma\gamma\rangle$. In this paper as an exercise, we study how to compute $\langle\gamma\zeta\zeta\rangle$, which is relatively easy. Firstly, we review the important computational method: {\sl the in-in formalism} (which is also called 'the Schwinger-Keldysh formalism\cite{ininpaper}').\vspace{3mm}

\leftline{\underline{\textbf{In-In formalism}}\,(This review is based on \cite{Collins} by Collins.)}\vspace{1mm}
In this method, we use the interaction picture, in which states evolve by the interaction Hamiltonian $H_{I}(t)$. Therefore the state at an arbitrary time $t$ can be described as
\begin{equation}\label{evolution}
\ket{0(t)}=U(t,-t')\ket{0(t')},
\end{equation}
where
\begin{equation}\label{timeevolution}
U(t,t')=T\left[\textrm{exp}\left[-i\int_{t'}^{t}dt''H_{I}(t'')\right]\right].
\end{equation}
This `$T$' means a time-ordering operator:
\begin{equation}
\begin{split}\label{timeordering}
T\left[A(t_1)B(t_2)\right]=\begin{cases}A(t_1)B(t_2)&(t_{1}>t_{2})\\
B(t_2)A(t_1)&(t_{1}<t_{2})\end{cases}.
\end{split}
\end{equation}
Using (\ref{evolution}), we can compute an expectation value of an operator $\mathcal{O}(t,\vx)$ at a time $t$:
\begin{eqnarray}
\langle\mathcal{O}(t,\vx)\rangle&=&\braket{0(t)|\mathcal{O}(t,\vx)|0(t)}\nonumber\\
&=&\braket{0|U^{\dagger}(t,-\infty)\mathcal{O}(t,\vx)U(t,-\infty)|0}\nonumber\\
&=&\braket{0|U^{\dagger}(t,-\infty)\left(U^{\dagger}(\infty,t)U(\infty,t)\right)\mathcal{O}(t,\vx)U(t,-\infty)|0}\nonumber\\
&=&\braket{0|U^{\dagger}(\infty,-\infty)U(\infty,t)\mathcal{O}(t,\vx)U(t,-\infty)|0}.\label{ininO}
\end{eqnarray}
Note that in the third line, the unitarity $1=U^{\dagger}U$ was inserted.\vspace{2mm}

Reading the fourth line from right to left, it follows that we are going from $t=-\infty$ to $\infty$, and then going back to $t= -\infty$. So we have to devide the time path into two parts as Figure \ref{ininpath}:

\begin{figure}[H]
  \centering
  \includegraphics[width=7cm]{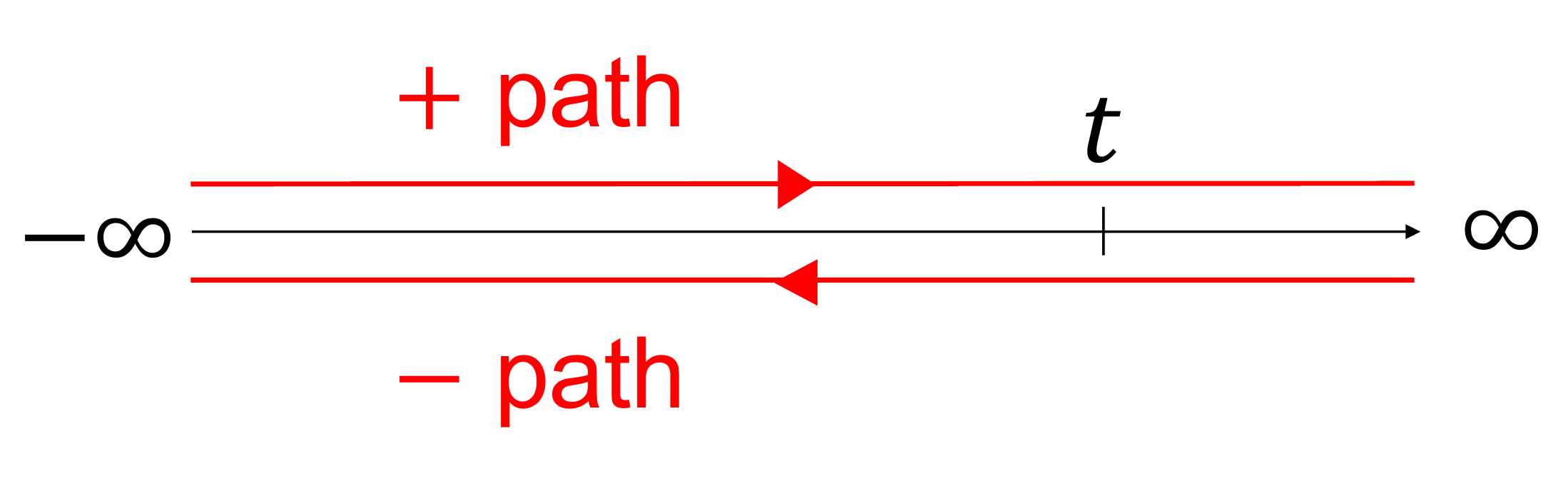}
  \caption{Time path in the in-in formalism}
  \label{ininpath}
\end{figure}
From now on, we label operators with `$+$' or `$-$' according to whether that operator is on the `$+$ path' or `$-$ path'. Notice that on the $+$ path, a time-ordering operator $T$ works as usual as (\ref{timeordering}), but on the $-$ path, the reversed result follows:
\begin{equation}
\begin{split}\label{timeordering}
T\left[A^{-}(t_1)B^{-}(t_2)\right]=\begin{cases}B^{-}(t_2)A^{-}(t_1)&(t_{1}>t_{2})\\
A^{-}(t_1)B^{-}(t_2)&(t_{1}<t_{2})\end{cases}.
\end{split}
\end{equation}

Considering these rules, the fourth line in (\ref{ininO}) can be rewritten as
\begin{eqnarray}
\langle\mathcal{O}(t,\vx)\rangle&=&\braket{0|T\left[\textrm{exp}\left[i\int_{-\infty}^{\infty}dt' H_{I}^{-}(t')\right]\right]T\left[\mathcal{O}^{+}(t,\vx)\textrm{exp}\left[-i\int_{-\infty}^{\infty}dt'H_{I}^{+}(t')\right]\right]|0}\nonumber\\
&=&\braket{0|T\left[\mathcal{O}^{+}(t,\vx)\textrm{exp}\left[-i\int_{-\infty}^{\infty}dt'\left(H_{I}^{+}(t')-H_{I}^{-}(t')\right)\right]\right]|0}.\label{ininformula}
\end{eqnarray}

Now we are ready to compute $\langle\gamma\zeta\zeta\rangle$.\vspace{3mm}

\leftline{\underline{\textbf{Computation of $\langle\gamma\zeta\zeta\rangle$}}}\vspace{2mm}
In the comoving gauge, the cubic action of $\gamma\zeta\zeta$ is
\begin{eqnarray}
S_{\gamma\zeta\zeta}&=&\int d^{4}x\Bigl[\epsilon a(t) \gamma_{ij}\partial_{i}\zeta\partial_{j}\zeta\nonumber\\
&+&\frac{1}{4}\epsilon^{2}a^{3}(t)\partial^{2}\gamma_{ij}(\partial_{i}\partial^{-2}\dot{\zeta})(\partial_{j}\partial^{-2}\dot{\zeta})+\frac{1}{2}\epsilon^{2}a^{3}(t)\dot{\gamma}_{ij}\partial_{i}\zeta(\partial_{j}\partial^{-2}\dot{\zeta})\nonumber\\
&+&f(\gamma,\zeta)\left.\frac{\delta L}{\delta\zeta}\right|_{1}+f_{ij}(\zeta)\left.\frac{\delta L}{\delta\gamma_{ij}}\right|_{1}\Bigl],\label{cubicactiongzz}
\end{eqnarray}
where
\begin{eqnarray}
f(\gamma,\zeta)&\equiv&\frac{1}{4H}\partial^{-2}(\dot{\gamma}_{ij}\partial_{i}\partial_{j}\zeta),\\
f_{ij}(\zeta)&\equiv&-\frac{\epsilon}{H}\left[(\partial_{i}\partial^{-2}\dot{\zeta})\partial_{j}\zeta+(\partial_{j}\partial^{-2}\dot{\zeta})\partial_{i}\zeta\right]+\frac{1}{H^{2}}a^{-2}(t)\partial_{i}\zeta\partial_{j}\zeta.
\end{eqnarray}
Notice that $\left.\frac{\delta L}{\delta\zeta}\right|_{1}$ and $\left.\frac{\delta L}{\delta\gamma_{ij}}\right|_{1}$ are the equations of motion of $\zeta$ and $\gamma$ respectively (the subsctipt `1' means `first order'). Then, performing the field redefinitions such as
\begin{eqnarray}
\zeta_{n}&\equiv&\zeta+f(\gamma,\zeta),\label{zetandefdef}\\
\tilde{\gamma}_{ij}&\equiv&\gamma_{ij}+f_{ij}(\zeta),\label{gammatdefdef}
\end{eqnarray}
the quadratic actions (\ref{quadaction}) and (\ref{quadactiongamma}) produce some cubic terms which eliminate the third line in (\ref{cubicactiongzz}):
\begin{eqnarray}
S_{\zeta\zeta}&=&S_{\zetan\zetan}-2\int d^{4}x\ \epsilon\,\left[a^{3}(t)\dot{\zeta}\dot{f}(\gamma,\zeta)-a(t)\partial_{i}\zeta\partial_{i}f(\gamma,\zeta)\right]\nonumber\\
&=&S_{\zetan\zetan}+\int d^{4}x\ 2\left[\frac{d}{dt}\left(\epsilon a^{3}(t)\dot{\zeta}\right)-\epsilon a(t)\partial^{2}\zeta\right]f(\gamma,\zeta)\nonumber\\
&=&S_{\zetan\zetan}+\int d^{4}x\left[-\left.\frac{\delta L}{\delta\zeta}\right|_{1}\right]f(\gamma,\zeta),
\end{eqnarray}

\begin{eqnarray}
S_{\gamma\gamma}&=&S_{\gammat\gammat}-\frac{1}{4}\int d^{4}x\left[a^{3}(t)\dot{\gamma}_{ij}\dot{f}_{ij}(\zeta)-a(t)\partial_{l}\gamma_{ij}\partial_{l}f_{ij}(\zeta)\right]\nonumber\\
&=&S_{\gammat\gammat}+\int d^{4}x\ \frac{1}{4}\left[\frac{d}{dt}\left(a^{3}(t)\dot{\gamma}_{ij}\right)-a(t)\partial^{2}\gamma_{ij}\right]f_{ij}(\zeta)\nonumber\\
&=&S_{\gammat\gammat}+\int d^{4}x\left[-\left.\frac{\delta L}{\delta\gamma_{ij}}\right|_{1}\right]f_{ij}(\zeta).
\end{eqnarray}
Note that $\zetan$ and $\gammat_{ij}$ in (\ref{zetandefdef}) and (\ref{gammatdefdef}) are the same as the fluctuations in the spatially-flat gauge, defined as (\ref{spatially}) and (\ref{zetandef}). As for the proof of this, refer to Appendix A in \cite{Mal}.\vspace{2mm}

Considering the field redefinitions (\ref{zetandefdef}) and (\ref{gammatdefdef}), the three point function $\langle\gamma\zeta\zeta\rangle$ becomes
\begin{equation}
\begin{split}\label{gzz}
\braket{0(t)|\gaij\zeta\zeta|0(t)}&=\braket{0(t)|\gaijt\zetan\zetan|0(t)}\\
&\qquad-\braket{0|f_{ij}(\zetan)\zetan\zetan|0}\\
&\qquad\quad-\bigl[\braket{0|\gaijt f(\gammat,\zetan)\zetan|0}\\
&\qquad\qquad+\braket{0|\gaijt\zetan f(\gammat,\zetan)|0}\bigl].
\end{split}
\end{equation}

In order to compute (\ref{gzz}), we show the two point functions of $\zeta$ and $\gamma$ below, which can be easily derived from (\ref{fourzeta}) and (\ref{fourgamma}):\vspace{2mm}

\begin{table}[H]
  \centering
  \caption{Two point functions of $\zeta$ and $\gamma$}
  \begin{tabular}{|l|}\hline
  \\
  $\displaystyle\braket{0|\zeta(t,\vx)\zeta(t',\vx')|0}=\int \frac{d^{3}k}{(2\pi)^{3}}e^{i\vk\cdot(\vx-\vx')}\left[\frac{H^{2}}{4\epsilon k^{3}}(1+ik\eta)(1-i k\eta')e^{i k(\eta'-\eta)}\right]$\\
  \\
  $\displaystyle\braket{0|\zeta(t,\vx)\dot{\zeta}(t',\vx')|0}=\int \frac{d^{3}k}{(2\pi)^{3}}e^{i\vk\cdot(\vx-\vx')}\left[-\frac{H^{3}}{4\epsilon k}(1+ik\eta)\eta'^{2}e^{i k(\eta'-\eta)}\right]$\\
  \\ \hline\hline
  \\
  $\displaystyle\braket{0|\gamma_{ij}(t,\vx)\gamma_{\alpha\beta}(t',\vx')|0}=\sum_{s}\int \frac{d^{3}q}{(2\pi)^{3}}e^{i\vq\cdot(\vx-\vx')}\epsilon_{ij}^{s}\epsilon_{\alpha\beta}^{s}\left[\frac{H^{2}}{q^{3}}(1+iq\eta)(1-i q\eta')e^{i q(\eta'-\eta)}\right]$\\
  \\
  $\displaystyle\braket{0|\gamma_{ij}(t,\vx)\dot{\gamma}_{\alpha\beta}(t',\vx')|0}=\sum_{s}\int \frac{d^{3}q}{(2\pi)^{3}}e^{i\vq\cdot(\vx-\vx')}\epsilon_{ij}^{s}\epsilon_{\alpha\beta}^{s}\left[-\frac{H^{3}}{q}(1+iq\eta)\eta'^{2}e^{i q(\eta'-\eta)}\right]$\\
  \\ \hline
  \end{tabular}\label{tipszg}
\end{table}
Notice that in Table \ref{tipszg}, the dot appeared in the left side means $d/dt'$, and the right side is shown using not $t$ but $\eta$.\vspace{3mm}

Using Table \ref{tipszg}, let us compute $\langle\gamma\zeta\zeta\rangle$. From now on, all expectation values $\langle\cdots\rangle$ will be represented mainly in the momentum space. In addition, note that we evaluate these quantities $\langle\cdots\rangle$ at $t\to\infty$, which is equivalent with $\eta\to0$.\vspace{3mm}

The second line in (\ref{gzz}) is
\begin{equation}
\begin{split}
&-\braket{0|f_{ij}(\zetan)\zetan(\vk_1)\zetan(\vk_2)|0}\\
&\quad=(2\pi)^{3}\delta^{3}(\vk_1+\vk_2+\vl)\\
&\qquad\times\Biggl[2\frac{\epsilon}{H}(k_1)_{i}(k_2)_{j}\biggl\{\frac{1}{(k_1)^2}\left(\biarii\right)\binashi\eta^2(1+ik_1\eta)(1+ik_2\eta)(1-ik_2\eta)\\
&\qquad\qquad\qquad\qquad\qquad\frac{1}{(k_2)^2}\left(\biari\right)\binashii\eta^2(1+ik_2\eta)(1+ik_1\eta)(1-ik_1\eta)\biggl\}\\
&\qquad-2\eta^{2}(-i)^{2}(k_{1})_{i}(k_{2})_{j}\binashii\binashi(1+ik_{1}\eta)(1-ik_{1}\eta)(1+ik_2\eta)(1-ik_2\eta)\Biggl],
\end{split}
\end{equation}
which vanishes when $\eta\to0$. Note that the vector $\vl$ is the momentum of $f_{ij}$. \vspace{3mm}

The third line in (\ref{gzz}) is
\begin{equation}
\begin{split}
&-\braket{0|\gammat^{s}(\vq) f(\gammat^{s'},\zetan)\zetan(\vk)|0}\\
&\qquad=(2\pi)^{3}\delta^{3}(\vq+\vk+\vl)\delta_{ss'}\\
&\qquad\quad\times\left[-\frac{1}{4H}\frac{\epsilon_{ij}^{s'}k_{i}k_{j}}{|\vq+\vk|^2}\left(\biariq\right)\binashik\eta^2(1+iq\eta)(1+ik\eta)(1-ik\eta)\right],
\end{split}
\end{equation}
which also vanishes when $\eta\to0$.\vspace{3mm}

The fourth line is equivalent with the third line, so the remaining term is just the first line. In order to compute it, we have to consider the time-evolution of the state $\ket{0(t)}$ as (\ref{evolution}) and (\ref{timeevolution}). With the help of the in-in formalism (\ref{ininformula}), the first line in (\ref{gzz}) becomes
\begin{equation}
\begin{split}
&\braket{0(t)|\gaijt(t,\vx)\zetan(t,\vy)\zetan(t,\vz)|0(t)}\\
&\qquad=\braket{0|T\left[\gaijt^{+}\zetan^{+}\zetan^{+}\textrm{exp}\left[-i\int_{-\infty}^{t}dt'\left(H_{I}^{+}(t')-H_{I}^{-}(t')\right)\right]\right]|0}\\
&\qquad=2\,\textrm{Im}\left[\int_{-\infty}^{t}dt'\braket{0|\gaijt\zetan\zetan H_{I}(t')|0}\right],
\end{split}\label{gzzmain}
\end{equation}
in which the interaction Hamiltonian is given by the cubic action (\ref{cubicactiongzz}):
\begin{eqnarray}
H_{I}(t)&=&-L_{I}(t)\nonumber\\
&=&-\epsilon a(t)\int d^{3}x\,\gaijt\partial_{i}\zetan\partial_{j}\zetan.
\end{eqnarray}
Note that we can neglect the second line in (\ref{cubicactiongzz}) because it is higher order in $\epsilon$. \vspace{3mm}

In the momentum space, (\ref{gzzmain}) becomes (dropping the tilde and the subscript $n$)
\begin{equation}
\begin{split}
&\braket{0(t)|\gamma^{s}(\vq)\zeta(\vk_1)\zeta(\vk_2)|0(t)}\\
&\qquad=(2\pi)^{3}\delta^{3}(\vq+\vk_1+\vk_2)\\
&\qquad\qquad\times\left(-4\epsilon\right)\binashiq\binashii\binashi\epsilon_{ij}^{s}(\vq)(-i)^2 (k_1)_{i}(k_2)_{j}\\
&\qquad\qquad\quad\times\textrm{Im}\left[\int_{-\infty}^0 \frac{d\eta'}{\eta'^2 H^2}(1-iq\eta')(1-ik_{1}\eta')(1-ik_{2}\eta')e^{i(q+k_1+k_2)\eta'}\right],
\end{split}\label{gzznosoft}
\end{equation}
in which the $\eta\to0$ limit is already taken.\vspace{3mm}

Although we can perform the integral in (\ref{gzznosoft}) now, let us take the $\vq\to0$ limit ({\sl soft-graviton}); it leads to the consistency relation which will be mentioned later. Under this limit, (\ref{gzznosoft}) becomes
\begin{equation}
\begin{split}
&\lim_{\vq\to0}\braket{0(t)|\gamma^{s}(\vq)\zeta(\vk_1)\zeta(\vk_2)|0(t)}\\
&\qquad=(2\pi)^{3}\delta^{3}(\vq+\vk_1+\vk_2)\\
&\qquad\qquad\times4\epsilon\binashiq\binashi\binashi\epsilon_{ij}^{s}(\vq)\left(-(k_2)_{i}(k_2)_{j}\right)\\
&\qquad\qquad\quad\times\textrm{Im}\left[\int_{-\infty}^0 \frac{d\eta'}{\eta'^2 H^2}(1-ik_{2}\eta')^2 e^{2ik_2 \eta'}\right].
\end{split}\label{gzzsoft}
\end{equation}
Note that under this limit, the delta function in (\ref{gzzsoft}) means $\vk_1=-\vk_2$, so that $k_1=k_2$ and $(k_1)_{i}=-(k_2)_{i}$ .\vspace{3mm}

The integral in (\ref{gzzsoft}) can be computed using the $i\epsilon$ prescription and the integration by parts (see Appendix \ref{app0}):
\begin{equation}
\textrm{Im}\left[\int_{-\infty}^0 \frac{d\eta'}{\eta'^2 H^2}(1-ik_{2}\eta')^2 e^{2ik_2 \eta'}\right]=-\frac{3}{2}\frac{k_2}{H^2}.
\end{equation}

Substituting this result into (\ref{gzzsoft}), we finally obtain
\begin{equation}
\begin{split}
&\lim_{\vq\to0}\braket{0(t)|\gamma^{s}(\vq)\zeta(\vk_1)\zeta(\vk_2)|0(t)}\\
&\qquad=(2\pi)^{3}\delta^{3}(\vq+\vk_1+\vk_2)\epsilon_{ij}^{s}(\vq)\frac{(k_2)_{i}(k_2)_{j}}{(k_2)^2}P_{\gamma}(q)P_{\zeta}(k_2)\times\frac{3}{2},
\end{split}\label{gzzsoftsoft}
\end{equation}
where $P_{\gamma}(q)$ and $P_{\zeta}(k)$ are the power spectra defined as (\ref{powersgamma}) and (\ref{powerszeta}) respectively.\vspace{3mm}

Compareing (\ref{gzzsoftsoft}) with (\ref{twopoint}), we can find the relation:
\begin{equation}\label{consistencygzz}
\lim_{\vq\to0}\frac{\langle\gamma^{s}(\vq)\zeta(\vk_1)\zeta(\vk_2)\rangle'}{P_{\gamma}(q)}=-\frac{1}{2}\epsilon_{ij}^{s}(\vq)(k_2)_{i}\frac{\partial}{\partial(k_2)^{j}}\langle\zeta(\vk_1)\zeta(\vk_2)\rangle'.
\end{equation}
(This $'$ means removing the delta function $\delta^3(\vq+\vk_1+\vk_2)$ or $\delta^3(\vk_1+\vk_2)$ from the original correlation function $\langle\cdots\rangle$.) This is one of the `{\bf consistency relations}', which relate $(n+1)$-point functions to $n$-point functions. Although we have computed only $\langle\gamma\zeta\zeta\rangle$, we can find other consistency relations if we compute other three point functions. For example as for $\langle\zeta\zeta\zeta\rangle$, it follows that
\begin{equation}
\lim_{\vk_3\to0}\frac{\langle\zeta(\vk_1)\zeta(\vk_2)\zeta(\vk_3)\rangle'}{P_{\zeta}(k_3)}=(1-n_{s})\langle\zeta(\vk_1)\zeta(\vk_2)\rangle',
\end{equation}
where $n_{s}$ is called `tilt', which is defined as $k^3 P_{\zeta}(k)\propto k^{n_{s}-1}$. These consistency relations can be also derived as Ward identities with the appropriate Noether charges\cite{Hinter}\cite{Ward2}.

\section{Effective Field Theory of Inflation}
In this section, we introduce another approach to describe the primordial fluctuations, constructed by Cheung {\sl et al.}\cite{Effective} Using this method, we will be able to expand the Maldacena's theory.
\subsection{Set-up}
In this method, we use `unitary gauge', in which a Lagrangian is allowed to change under time diffeomorphism while invariant under spatial diffeomorphism. Assuming a flat FRW background, an action can be given by
\begin{equation}\label{eftaction}
S=\int d^4 x\sqrt{-g}\left[\frac{1}{2}R+\dot{H}(t)g^{00}-\left(3H^2(t)+\dot{H}(t)\right)+\cdots\right],
\end{equation} 
where the first three terms satisfy the Einstein's equations and $\cdots$ are terms which deviate from this background containing $\delta g^{00}$ or $\delta K_{\mu\nu}$ ($K_{\mu\nu}$: extrinsic curvature). Note that the first term is invariant under all diffeomorphism because $R$ is a scalar, but the second and the third terms change under time diffeomorphism. Especially as for $g^{00}$, when we take $t=:\tilde{t}+\pi(\tilde{t},\vx)$,
\begin{eqnarray}\label{eftg00}
g^{00}&=&\frac{\partial x^0}{\partial\tilde{x}^{\mu}}\frac{\partial x^0}{\partial\tilde{x}^{\nu}}\tilde{g}^{\mu\nu}\nonumber\\
&=&\bigl\{(1+\pidot)\delta_{\mu}^0+(\partial_{i}\pi)\delta_{\mu}^{i}\bigl\}\bigl\{(1+\pidot)\delta_{\nu}^0+(\partial_{i}\pi)\delta_{\nu}^{i}\bigl\}\tilde{g}^{\mu\nu}\nonumber\\
&=&(1+\pidot)^2\gtil^{00}+2(1+\pidot)(\parti\pi)\gtil^{0i}+(\parti\pi)(\partj\pi)\gtil^{ij},
\end{eqnarray}
where the dot means $d/d\tilde{t}$. In the EFT method, we identify the $\pi$, `Nambu-Goldstone boson', with a primordial scalar fluctuation, assuming that the spatial metric $\tilde{h}_{ij}$ has only tensor fluctuations. Therefore, it follows that by performing the time diffeomorphism $t=\ttil+\pi$, we have moved to the spatially-flat gauge (\ref{spatially}) in the Maldacena's theory, in which $\varphi$ corresponds to $\pi$.\vspace{3mm}

The readers may be worried about the broken gauge invariance, but it can be repaired by assigning to $\pi$ a transformation rule such as
\begin{equation}
\pi\to\tilde{\pi}(\ttil,\vx)=\pi(t,\vx)+\pi(\ttil,\vx),
\end{equation}
when $t\to\ttil=t-\pi(\ttil,\vx)$. Actually, this results in $t+\pi(t,\vx)=\ttil+\tilde{\pi}(\ttil,\vx)$; the form is invariant. This technique to restore the gauge invariance is called `St\"{u}ckelberg trick'.

\subsection{Relation to the Maldacena's theory}
In this subsection, we will confirm that the EFT method actually derives the Maldacena's theory. According to the ADM formalism, an inverse metric can be written as
\begin{equation}\label{adminvg}
\gtil^{\mu\nu}=
\frac{1}{N^2}\left(
\begin{array}{ccc}
-1&N^{i}\\
N^{j}&N^{2}\tilde{h}^{ij}-N^{i}N^{j}
\end{array}
\right).
\end{equation}
Recall that we assume $\tilde{h}_{ij}$ is the same as (\ref{spatially}). Substituting (\ref{adminvg}) and (\ref{eftg00}) into the leading terms in the action (\ref{eftaction}), and using $\sqrt{-g}=N\sqrt{h}$ and $R=R^{(3)}+N^{-2}(E_{ij}E^{ij}-E^2)$, it follows that
\begin{equation}
\begin{split}
S=\int d^4 \tilde{x}\,a^3 (\ttil)\biggl[\frac{1}{2}&NR^{(3)}+\frac{1}{2}N^{-1}(E_{ij}E^{ij}-E^2)\\
&+\frac{d}{dt}H(\ttil+\pi)\Bigl\{-N^{-1}(1+\pidot)^2+2N^{-1}(1+\pidot)(\parti\pi)N^{i}\\
&\qquad\qquad\qquad+N\tilde{h}^{ij}(\parti\pi)(\partj\pi)-N^{-1}N^{i}N^{j}(\parti\pi)(\partj\pi)\Bigl\}\\
&-N\left(3H^2 (\ttil+\pi)+\frac{d}{dt}H(\ttil+\pi)\right)\biggl],
\end{split}\label{eftaction2}
\end{equation}
where the dot means $d/d\ttil$. We can compute $R^{(3)}$ and $E_{ij}E^{ij}-E^2$ using $h_{ij}$ as (\ref{spatially}) (dropping the gamma's tilde):
\begin{equation}
R^{(3)}=-\frac{1}{4}a^{-2}(\ttil)\partl\gaij\partl\gaij+\mathcal{O}(\gamma^{3}),
\end{equation}
\begin{equation}
\begin{split}
E_{ij}E^{ij}-E^2=-6&H^2+4H\parti N^{i}\\
&+\frac{1}{2}\parti N^{j}\left(\parti N^{j}+\partj N^{i}\right)-(\parti N^{i})^2\\
&-\frac{1}{2}\left(\parti N^{j}+\partj N^{i}\right)\dot{\gamma}_{ij}\\
&+\frac{1}{4}\gadij\gadij-\frac{1}{2}\gadij\partl\gaij N^{l}\\
&+\partl\gaij N^{l}\parti N^{j}+\mathcal{O}(\gamma^3).
\end{split}
\end{equation}
In addition, be careful of $d H(\ttil+\pi)/dt$ in (\ref{eftaction2}), because there appear both $t$ and $\ttil$. We can rewrite it to the form using only $\ttil$:
\begin{equation}
\begin{split}
\frac{d}{dt}H(\ttil+\pi)&=\frac{d\ttil}{dt}\frac{d}{d\ttil}H(\ttil+\pi)\\
&=\left[1-\frac{d\ttil}{dt}\pidot\right]\dot{H}(\ttil+\pi)\\
&=\left[1-\Bigl\{1-\frac{d\ttil}{dt}\pidot\Bigl\}\pidot\right]\dot{H}(\ttil+\pi)\\
&=\cdots\\
&=\frac{\dot{H}(\ttil+\pi)}{1+\pidot}.
\end{split}
\end{equation}

In order to determine the lapse $N$ and the shift $N^{i}$, we have to solve the constraint equations
\begin{equation}\label{eftconstraint}
\frac{\delta S}{\delta N}=0,\quad\frac{\delta S}{\delta N^{j}}=0,
\end{equation}
which transform to
\begin{equation}
\begin{split}
0=R^{(3)}&-N^{-2}(E_{ij}E^{ij}-E^2)\\
&+2\frac{\dot{H}(\ttil+\pi)}{1+\pidot}\Bigl\{N^{-2}(1+\pidot)^2-2N^{-2}(1+\pidot)(\parti\pi)N^{i}\\
&\qquad\qquad\qquad\quad +\tilde{h}^{ij}(\parti\pi)(\partj\pi)+N^{-2}N^{i}N^{j}(\parti\pi)(\partj\pi)\Bigl\}\\
&-6H^2 (\ttil+\pi)-2\frac{\dot{H}(\ttil+\pi)}{1+\pidot}
\end{split}\label{efthami}
\end{equation}
and
\begin{equation}
\begin{split}
0=\nabla_{i}\bigl[&N^{-1}(E^{i}_{j}-\delta^{i}_{j}E)\bigl]\\
&+2\frac{\dot{H}(\ttil+\pi)}{1+\pidot}\Bigl\{N^{-1}(1+\pidot)(\partj\pi)-N^{-1}N^{i}(\parti\pi)(\partj\pi)\Bigl\}.
\end{split}\label{eftmome}
\end{equation}
Setting $N=:1+N_1$, and using a Taylor series for $H(\ttil+\pi)$, we can solve them to the first order:
\begin{eqnarray}
N_1&=&-\epsilon\,(-H\pi)\label{n1eft},\\
\parti N^{i}&=&\epsilon\frac{d}{d\ttil}(-H\pi).\label{nieft}
\end{eqnarray}
Note that $\epsilon=-\dot{H}/H^2$. By comparing (\ref{n1eft}) and (\ref{nieft}) with (\ref{n1mal}) and (\ref{nimal}), it follows that
\begin{equation}\label{zetapi}
\zetan=-H\pi.
\end{equation}
Actually, if we substitute $\pi=-\zetan/H$ into (\ref{eftaction2}), the same action as Maldacena's can be obtained after some integrations by parts. Note that we have just examined the leading terms in the EFT action (\ref{eftaction}). Therefore, if we consider the terms $\cdots$ in (\ref{eftaction}), we can expand the Maldacena's theory, which leads to the next section.

\section{Application of the EFT method}
In this section, we use the EFT method to introduce another scalar field $\sigma$, whose mass is the order of the Hubble parameter, into the Maldacena's theory. Such a model has been already improved by Chen and Wang as `quasi-single field inflation\cite{Quasi}', and the EFT approach to it has been also constructed by Noumi {\sl et al.}\cite{Noumi} The interesting point is that there occurs a $\zeta\sigma$ coupling, which corrects the power spectrum of $\zeta$. In this paper, by using the EFT method, we will see that a coupling with a graviton can also appear, {\sl `$\gamma\zeta\sigma$ coupling'}, and compute correlation functions $\langle\gamma\zeta\zeta\rangle$ and $\langle\gamma\gamma\zeta\zeta\rangle$ including this coupling in the soft-graviton limit.
Finally, we generalize this theory; we consider couplings of $\zeta$, $\sigma$ and $N$ soft-gravitons,  compute $\langle\gamma^{s_1}\cdots\gamma^{s_{N}}\zeta\zeta\rangle$ including these couplings, and discuss the several features.

\subsection{Review of `Quasi-single field inflation'}
Before discussing the EFT approach, let us review the quasi-single field inflation\cite{Quasi}.
In this theory, we introduce the polar coordinates as below; the inflaton moves along the tangential direction ($\theta$), while another scalar field moves along the radial direction ($\sigma$):
\begin{figure}[H]
  \centering
  \includegraphics[width=7cm]{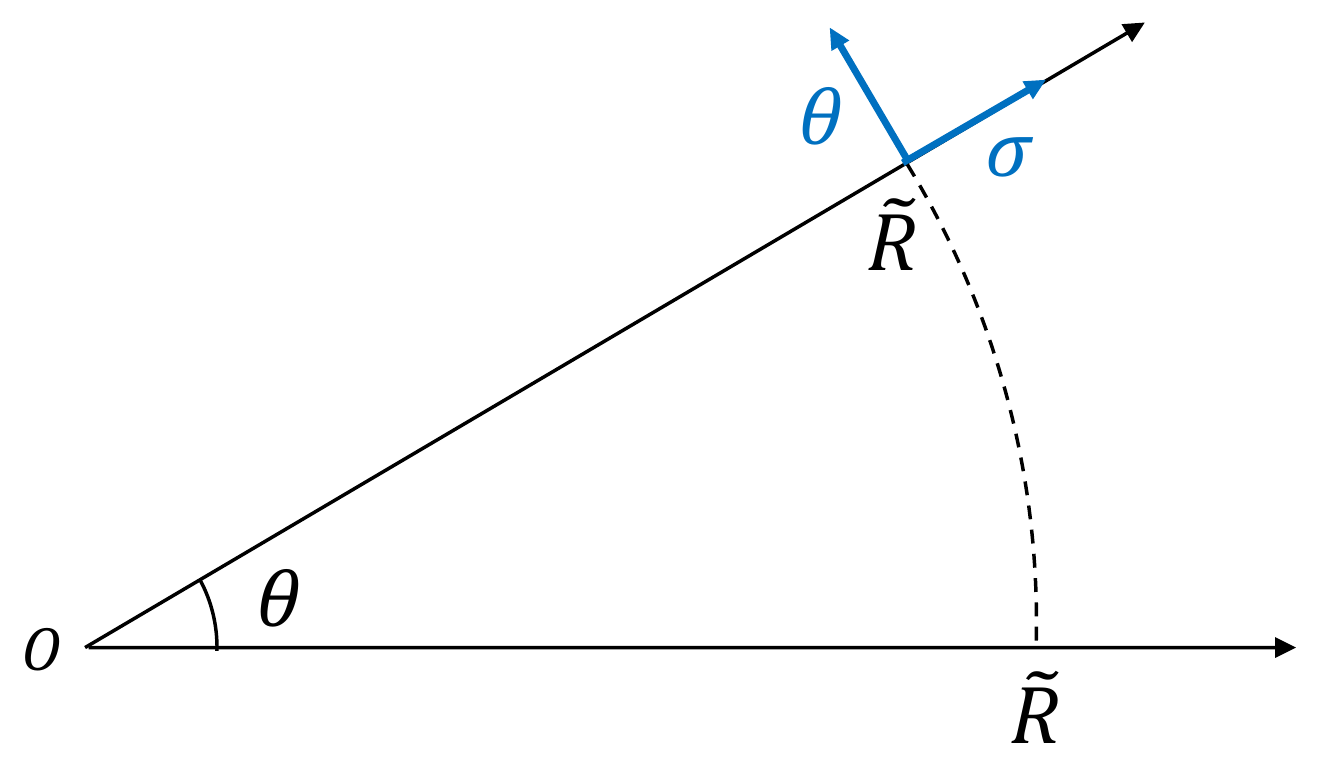}
  \caption{The polar coordinates in the quasi-single field inflation}
  \label{quasifigure}
\end{figure}
Then, the action of them can be described as
\begin{equation}
\begin{split}\label{Quasiaction}
S_{m}=\int d^4 x\sqrt{-g}\biggl[-\frac{1}{2}&(\tilde{R}+\chi)^2 g^{\mu\nu}\partial_{\mu}\theta\partial_{\nu}\theta\\
&-\frac{1}{2}g^{\mu\nu}\partial_{\mu}\chi\partial_{\nu}\chi-V_{\textrm{sr}}(\theta)-V(\chi)\biggl],
\end{split}
\end{equation}
where $V_{\textrm{sr}}(\theta)$ is a usual slow-roll potential, and $V(\chi)$ is a potential of another scalar field $\chi$.\vspace{2mm}

At first, let us consider the homogeneous and isotropic universe. Setting $\theta=\theta_0(t)$ and $\chi=\chi_0(t)=\chi_0$ (const.), and using the action (\ref{Quasiaction}) for the Einstein's equations, we obtain
\begin{eqnarray}
3H^2&=&\frac{1}{2}R^2\dot{\theta}_0^2+V(\chi_0)+V_{\textrm{sr}}(\theta_0),\label{relationQuasione}\\
-2\dot{H}&=&R^2\dot{\theta}_0^2\label{Quasihdot},\label{relationQuasitwo}
\end{eqnarray}
where $R\equiv\tilde{R}+\chi_0$. Then, using the action (\ref{Quasiaction}) to derive the equations of motion for $\theta_0(t)$ and $\chi_0(t)$, it follows that
\begin{eqnarray}
R^2\ddot{\theta}_0+3HR^2\dot{\theta}_0+V'_{\textrm{sr}}(\theta_0)&=&0,\\
R\dot{\theta}_0^2-V'(\chi_0)&=&0.\label{relationQuasithree}
\end{eqnarray}

Next, let us add the fluctuations to the inflaton field ($R\theta$) and $\chi$. We use `{\sl the uniform inflaton gauge}':
\begin{equation}
\begin{split}\label{uniforminflaton}
&\theta=\theta_0(t), \quad\chi=\chi_0+\sigma(t,\vx),\\
&h_{ij}=a^2(t)e^{2\zeta}\delta_{ij},
\end{split}
\end{equation}
where $\chi_0$ is constant.
Note that we neglect any gravitons for simplicity.\vspace{2mm}

From now on, we will use the ADM formalism again. Using the ADM metric (\ref{ADMmetric}), the action becomes
\begin{equation}\label{QuasiADMaction}
S=\frac{1}{2}\int d^4 x\sqrt{h}\left[N\left(R^{(3)}+2\mathcal{L}_{m}\right)+N^{-1}\left(E_{ij}E^{ij}-E^2\right)\right],
\end{equation}
where $S_{m}=:\int d^4 x\sqrt{-g}\,\mathcal{L}_{m}$ is (\ref{Quasiaction}).
Then, setting $N=:1+N_1$ and solving the constraint equations (\ref{constraints}) to the first order, we obtain
\begin{eqnarray}
N_1&=&\frac{\dot{\zeta}}{H},\label{Quasiconstone}\\
\parti N^{i}&=&-a^{-2}\frac{1}{H}\partial^2\zeta+\epsilon\left(\zetadot-\frac{2H}{R}\sigma\right).\label{Quasiconsttwo}
\end{eqnarray}
Note that $\epsilon\equiv-\dot{H}/H^2=R^2\dot{\theta}_0^2/2H^2$, which follows from (\ref{Quasihdot}).
If we set $\sigma=0$, these results are consistent with (\ref{Quasirefone}) and (\ref{Quasireftwo}) in the comoving gauge in the Maldacena's theory.\vspace{2mm}

Substituting (\ref{Quasiconstone}) and (\ref{Quasiconsttwo}) into the action (\ref{QuasiADMaction}) and performing some integrations by parts, we obtain the quadratic action $S_2$:
\begin{equation}
\begin{split}
S_2=\int d^4 x\biggl[&\epsilon \left\{a^3(t)\zetadot^2-a(t)(\partial\zeta)^2\right\}\\
&+\frac{1}{2}\left\{a^3(t)\dot{\sigma}^2-a(t)(\partial\sigma)^2-a^3(t)\left(V''(\chi_0)-\dot{\theta}_0^2\right)\sigma^2\right\}\\
&-2a^3(t)\frac{R\dot{\theta}_0^2}{H}\dot{\zeta}\,\sigma\biggl].
\end{split}
\end{equation}
The first line is the $\zeta$'s action, which is equivalent with (\ref{quadaction}).
The second line is the $\sigma$'s action, in which the last term is interpreted as the mass term; $m^2\equiv V''(\chi_0)-\dot{\theta}_0^2\sim\mathcal{O}(H^2)$.
Finally, the third line is the interaction term, which produces a $\zeta\sigma$ coupling, and corrects the power spectrum of $\zeta$.
We will discuss the explicit computation later, using the EFT method.

\subsection{Set-up}
We will use the EFT method from now on.
Firstly, we construct the action in the unitary gauge, which we studied in Section 3:
\begin{equation}\label{eftaction10}
S=S_{0}+S_{\sigma}+S_{I},
\end{equation}
where
\begin{equation}\label{eftaction0}
S_{0}=\int d^4 x\sqrt{-g}\left[\frac{1}{2}R+\dot{H}(t)g^{00}-\left(3H^2(t)+\dot{H}(t)\right)\right],
\end{equation}
\begin{equation}\label{sigmaaction}
S_{\sigma}=\int d^4 x\sqrt{-g}\left[-\frac{1}{2}g^{\mu\nu}\partial_{\mu}\sigma\partial_{\nu}\sigma-\frac{1}{2}m^2 \sigma^2 +\cdots \right],
\end{equation}
\begin{equation}\label{eftaction3}
S_{I}=\int d^4 x\sqrt{-g}\left[c_1 \delta g^{00}\sigma+c_2 (\delta g^{00})^2 \sigma+c_3 \delta g^{00}\sigma^2 \underline{+c_4 \delta g^{00}g^{\mu\nu}\partial_{\mu}\partial_{\nu}\sigma}\right].
\end{equation}

Firstly, $S_0$ shows the leading terms in the EFT action (\ref{eftaction}). Secondly, $S_{\sigma}$ is the $\sigma$'s action, assuming that $\sigma(t,\vx)$ is a real scalar field. Note that $m$ is the mass of $\sigma$, which is the order of the Hubble parameter, and $\cdots$ in (\ref{sigmaaction}) correspond to the other terms from $\sigma$'s potential $V(\sigma)$.\vspace{3mm}

Finally, $S_{I}$ shows the correction terms $\cdots$ in the EFT action (\ref{eftaction}), containing $\delta g^{00}(=g^{00}+1)$ and $\sigma$. The first three terms produce the terms which contain only $\zeta$ and $\sigma$ to the third order. In the unitary gauge, we can write much more terms such as $\delta g^{00}\partial_{\mu}\sigma\partial^{\mu}\sigma$ or $(\delta g^{00})^2 \partial^0 \sigma$, but we omit such other terms because they don't affect our computational results, as we will see later.
The fourth term, which is underlined in (\ref{eftaction3}), is the most important part; it produces a $\gamma\zeta\sigma$ coupling, assuming that $g^{\mu\nu}$ produces just one graviton. We assume that all of the coefficients $c_1$, $c_2$, $c_3$ and $c_4$ are constant.\vspace{3mm}

Then, we have to perform the time diffeomorphism $t=:\ttil+\pi(\ttil,\vx)$, substitute (\ref{eftg00}) for $g^{00}$ and $\delta g^{00}=g^{00}+1$ in (\ref{eftaction0}) and (\ref{eftaction3}), and solve the constraint equations (\ref{eftconstraint}). To the first order, the only change from Section 3 is to add the term `$+4c_1 \sigma$' to the right side in (\ref{efthami}). We can easily solve them (dropping any tildes):
\begin{eqnarray}
N_1&=&-\epsilon(-H\pi),\\
\parti N^{i}&=&\epsilon\frac{d}{dt}(-H\pi)+c_1\frac{\sigma}{H}.
\end{eqnarray}
Therefore, it follows that $\sigma$ changes the shift $N^{i}$. Then, substituting these solutions into the action (\ref{eftaction10}), and rewriting $\pi$ as $\pi=-\zeta/H$ which follows from (\ref{zetapi}) (dropping the subscript `$n$'), we can finally obtain the action including $\gamma$, $\zeta$ and $\sigma$.\vspace{3mm}

As for the quadratic action, $S_{\zeta\zeta}$ and $S_{\gamma\gamma}$ do not change from (\ref{quadaction}) and (\ref{quadactiongamma}) respectively. The new terms including $\sigma$ are
\begin{eqnarray}
S_{\sigma\sigma}&=&\int d^4 x\left[\frac{1}{2}a^3 (t)\sigdot^2-\frac{1}{2}a(t)(\partial\sigma)^2-\frac{1}{2}a^3(t)m^2\sigma^2 \right]\label{quadsigma},\\
S_{\zeta\sigma}&=&\int d^4 x\left[2\frac{c_1}{H}a^3(t)\zetadot\sigma\right].\label{quadzesig}
\end{eqnarray}

As for the cubic action, a lot of new terms occur such as $\zeta\sigma\sigma$ or $\zeta\zeta\sigma$ 
, but as mentioned before, we will just examine a $\gamma\zeta\sigma$ coupling which is produced from the fourth term in (\ref{eftaction3}):
\begin{equation}\label{cubicgzsig}
S_{\gamma\zeta\sigma}=\int d^4 x\left[-2\frac{c_4}{H}a(t)\zetadot\gaij\parti\partj\sigma\right].
\end{equation}\vspace{2mm}

\leftline{\underline{{\bf Relation to `Quasi-single field inflation'}}}\vspace{2mm}

Let us briefly check how this set-up relates to the quasi-single field inflation.
We use the uniform inflaton gauge (\ref{uniforminflaton}) again.
Then, the action (\ref{Quasiaction}) becomes
\begin{equation}
\begin{split}\label{actionrelationQuasi}
S_{m}=\int d^4 x\sqrt{-g}\biggl[-\frac{1}{2}&\left(R+\sigma\right)^2 g^{00}\,\dot{\theta}_0^2\\
&-\frac{1}{2}g^{\mu\nu}\partial_{\mu}\sigma\partial_{\nu}\sigma-V_{\textrm{sr}}(\theta_0)-V(\chi_0+\sigma)\biggl],
\end{split}
\end{equation}
where $R\equiv\tilde{R}+\chi_0$.
Then, expanding the potential $V(\chi)$ as
\begin{equation} 
V(\chi_0+\sigma)=V(\chi_0)+V'(\chi_0)\sigma+\frac{1}{2}V''(\chi_0)\sigma^2+\frac{1}{6}V'''(\chi_0)\sigma^3+\cdots,
\end{equation}
and substituting (\ref{relationQuasione}), (\ref{relationQuasitwo}) and (\ref{relationQuasithree}) into (\ref{actionrelationQuasi}), it becomes
\begin{equation}
\begin{split}
S_{m}=\int &d^4 x\sqrt{-g}\biggl[\dot{H}(t)g^{00}-\left(3H^2(t)+\dot{H}(t)\right)\\
&-\frac{1}{2}g^{\mu\nu}\partial_{\mu}\sigma\partial_{\nu}\sigma-\frac{1}{2}\left(V''(\chi_0)-\dot{\theta}_0^2\right)\sigma^2-\frac{1}{6}V'''(\chi_0)\sigma^3-\cdots\\
&\qquad\qquad-R\dot{\theta}_0^2\left(g^{00}+1\right)\sigma-\frac{1}{2}\dot{\theta}_0^2\left(g^{00}+1\right)\sigma^2\biggl].
\end{split}
\end{equation}
The first line corresponds to $S_0$ (\ref{eftaction0}) excluding the gravity term.
The second line corresponds to $S_\sigma$ (\ref{sigmaaction}), where $m^2\equiv V''(\chi_0)-\dot{\theta}_0^2$.
Finally, the third line corresponds to $S_{I}$ (\ref{eftaction3}), setting
\begin{equation}
c_1=-R\dot{\theta}_0^2,\quad c_2=0,\quad c_3=-\frac{1}{2}\dot{\theta}_0^2,\quad c_4=0.
\end{equation}
Therefore, it follows that the quasi-single field inflation is just an example produced by the EFT method.

\subsection{Quantization of $\sigma$}
The way to quantize $\sigma$ is just the same as $\zeta$. Firstly, we rewrite the quadratic action of $\sigma$ (\ref{quadsigma}) by using the conformal time $\eta$ and $a(t)=-1/\eta H$, and derive the equation of motion:
\begin{equation}\label{sigmaeom}
\sigma''-\frac{2}{\eta}\sigma'-\partial^2 \sigma+\frac{m^2}{\eta^2 H^2}\sigma=0,
\end{equation}
where $'$ means $d/d\eta$. Then, we represent $\sigma$ using the Fourier transform:
\begin{equation}\label{sigmafour}
\sigma(\eta,\vx)=\int\frac{d^3{k}}{(2\pi)^{3}}\left[v_{k}(\eta)a_{\scriptsize{\vk}}e^{i\vk\cdot\vx}+v_{k}^{*}(\eta)a_{\scriptsize{\vk}}^{\dagger}e^{-i\vk\cdot\vx}\right],
\end{equation}
where $v_{k}$ and $v_{k}^{*}$ are the solutions of E.O.M (\ref{sigmaeom}), and $a_{\vk}$ and $a_{\vk}^{\dagger}$ are the annihilation and creation operators respectively. The solutions to (\ref{sigmaeom}) are already known (for example, refer to \cite{Higuchi} by Higuchi):
\begin{eqnarray}
v_{k}(\eta)&=&-i e^{i(\nu+\frac{1}{2})\frac{\pi}{2}}\frac{\sqrt{\pi}}{2}H(-\eta)^{3/2}\hankel(-k\eta)\label{sigmavk},\\
v_{k}^{*}(\eta)&=&i e^{-i(\nu^{*}+\frac{1}{2})\frac{\pi}{2}}\frac{\sqrt{\pi}}{2}H(-\eta)^{3/2}\hankelc(-k\eta)\label{sigmavkc},
\end{eqnarray}
where
\begin{equation}
\nu\equiv\sqrt{\frac{9}{4}-\frac{m^2}{H^2}},
\end{equation}
and $\hankel$ is a Hankel function of the first kind. Using (\ref{sigmafour}) with (\ref{sigmavk}) and (\ref{sigmavkc}), we can compute the two point function of $\sigma$:
\begin{table}[H]
  \centering
  \caption{Two point function of $\sigma$}
  \begin{tabular}{|l|}\hline
  \\
  $\displaystyle\braket{0|\sigma(\eta,\vx)\sigma(\eta',\vx')|0}=\int \frac{d^{3}k}{(2\pi)^{3}}e^{i\vk\cdot(\vx-\vx')}\left[e^{-\pi\textrm{Im}(\nu)}\frac{\pi}{4}H^2 (\eta\eta')^{\frac{3}{2}}\hankel(-k\eta)\hankelc(-k\eta')\right]$\\
  \\ \hline
  \end{tabular}\label{tipssigma}
\end{table}

\subsection{Computation of $\langle\zeta\zeta\rangle$, $\langle\gamma\zeta\zeta\rangle$ and $\langle\gamma\gamma\zeta\zeta\rangle$ with $\sigma$}
We are interested in how $\sigma$ affects the correlation functions of $\zeta$ and $\gamma$. In this paper, we will compute $\zetzet$, $\gamzetzet$ and $\gamgamzetzet$ affected by $\sigma$ as the diagrams below:
\begin{figure}[H]
  \centering
  \includegraphics[width=7cm]{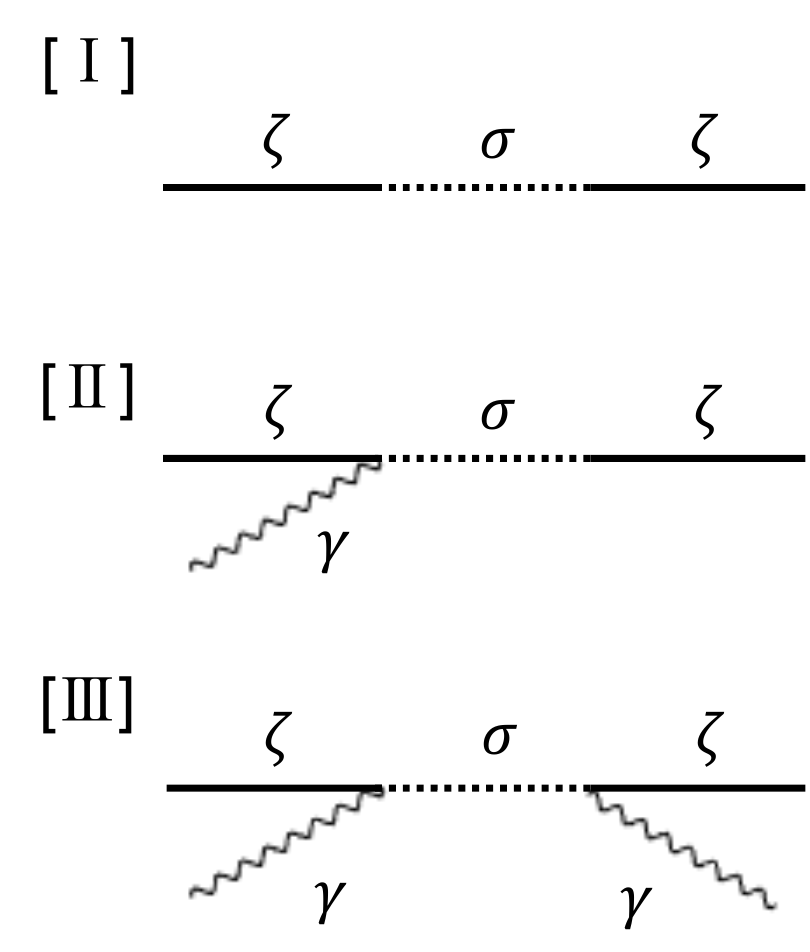}
  \caption{The diagrams of [I]\,$\zetzet$, [I\hspace{-.1em}I]\,$\gamzetzet$ and [I\hspace{-.1em}I\hspace{-.1em}I]\,$\gamgamzetzet$}
  \label{diagrams}
\end{figure}
The diagram [I] was already computed by Chen and Wang in their quasi-single field model\cite{chenwang}. In the following computation, we will generalize their method to compute also [I\hspace{-.1em}I] and [I\hspace{-.1em}I\hspace{-.1em}I].\vspace{3mm}

Note that these diagrams have two vertices respectively, so using the in-in formalism (\ref{ininformula}), the correlation function of $\mathcal{O}(t,\vx)$ ($=$$\zeta\zeta$, $\gamma\zeta\zeta$ and $\gamma\gamma\zeta\zeta$) is given by
\begin{equation}
\begin{split}
&\braket{0(t)|\mathcal{O}(t,\vx)|0(t)}\\
&\qquad=\int_{-\infty}^{t}dt_1\int_{-\infty}^{t}dt_2 \braket{0|H_{I}(t_1)\mathcal{O}(t,\vx)H_{I}(t_2)|0}\\
&\qquad\qquad-2\textrm{Re}\left[\int_{-\infty}^{t}dt_1\int_{-\infty}^{t_1}dt_2 \braket{0|\mathcal{O}(t,\vx)H_{I}(t_1)H_{I}(t_2)|0}\right].
\end{split}\label{strategy}
\end{equation}

\leftline{\underline{{\bf [I] Computation of} ${\bm \zetzet}$}}\vspace{2mm}
The interaction Hamiltonian $H_{I}(t)=-L_{I}(t)$ is given by the action (\ref{quadzesig}):
\begin{equation}
H_{I}(t)=-2\frac{c_1}{H}a^3 (t)\int d^3 x \,\zetadot\sigma.
\end{equation}

Using the two point functions given by Table \ref{tipszg} and \ref{tipssigma} into (\ref{strategy}) in the momentum space,
\begin{equation}
\begin{split}
&\langle\zeta(\vk_1)\zeta(\vk_2)\rangle\\
&=(2\pi)^3\delta^3 (\vk_1+\vk_2)\\
&\qquad\times\left(-2\frac{c_1}{H}\right)^2\left(\biarii\right)\left(\biari\right)\sigmatip\,\underline{\times\,2!}\\
&\qquad\times\Biggl[\int_{-\infty}^{0}\frac{d\eta_1}{\eta_1^4H^4}\int_{-\infty}^{0}\frac{d\eta_2}{\eta_2^4H^4}\eta_1^2e^{-ik_1\eta_1}\eta_2^2e^{ik_2\eta_2}(\eta_1\eta_2)^{\frac{3}{2}}\hankel(-k_2\eta_1)\hankelc(-k_2\eta_2)\\
&\qquad-2\textrm{Re}\left[\int_{-\infty}^{0}\frac{d\eta_1}{\eta_1^4H^4}\int_{-\infty}^{\eta_1}\frac{d\eta_2}{\eta_2^4H^4}\eta_1^2e^{ik_1\eta_1}\eta_2^2e^{ik_2\eta_2}(\eta_1\eta_2)^{\frac{3}{2}}\hankel(-k_2\eta_1)\hankelc(-k_2\eta_2)\right]\Biggl].
\end{split}\label{zetzetmiddle}
\end{equation}
Note that we evaluate the correlation functions at $t\to\infty$, which corresponds to $\eta\to0$.
Then, the underlined factor `$\times\,2!$' is a combinatorial factor for the diagram [I] in Figure \ref{diagrams}.\vspace{3mm}

Then, setting $x\equiv-k_2\eta_1$ and $y\equiv-k_2\eta_2$, (\ref{zetzetmiddle}) becomes
\begin{equation}
\begin{split}
&\langle\zeta(\vk_1)\zeta(\vk_2)\rangle\\
&=(2\pi)^3\delta^3 (\vk_1+\vk_2)\ P_{\zeta}(k_2)\ \frac{\pi c_1^2}{2\epsilon H^4}e^{-\pi \textrm{Im}(\nu)}\\
&\qquad\times\Biggl[\left|\int_{0}^{\infty}dx\ x^{-\frac{1}{2}}e^{ix}\hankel(x)\right|^2\\
&\qquad\quad-2\textrm{Re}\left[\int_{0}^{\infty}dx\ x^{-\frac{1}{2}}e^{-ix}\hankel(x)\int_{x}^{\infty}dy\ y^{-\frac{1}{2}}e^{-iy}\hankelc(y)\right]\Biggl],
\end{split}\label{zzresult}
\end{equation}
where $P_{\zeta}(k)=H^2/4\epsilon k^3$ is the power spectrum. These integrations will be performed later.\vspace{3mm}

\leftline{\underline{{\bf [I\hspace{-.1em}I] Computation of} ${\bm \gamzetzet}$ {\bf in the soft-graviton limit} ${\bm \vq\to0}$}}\vspace{2mm}

The interaction Hamiltonian is given by (\ref{quadzesig}) and (\ref{cubicgzsig}):
\begin{equation}
H_{I}(t)=-2\frac{c_1}{H}a^3 (t)\int d^3 x \,\zetadot\sigma+2\frac{c_4}{H}a(t)\int d^3 x \,\zetadot\gaij\parti\partj\sigma.
\end{equation}

Using the two point functions given by Table \ref{tipszg} and \ref{tipssigma} into (\ref{strategy}) in the momentum space,
\begin{equation}
\begin{split}
&\langle\gamma^{s}(\vq)\zeta(\vk_1)\zeta(\vk_2)\rangle\\
&=(2\pi)^3\delta^3 (\vq+\vk_1+\vk_2)\ \epsilon_{ij}^{s}(\vq)(-i)^2 (k_2)_{i}(k_2)_{j}\\
&\qquad\times\left(-4\frac{c_1 c_4}{H^2}\right)\binashiq\left(\biarii\right)\left(\biari\right)\sigmatip\,\underline{\times\,2!}\\
&\qquad\times\Biggl[2\textrm{Re}\left[\int_{-\infty}^{0}\frac{d\eta_1}{\eta_1^4H^4}\int_{-\infty}^{0}\frac{d\eta_2}{\eta_2^2H^2}\eta_1^2e^{-ik_1\eta_1}\eta_2^2e^{ik_2\eta_2}(\eta_1\eta_2)^{\frac{3}{2}}\hankel(-k_2\eta_1)\hankelc(-k_2\eta_2)\right]\\
&\quad\qquad-2\textrm{Re}\left[\int_{-\infty}^{0}\frac{d\eta_1}{\eta_1^4H^4}\int_{-\infty}^{\eta_1}\frac{d\eta_2}{\eta_2^2H^2}\eta_1^2e^{ik_1\eta_1}\eta_2^2e^{ik_2\eta_2}(\eta_1\eta_2)^{\frac{3}{2}}\hankel(-k_2\eta_1)\hankelc(-k_2\eta_2)\right]\\
&\quad\qquad-2\textrm{Re}\left[\int_{-\infty}^{0}\frac{d\eta_1}{\eta_1^2H^2}\int_{-\infty}^{\eta_1}\frac{d\eta_2}{\eta_2^4H^4}\eta_1^2e^{ik_1\eta_1}\eta_2^2e^{ik_2\eta_2}(\eta_1\eta_2)^{\frac{3}{2}}\hankel(-k_2\eta_1)\hankelc(-k_2\eta_2)\right]\Biggl].
\end{split}\label{gzzmiddle}
\end{equation}
Note that we are taking the $\vq\to0$ limit, so $q$ does not appear in the integrals.
Then, the underlined factor `$\times\,2!$' is a combinatorial factor for the diagram [I\hspace{-.1em}I] in Figure \ref{diagrams}.\vspace{3mm}

Setting $x\equiv-k_2\eta_1$ and $y\equiv-k_2\eta_2$, (\ref{gzzmiddle}) becomes
\begin{equation}
\begin{split}
&\langle\gamma^{s}(\vq)\zeta(\vk_1)\zeta(\vk_2)\rangle\\
&=(2\pi)^3\delta^3 (\vq+\vk_1+\vk_2)\ \epsilon_{ij}^{s}(\vq)\frac{(k_2)_{i}(k_2)_{j}}{(k_2)^2}P_{\gamma}(q)P_{\zeta}(k_2) \frac{\pi c_1 c_4}{\epsilon H^2}e^{-\pi \textrm{Im}(\nu)}\\
&\qquad\times\Biggl[\textrm{Re}\left[\int_{0}^{\infty}dx\ x^{-\frac{1}{2}}e^{ix}\hankel(x)\int_{0}^{\infty}dy\ y^{\frac{3}{2}}e^{-iy}\hankelc(y)\right]\\
&\qquad\qquad-\textrm{Re}\left[\int_{0}^{\infty}dx\ x^{-\frac{1}{2}}e^{-ix}\hankel(x)\int_{x}^{\infty}dy\ y^{\frac{3}{2}}e^{-iy}\hankelc(y)\right]\\
&\qquad\qquad-\textrm{Re}\left[\int_{0}^{\infty}dx\ x^{\frac{3}{2}}e^{-ix}\hankel(x)\int_{x}^{\infty}dy\ y^{-\frac{1}{2}}e^{-iy}\hankelc(y)\right]\Biggl],
\end{split}\label{gzzresult}
\end{equation}
where $P_{\gamma}(q)=H^2/q^3$ and $P_{\zeta}(k)=H^2/4\epsilon k^3$ are the power spectra.\vspace{3mm}

\leftline{\underline{{\bf [I\hspace{-.1em}I\hspace{-.1em}I] Computation of} ${\bm \gamgamzetzet}$ {\bf in the soft-graviton limit} ${\bm \vq_1,\vq_2\to0}$}}\vspace{2mm}
The interaction Hamiltonian is given only by (\ref{cubicgzsig}):
\begin{equation}
H_{I}(t)=2\frac{c_4}{H}a(t)\int d^3 x \,\zetadot\gaij\parti\partj\sigma.
\end{equation}

Using the two point functions given by Table \ref{tipszg} and \ref{tipssigma} into (\ref{strategy}) in the momentum space,
\begin{equation}
\begin{split}
&\langle\gamma^{s_1}(\vq_1)\gamma^{s_2}(\vq_2)\zeta(\vk_1)\zeta(\vk_2)\rangle\\
&=(2\pi)^3\delta^3 (\vq_1+\vq_2+\vk_1+\vk_2)\ \epsilon_{ij}^{s_1}(\vq_1)(-i)^2 (k_1)_{i}(k_1)_{j}\ \epsilon_{\alpha\beta}^{s_2}(\vq_2)(-i)^2 (k_2)_{\alpha}(k_2)_{\beta}\\
&\qquad\times\left(2\frac{c_4}{H}\right)^2\frac{H^2}{(q_1)^3}\frac{H^2}{(q_2)^3}\left(\biarii\right)\left(\biari\right)\sigmatip\,\underline{\times\,2!\times2!}\\
&\qquad\times\Biggl[\int_{-\infty}^{0}\frac{d\eta_1}{\eta_1^2H^2}\int_{-\infty}^{0}\frac{d\eta_2}{\eta_2^2H^2}\eta_1^2e^{-ik_1\eta_1}\eta_2^2e^{ik_2\eta_2}(\eta_1\eta_2)^{\frac{3}{2}}\hankel(-k_2\eta_1)\hankelc(-k_2\eta_2)\\
&\qquad-2\textrm{Re}\left[\int_{-\infty}^{0}\frac{d\eta_1}{\eta_1^2H^2}\int_{-\infty}^{\eta_1}\frac{d\eta_2}{\eta_2^2H^2}\eta_1^2e^{ik_1\eta_1}\eta_2^2e^{ik_2\eta_2}(\eta_1\eta_2)^{\frac{3}{2}}\hankel(-k_2\eta_1)\hankelc(-k_2\eta_2)\right]\Biggl].
\end{split}\label{ggzzmiddle}
\end{equation}
Note that we are taking the double-soft limit $\vq_1, \vq_2 \to 0$.
Then, the underlined factor `$\times\,2!\times2!$' is a combinatorial factor for the diagram [I\hspace{-.1em}I\hspace{-.1em}I] in Figure \ref{diagrams}.\vspace{3mm}

Setting $x\equiv-k_2\eta_1$ and $y\equiv-k_2\eta_2$, (\ref{ggzzmiddle}) becomes
\begin{equation}
\begin{split}
&\langle\gamma^{s_1}(\vq_1)\gamma^{s_2}(\vq_2)\zeta(\vk_1)\zeta(\vk_2)\rangle\\
&=(2\pi)^3\delta^3 (\vq_1+\vq_2+\vk_1+\vk_2)\\
&\qquad\times\epsilon_{ij}^{s_1}(\vq_1)\frac{(k_2)_{i}(k_2)_{j}}{(k_2)^2}\epsilon_{\alpha\beta}^{s_2}(\vq_2)\frac{(k_2)_{\alpha}(k_2)_{\beta}}{(k_2)^2}P_{\gamma}(q_1)P_{\gamma}(q_2)P_{\zeta}(k_2)\frac{\pi c_4^2}{\epsilon}e^{-\pi \textrm{Im}(\nu)}\\
&\qquad\times\Biggl[\left|\int_{0}^{\infty}dx\ x^{\frac{3}{2}}e^{ix}\hankel(x)\right|^2\\
&\qquad\qquad-2\textrm{Re}\left[\int_{0}^{\infty}dx\ x^{\frac{3}{2}}e^{-ix}\hankel(x)\int_{x}^{\infty}dy\ y^{\frac{3}{2}}e^{-iy}\hankelc(y)\right]\Biggl],
\end{split}\label{ggzzresult}
\end{equation}
where $P_{\gamma}(q)=H^2/q^3$ and $P_{\zeta}(k)=H^2/4\epsilon k^3$ are the power spectra.

\subsubsection{Computation of the integrals}
We have to compute the integrals in (\ref{zzresult}), (\ref{gzzresult}) and (\ref{ggzzresult}), which are
\begin{equation}\label{easyint}
\int_{0}^{\infty}dx\ x^{l} e^{ix}\hankelm(x),
\end{equation}
and
\begin{equation}\label{diffint}
\int_{0}^{\infty}dx\ x^{m} e^{-ix}\hankelm(x)\int_{x}^{\infty}dy\ y^{l} e^{-iy}\hankelmc(y).
\end{equation}
Note that $l$ and $m$ are $-1/2$ or $3/2$, and $\nu=:i\mu=i\sqrt{m^2/H^2-9/4}$\footnote{Be careful not to confuse $m$ appeared in (\ref{diffint}) with the mass of $\sigma$.}.
From now on, we assume that $\mu$ is real, which means $m\geq3H/2$.\vspace{3mm}

\leftline{\underline{{\bf Computation of (\ref{easyint})}}}\vspace{2mm}
 
 Firstly, we perform the indefinite integration by Mathematica 11:
 \begin{equation}
 \begin{split}
 \int &dx\ x^{l} e^{ix}\hankelm(x)\\
 &=x^{1+l-i\mu}\frac{-2^{i\mu}\csch(\pi\mu)}{(1+l-i\mu)\Gamma(1-i\mu)}\,{}_2F_2(a_1, a_1+\frac{1}{2}+l; 2a_1, a_1+\frac{3}{2}+l; z)\\
 &+x^{1+l+i\mu}\frac{2^{-i\mu}(1+\coth(\pi\mu))}{(1+l+i\mu)\Gamma(1+i\mu)}\,{}_2F_2(a_2, a_2+\frac{1}{2}+l; 2a_2, a_2+\frac{3}{2}+l; z),
 \end{split}\label{indef1}
 \end{equation}
where $a_1\equiv1/2-i\mu$, $a_2\equiv1/2+i\mu$ and $z\equiv2ix$. Note that ${}_{p}F_{q}$ is a generalized hypergeometric function:
 \begin{equation}
 {}_{p}F_{q}(a_1,\cdots,a_{p};b_1,\cdots,b_{q};z)=\sum_{n=0}^{\infty}\frac{(a_1)_{n}\cdots(a_{p})_{n}}{(b_1)_{n}\cdots(b_{q})_{n}}\frac{z^{n}}{n!},
 \end{equation}
 in which $(a)_{n}\equiv a(a+1)\cdots(a+n-1)$ when $n\geq1$ and $(a)_0=1$. Notice that $\csch(\pi\mu)=1/\sinh(\pi\mu)$.\vspace{3mm}
 
When $x=0$, (\ref{indef1}) vanishes.

When $x\to\infty$, we have to examine the asymptotic behavior of $\pfq$. The leading terms can be obtained by Mathematica 11:

(the command is `$\textrm{Series}[\pfq(\cdots; z),\{z,\infty,0\}]$')
\begin{equation}
\begin{split}
&\lim_{z\to\infty}\pfq(a, a+\frac{1}{2}+l; 2a, a+\frac{3}{2}+l; z)\\
&=z^{-\frac{1}{2}-a-l}(-1)^{\frac{3}{2}-a-l}2^{-2+2a}\frac{(1+2a+2l)}{\sqrt{\pi}}\frac{\Gamma(\frac{1}{2}+a)\Gamma(-\frac{1}{2}-l)\Gamma(\frac{1}{2}+a+l)}{\Gamma(-\frac{1}{2}+a-l)}\\
&\qquad+e^{z}z^{-1-a}(\cdots)+z^{-a}(\cdots),
\end{split}\label{asymp1}
\end{equation}
where both $(\cdots)$ are functions of $a$ and $l$. The term $e^{z}z^{-1-a}(\cdots)$ can be eliminated using the $i\epsilon$ prescription. In addition, the terms $z^{-a_1}(\cdots)$ and $z^{-a_2}(\cdots)$ are canceled with each other in (\ref{indef1}) (see Appendix \ref{app1}). Therefore, we need to consider only the first term in (\ref{asymp1}). Substituting it into (\ref{indef1}) and performing some calculations (see Appendix \ref{app2}), the result below can be obtained:
\begin{table}[H]
  \centering
  \caption{The result of (\ref{easyint})}
  \begin{tabular}{|l|}\hline
  \\
  $\displaystyle\int_{0}^{\infty}dx\ x^{l} e^{ix}\hankelm(x)=e^{\frac{\pi\mu}{2}}\frac{(i/2)^{l}}{\sqrt\pi}\frac{\Gamma(1+l-i\mu)\Gamma(1+l+i\mu)}{\Gamma(l+\frac{3}{2})}$\\
  \\ \hline
  \end{tabular}\label{resulteasy}
\end{table}

\leftline{\underline{{\bf Computation of (\ref{diffint})}}}\vspace{2mm}
We use the `resummation' trick\cite{chenwang}. Firstly, we rewrite (\ref{diffint}) using the definition of the Hankel function as
\begin{equation}
\begin{split}
\int_{0}^{\infty}dx\ &x^{m} e^{-ix}\hankelm(x)\,\mathcal{I}(x)\\
&=\left[1+\coth(\pi\mu)\right]\int_{0}^{\infty}dx\ x^{m}e^{-ix}\bessel(x)\,\mathcal{I}(x)\\
&\qquad-\csch(\pi\mu)\int_{0}^{\infty}dx\ x^{m}e^{-ix}\besselc(x)\,\mathcal{I}(x),
\end{split}\label{hankelmid}
\end{equation}
where
\begin{equation}
\mathcal{I}(x)\equiv\int_{x}^{\infty}dy\ y^{l} e^{-iy}\hankelmc(y).
\end{equation}

Then, using the series expansions
\begin{equation}
\bessel(x)=\sum_{n=0}^{\infty}a_{2n}x^{2n}\qquad\textrm{where}\quad a_{2n}=x^{i\mu}\,2^{-i\mu}\frac{(-1)^{n}\,2^{-2n}}{n!\,\Gamma(1+n+i\mu)},
\end{equation}
\begin{equation}
e^{-ix}=\sum_{n=0}^{\infty}b_{n}x^{n}\qquad\textrm{where}\quad b_{n}=\frac{(-i)^{n}}{n!},
\end{equation}
the function appeared in (\ref{hankelmid}) becomes
\begin{eqnarray}
x^{m}e^{-ix}\bessel(x)&=&\sum_{n=0}^{\infty}\left(\sum_{k=0}^{n}a_{2k}b_{n-2k}\right)x^{n+m}\nonumber\\
&=&\sum_{n=0}^{\infty}\frac{(-i)^{n}\,2^{n+i\mu}\Gamma(\frac{1}{2}+n+i\mu)}{\sqrt{\pi}\,n!\,\Gamma(1+n+2i\mu)}x^{n+m+i\mu}\label{suiko1}\\
&=:&\sum_{n=0}^{\infty}\ \cnp\ x^{n+m+i\mu}.\nonumber
\end{eqnarray}
Note that the summation in the first line above is performed by Mathematica 11. Similarly,
\begin{eqnarray}
x^{m}e^{-ix}\besselc(x)&=&\sum_{n=0}^{\infty}\frac{(-i)^{n}\,2^{n-i\mu}\Gamma(\frac{1}{2}+n-i\mu)}{\sqrt{\pi}\,n!\,\Gamma(1+n-2i\mu)}x^{n+m-i\mu}\label{suiko2}\\
&=:&\sum_{n=0}^{\infty}\ \cnm\ x^{n+m-i\mu}.\nonumber
\end{eqnarray}

Substituting (\ref{suiko1}) and (\ref{suiko2}) into (\ref{hankelmid}), it follows that
\begin{equation}
\begin{split}
\int_{0}^{\infty}dx\ x^{m} e^{-ix}&\hankelm(x)\,\mathcal{I}(x)\\
&=\left[1+\coth(\pi\mu)\right]\sum_{n=0}^{\infty} \cnp\int_{0}^{\infty}dx\ x^{n+m+i\mu}\ \mathcal{I}(x)\\
&\qquad-\csch(\pi\mu)\sum_{n=0}^{\infty}\cnm\int_{0}^{\infty}dx\ x^{n+m-i\mu}\ \mathcal{I}(x).
\end{split}\label{tochu10}
\end{equation}

Next, we evaluate the definite integral $\mathcal{I}(x)$ by taking the complex conjugate of the results in Table \ref{resulteasy} and (\ref{indef1}):
\begin{equation}
\begin{split}
\mathcal{I}(x)&\equiv\int_{x}^{\infty}dy\ y^{l} e^{-iy}\hankelmc(y)\\
&=e^{\frac{\pi\mu}{2}}\frac{(-i/2)^{l}}{\sqrt\pi}\frac{\Gamma(1+l+i\mu)\Gamma(1+l-i\mu)}{\Gamma(l+\frac{3}{2})}\\
&\quad+x^{1+l+i\mu}\frac{2^{-i\mu}\,\csch(\pi\mu)}{(1+l+i\mu)\Gamma(1+i\mu)}\,{}_2F_2(a_2, a_2+\frac{1}{2}+l; 2a_2, a_2+\frac{3}{2}+l; -2ix)\\
&\quad-x^{1+l-i\mu}\frac{2^{i\mu}(1+\coth(\pi\mu))}{(1+l-i\mu)\Gamma(1-i\mu)}\,{}_2F_2(a_1, a_1+\frac{1}{2}+l; 2a_1, a_1+\frac{3}{2}+l; -2ix).
\end{split}
\end{equation}
Using this,
\begin{equation}
\begin{split}
\int_{0}^{\infty}&dx\ x^{n+m+i\mu}\ \mathcal{I}(x)\\
&=e^{\frac{\pi\mu}{2}}\frac{(-i/2)^{l}}{\sqrt\pi}\frac{\Gamma(1+l+i\mu)\Gamma(1+l-i\mu)}{\Gamma(l+\frac{3}{2})}\left.\frac{x^{1+m+n+i\mu}}{1+m+n+i\mu}\right|_{x\to\infty}\\
&\quad+\frac{2^{-i\mu}\,\csch(\pi\mu)}{(1+l+i\mu)\Gamma(1+i\mu)}\int_{0}^{\infty}dx\ x^{1+l+m+n+2i\mu}\pfq(a_2,\cdots; -2ix)\\
&\quad-\frac{2^{i\mu}(1+\coth(\pi\mu))}{(1+l-i\mu)\Gamma(1-i\mu)}\int_{0}^{\infty}dx\ x^{1+l+m+n}\pfq(a_1,\cdots; -2ix).
\end{split}\label{tochu}
\end{equation}

In order to compute it, we firstly evaluate the integrals in (\ref{tochu}). The indefinite integral is
\begin{equation}
\begin{split}
\int dx\ &x^{p+n}\pfq(a, a+\frac{1}{2}+l; 2a, a+\frac{3}{2}+l; -2ix)\\
&=\frac{(1+2a+2l)\,x^{1+p+n}\,\pfq(a, 1+p+n; 2a, 2+p+n; -2ix)}{(1+p+n)(-1+2a-2p+2l-2n)}\\
&\quad+\frac{2x^{1+p+n}\,\pfq(a, a+\frac{1}{2}+l; 2a, a+\frac{3}{2}+l; -2ix)}{1-2a+2p-2l+2n},
\end{split}\label{tochu2}
\end{equation}
where $p\equiv1+l+m+2i\mu$ for the second term in (\ref{tochu}) while $p\equiv1+l+m$ for the third term in (\ref{tochu}).\vspace{3mm}

When $x=0$, (\ref{tochu2}) vanishes.

When $x\to\infty$, we have to evaluate the asymptotic behavior of the hypergeometric functions as before. We can use (\ref{asymp1}) for the second term in (\ref{tochu2}), and after some calulations (see Appendix \ref{app3}), the first term in (\ref{tochu}) is eliminated. In other words, the second term in (\ref{tochu2}) eliminates the first term in (\ref{tochu}).\vspace{3mm}

As for the first term in (\ref{tochu2}), the asymptotic behavior is
\begin{equation}
\begin{split}
&\lim_{z\to\infty}\pfq(a, 1+p+n; 2a, 2+p+n; z)\\
&=z^{-1-p-n}(-1)^{1-p-n}2^{-1+2a}\frac{(1+p+n)}{\sqrt{\pi}}\frac{\Gamma(\frac{1}{2}+a)\Gamma(-1+a-p-n)\Gamma(1+p+n)}{\Gamma(-1+2a-p-n)}\\
&\qquad+e^{z}z^{-1-a}(\cdots)+z^{-a}(\cdots),
\end{split}\label{asymp2}
\end{equation}
in which we can neglect the second and the third terms for the same reason as (\ref{asymp1}).\vspace{3mm}

When substituting (\ref{asymp2}) into (\ref{tochu2}), and then substituting (\ref{tochu2}) into (\ref{tochu}), the second term in (\ref{tochu}) vanishes. This is because the denominator of (\ref{asymp2}) becomes
\begin{equation}
\Gamma(-1+2a_2-p-n)=\Gamma(-(1+l+m+n)),
\end{equation}
which is divergent since $1+l+m+n$ is an integer greater than or equal to zero (recall that $l$ and $m$ are $-1/2$ or $3/2$). \vspace{3mm}

Finally, only the third term in (\ref{tochu}) remains, and just by using (\ref{asymp2}), we obtain
\begin{equation}
\begin{split}
\cnp&\int_{0}^{\infty}dx\ x^{n+m+i\mu}\ \mathcal{I}(x)\\
&=\frac{(-1)^{-2(l+m)}(2i)^{-2-l-m}}{\pi\,\sinh(\pi\mu)}\frac{e^{\pi\mu}}{1+m+n+i\mu}\\
&\qquad\quad\times(-1)^{n}\frac{\Gamma(2+l+m+n)}{\Gamma(n+1)}\frac{\Gamma(\frac{1}{2}+n+i\mu)\Gamma(-\frac{3}{2}-l-m-n-i\mu)}{\Gamma(1+n+2i\mu)\Gamma(-1-l-m-n-2i\mu)}.
\end{split}\label{tochup}
\end{equation}

Similarly, we can compute
\begin{equation}
\begin{split}
\cnm&\int_{0}^{\infty}dx\ x^{n+m-i\mu}\ \mathcal{I}(x)\\
&=\frac{(-1)^{-2(l+m)}(2i)^{-2-l-m}}{\pi\,\sinh(\pi\mu)}\frac{-1}{1+m+n-i\mu}\\
&\qquad\quad\times(-1)^{n}\frac{\Gamma(2+l+m+n)}{\Gamma(n+1)}\frac{\Gamma(\frac{1}{2}+n-i\mu)\Gamma(-\frac{3}{2}-l-m-n+i\mu)}{\Gamma(1+n-2i\mu)\Gamma(-1-l-m-n+2i\mu)}.
\end{split}\label{tochum}
\end{equation}

Using (\ref{tochup}) and (\ref{tochum}) into (\ref{tochu10}), the result below follows:
\begin{table}[H]
  \centering
  \caption{The result of (\ref{diffint})}
  \begin{tabular}{|l|}\hline
  \\
  $\displaystyle\int_{0}^{\infty}dx\ x^{m} e^{-ix}\hankelm(x)\int_{x}^{\infty}dy\ y^{l} e^{-iy}\hankelmc(y)$\\
  \\
  $\displaystyle\qquad=\frac{(-1)^{-2(l+m)}(2i)^{-2-l-m}}{\pi\,\sinh^2(\pi\mu)}$\\
  \\
  $\displaystyle\qquad\qquad\times\sum_{n=0}^{\infty}(-1)^{n}\frac{\Gamma(2+l+m+n)}{\Gamma(n+1)}$\\
  \\
  $\displaystyle\qquad\qquad\qquad\times\Biggl[\frac{e^{2\pi\mu}}{1+m+n+i\mu}\frac{\Gamma(\frac{1}{2}+n+i\mu)\Gamma(-\frac{3}{2}-l-m-n-i\mu)}{\Gamma(1+n+2i\mu)\Gamma(-1-l-m-n-2i\mu)}$\\
  \\
  $\displaystyle\qquad\qquad\qquad\quad+\frac{1}{1+m+n-i\mu}\frac{\Gamma(\frac{1}{2}+n-i\mu)\Gamma(-\frac{3}{2}-l-m-n+i\mu)}{\Gamma(1+n-2i\mu)\Gamma(-1-l-m-n+2i\mu)}\Biggl]$\\
  \\ \hline
  \end{tabular}\label{resultdiff}
\end{table}
(Recall that this result follows when $l+m$ is an integer greater than or equal to $-1$.)

\subsubsection{Final results}
Using Table \ref{resulteasy} and \ref{resultdiff} for the integrals in (\ref{zzresult}), (\ref{gzzresult}) and (\ref{ggzzresult}), we obtain the final results (see Appendix \ref{app4} for the results of the integrals):
\begin{equation}
\begin{split}
\langle&\zeta(\vk_1)\zeta(\vk_2)\rangle\qquad=\qquad(2\pi)^3\delta^3 (\vk_1+\vk_2)\,P_{\zeta}(k_2)\,\frac{c_1^2}{\epsilon H^4}F_1(m),
\end{split}\label{f1ori}
\end{equation}
\begin{equation}
\begin{split}
\langle\gamma^{s}&(\vq)\zeta(\vk_1)\zeta(\vk_2)\rangle\\
&=(2\pi)^3\delta^3 (\vq+\vk_1+\vk_2)\ \epsilon_{ij}^{s}(\vq)\frac{(k_2)_{i}(k_2)_{j}}{(k_2)^2}P_{\gamma}(q)P_{\zeta}(k_2) \frac{c_1 c_4}{\epsilon H^2}\,F_2(m),
\end{split}\label{f2ori}
\end{equation}
\begin{equation}
\begin{split}
&\langle\gamma^{s_1}(\vq_1)\gamma^{s_2}(\vq_2)\zeta(\vk_1)\zeta(\vk_2)\rangle\\
&\quad=(2\pi)^3\delta^3 (\vq_1+\vq_2+\vk_1+\vk_2)\\
&\qquad\quad\times\epsilon_{ij}^{s_1}(\vq_1)\frac{(k_2)_{i}(k_2)_{j}}{(k_2)^2}\epsilon_{\alpha\beta}^{s_2}(\vq_2)\frac{(k_2)_{\alpha}(k_2)_{\beta}}{(k_2)^2}P_{\gamma}(q_1)P_{\gamma}(q_2)P_{\zeta}(k_2)\frac{c_4^2}{\epsilon}\,F_3(m),
\end{split}\label{f3ori}
\end{equation}
where
\begin{equation}
\begin{split}
F_1(m)\equiv&\ \frac{\pi^2}{\cosh^2(\pi\mu)}\\
&\quad-\frac{1}{\sinh(\pi\mu)}\textrm{Re}\left[\sum_{n=0}^{\infty}(-1)^{n}\Biggl\{\frac{e^{\pi\mu}}{(\frac{1}{2}+n+i\mu)^2}-\frac{e^{-\pi\mu}}{(\frac{1}{2}+n-i\mu)^2}\Biggl\}\right],
\end{split} \label{f1m}
\end{equation}
\begin{equation}
\begin{split}
F_2(m)\equiv&-\frac{1}{4}\left(\frac{1}{4}+\mu^2\right)\left(\frac{9}{4}+\mu^2\right)\frac{\pi^2}{\cosh^2(\pi\mu)}\\
&\quad\quad+\frac{1}{2\,\sinh(\pi\mu)}\,\textrm{Re}\Biggl[\,\sum_{n=0}^{\infty}(-1)^{n}(n+1)(n+2)\\
&\qquad\qquad\qquad\qquad\qquad\times\Biggl\{e^{\pi\mu}\frac{(1+n+2i\mu)(2+n+2i\mu)}{(\frac{1}{2}+n+i\mu)^2(\frac{5}{2}+n+i\mu)^2}\\
&\qquad\qquad\qquad\qquad\qquad\qquad-e^{-\pi\mu}\frac{(1+n-2i\mu)(2+n-2i\mu)}{(\frac{1}{2}+n-i\mu)^2(\frac{5}{2}+n-i\mu)^2}\Biggl\}\Biggl],
\end{split}\label{f2muzui}
\end{equation}
\begin{equation}
\begin{split}
F_3&(m)\equiv\\
&\frac{1}{32}\left(\frac{1}{4}+\mu^2\right)^2\left(\frac{9}{4}+\mu^2\right)^2\frac{\pi^2}{\cosh^2(\pi\mu)}\\
&-\frac{1}{8\,\sinh(\pi\mu)}\\
&\quad\quad\times\textrm{Re}\Biggl[\,\sum_{n=0}^{\infty}(-1)^{n}(n+1)(n+2)(n+3)(n+4)\\
&\qquad\times\Biggl\{e^{\pi\mu}\frac{(1+n+2i\mu)(2+n+2i\mu)(3+n+2i\mu)(4+n+2i\mu)}{(\frac{1}{2}+n+i\mu)(\frac{3}{2}+n+i\mu)(\frac{5}{2}+n+i\mu)^2(\frac{7}{2}+n+i\mu)(\frac{9}{2}+n+i\mu)}\\
&\quad\qquad-e^{-\pi\mu}\frac{(1+n-2i\mu)(2+n-2i\mu)(3+n-2i\mu)(4+n-2i\mu)}{(\frac{1}{2}+n-i\mu)(\frac{3}{2}+n-i\mu)(\frac{5}{2}+n-i\mu)^2(\frac{7}{2}+n-i\mu)(\frac{9}{2}+n-i\mu)}\Biggl\}\Biggl].
\end{split}\label{f3muzui}
\end{equation}

These summations can be performed by Mathematica 11\footnote{The analytical results for $F_2(m)$ and $F_3(m)$ include a lot of `HurwitzLerchPhi', `HurwitzZeta' and `HypergeometricPFQ', while only `HurwitzZeta' for $F_1(m)$.}. Especially the result of (\ref{f1m}) is consistent with the function which Chen and Wang obtained\cite{chenwang}. In addition, if we replace $\mu$ with $-i\nu=-i\sqrt{9/4-m^2/H^2}$ (analytical continuation), we can plot $F_1(m)$, $F_2(m)$ and $F_3(m)$ in $m>0$:
\begin{figure}[H]
  \centering
  \includegraphics[width=9cm]{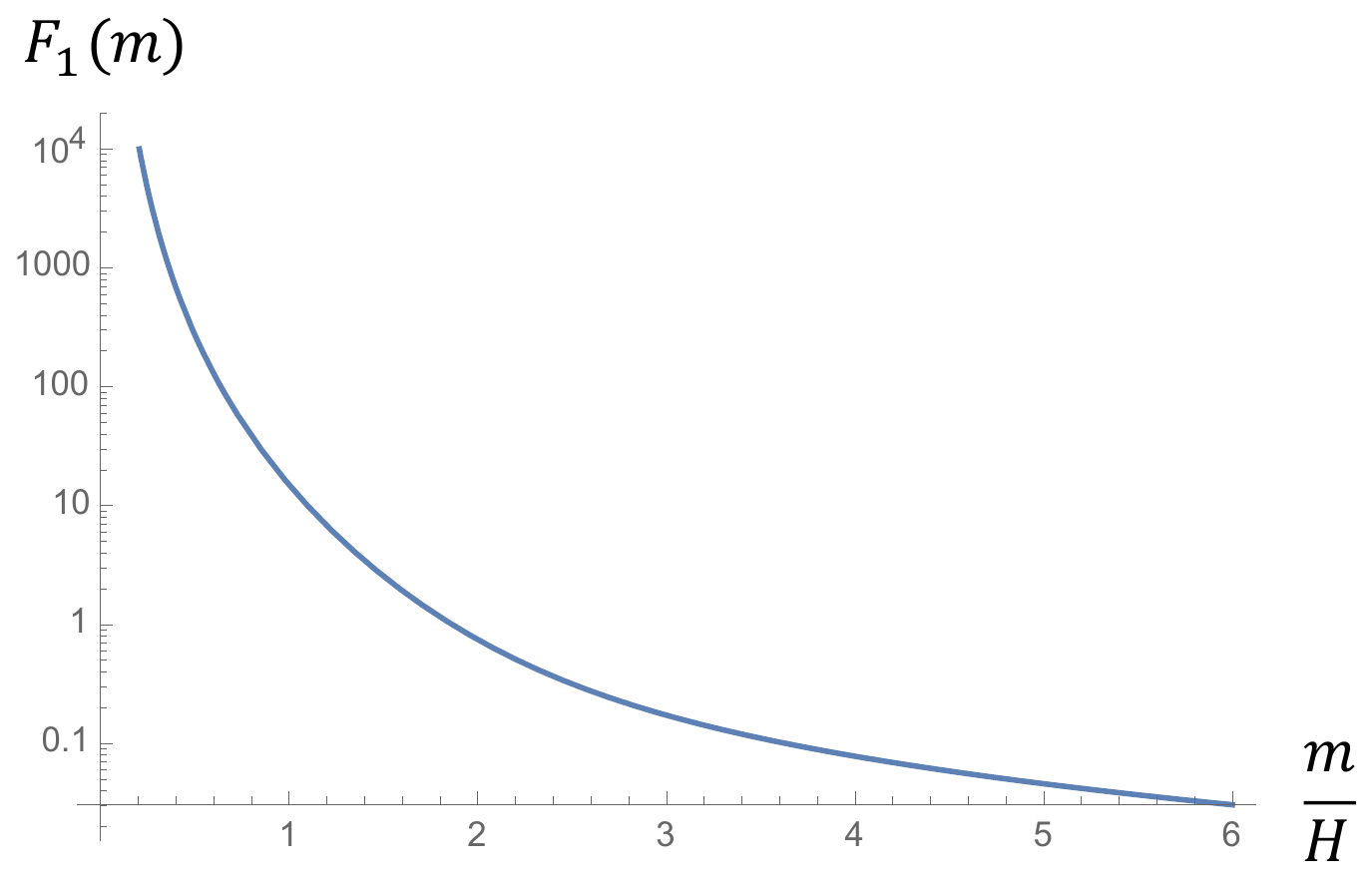}
  \caption{The plot for $F_1(m)$ in $m>0$}
  \label{fonem}
\end{figure}
\begin{figure}[H]
  \centering
  \includegraphics[width=9cm]{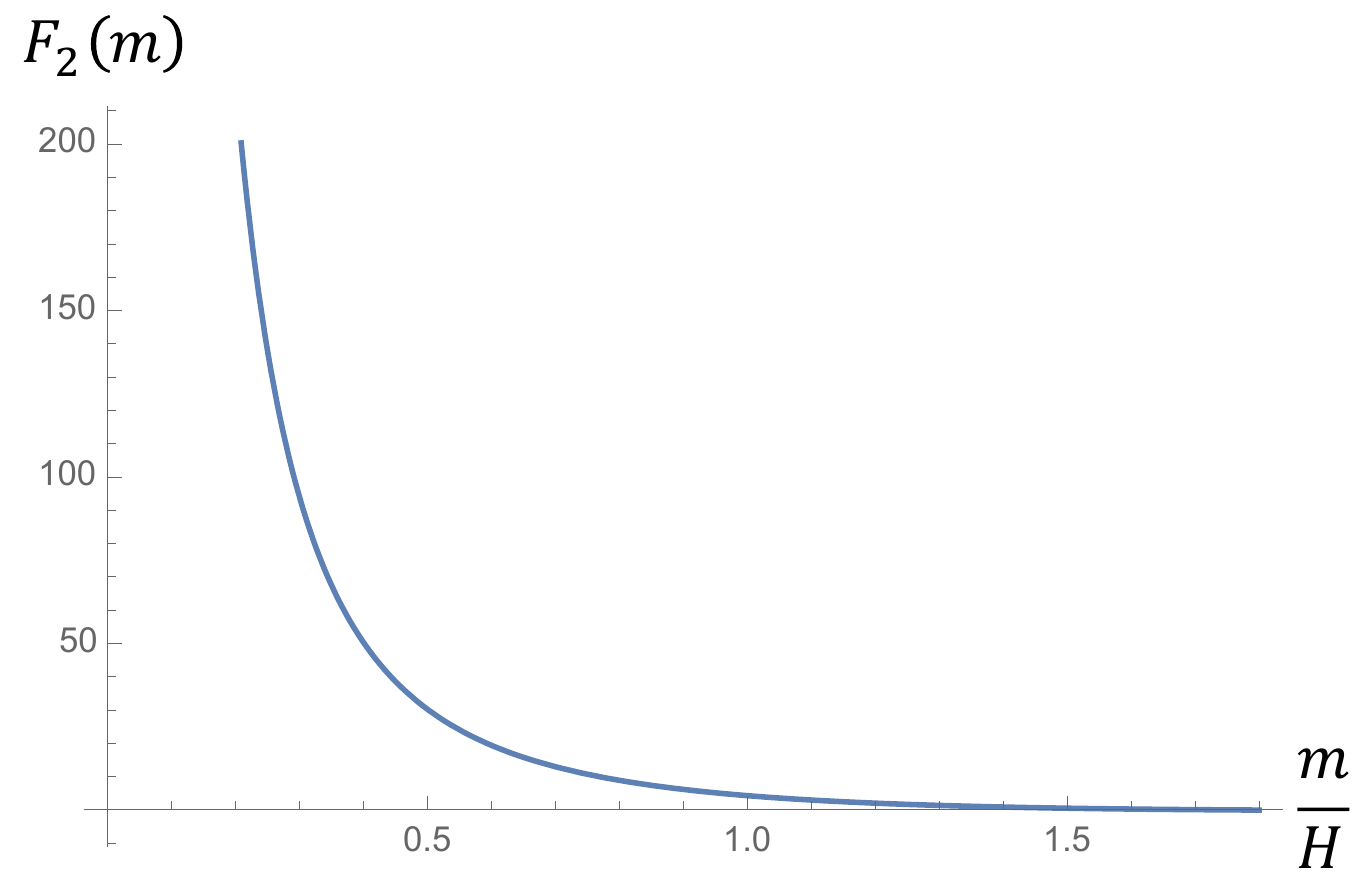}
  \caption{The plot for $F_2(m)$ in $m>0$}
  \label{ftwom}
\end{figure}
\begin{figure}[H]
  \centering
  \includegraphics[width=9cm]{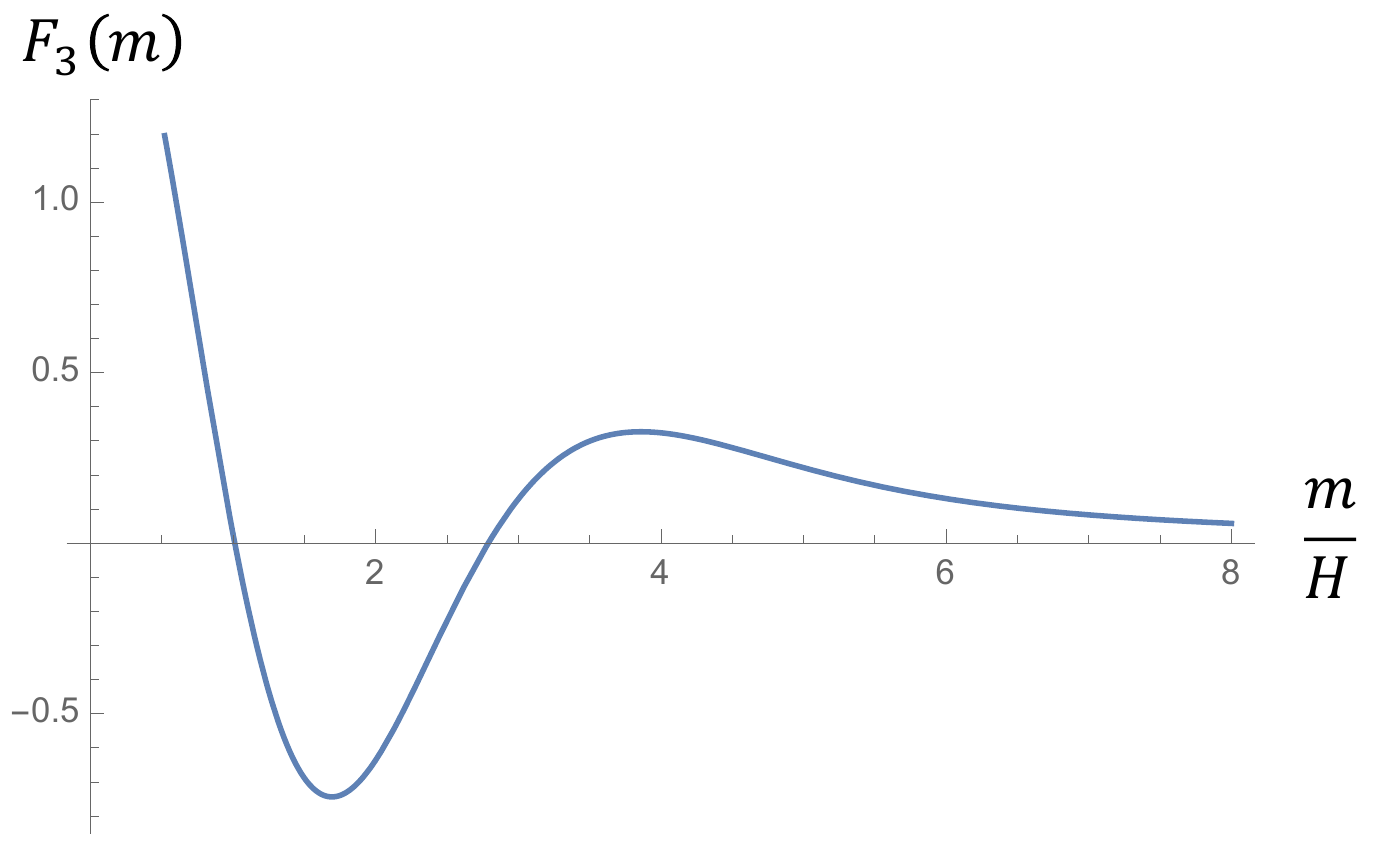}
  \caption{The plot for $F_3(m)$ in $m>0$}
  \label{fthreem}
\end{figure}
\begin{figure}[H]
  \centering
  \includegraphics[width=9cm]{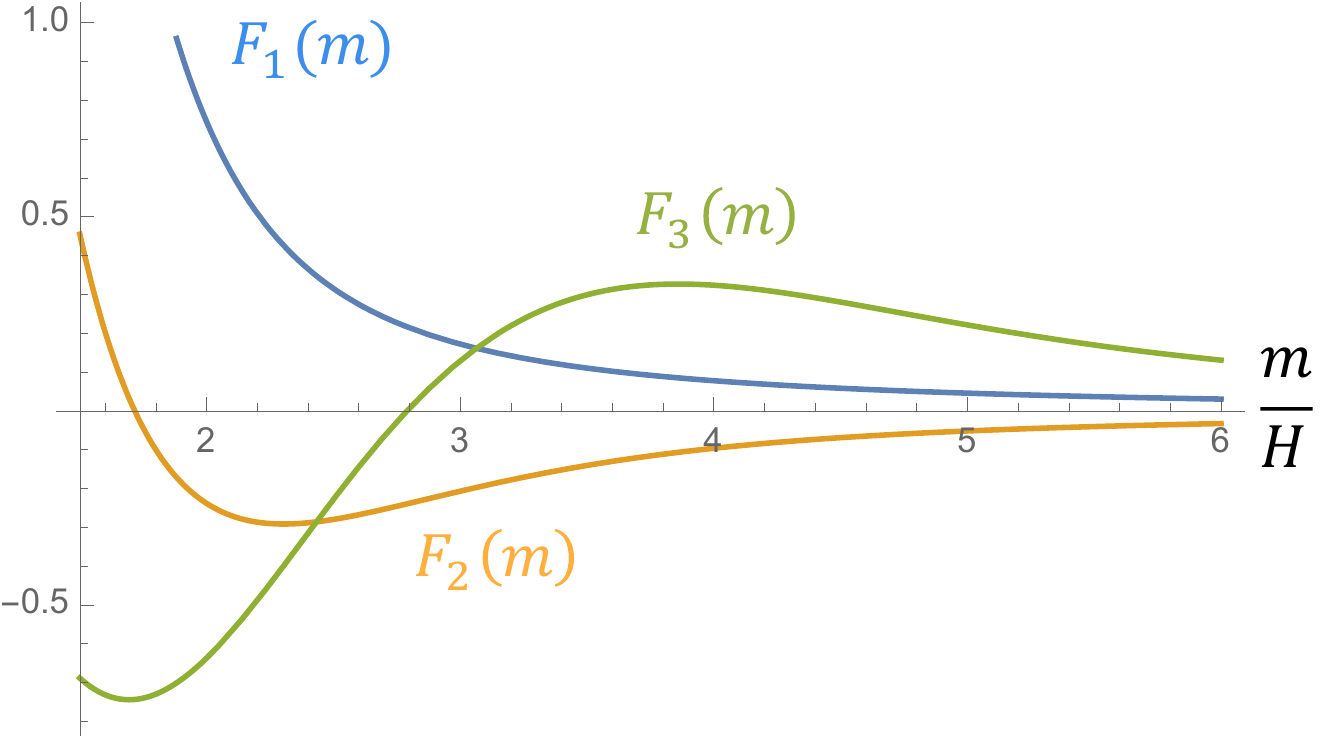}
  \caption{The plots for $F_1(m)$, $F_2(m)$ and $F_3(m)$ in $m\geq1.5H$}
  \label{comparefs}
\end{figure}

While all of the functions approach zero when $m\to\infty$, their behaviors are completely different as Figure \ref{comparefs} shows; they have as many local maximums or local minimums as gravitons. \vspace{1mm}

Now, in order to examine the relations among the functions, let us define the new functions:
\begin{equation}
G_{12}(m)\equiv\frac{F_1(m)}{F_2(m)},\quad G_{23}(m)\equiv\frac{F_2(m)}{F_3(m)}.
\end{equation}
Then we can perform the numerical analysis of these functions by Mathematica 11 as below:
\begin{table}[H]
  \centering
  \caption{The numerical analysis of $G_{12}(m)$ and $G_{23}(m)$}
  \begin{tabular}{|c||c|c|c|c|c|c|}\hline
  $\displaystyle m/H$&$\displaystyle 5.0$&$\displaystyle 10.0$&$\displaystyle 15.0$&$\displaystyle 20.0$&$\displaystyle 25.0$&$\displaystyle 30.0$\\ \hline\hline
  $\displaystyle G_{12}(m)$&$\displaystyle -0.88031$&$\displaystyle -0.978412$&$\displaystyle -0.99082$&$\displaystyle -0.99491$&$\displaystyle -0.996764$&$\displaystyle -0.99776$\\ \hline
  $\displaystyle G_{23}(m)$&$\displaystyle -0.233967$&$\displaystyle -0.310469$&$\displaystyle -0.323965$&$\displaystyle -0.328174$&$\displaystyle -0.330123$&$\displaystyle -0.331035$\\ \hline
  \end{tabular}\label{numericalresult}
\end{table}
This result suggests that $G_{12}(m)\to-1$ and $G_{23}(m)\to-1/3$ when $m\to\infty$.

\subsubsection{$m\to\infty$ limit}\label{mtoinftysec}
It is difficult to derive the asymptotic forms of the functions $F_1(m)$, $F_2(m)$ and $F_3(m)$ when $m\to\infty$, so we take an alternative way in order to examine the asymptotic behaviors of these functions.\vspace{2mm}

If we assume that $\sigma$ can be integrated out when $m\to\infty,$ the diagrams in Figure \ref{diagrams} can be simplified as below:
\begin{figure}[H]
  \centering
  \includegraphics[width=10cm]{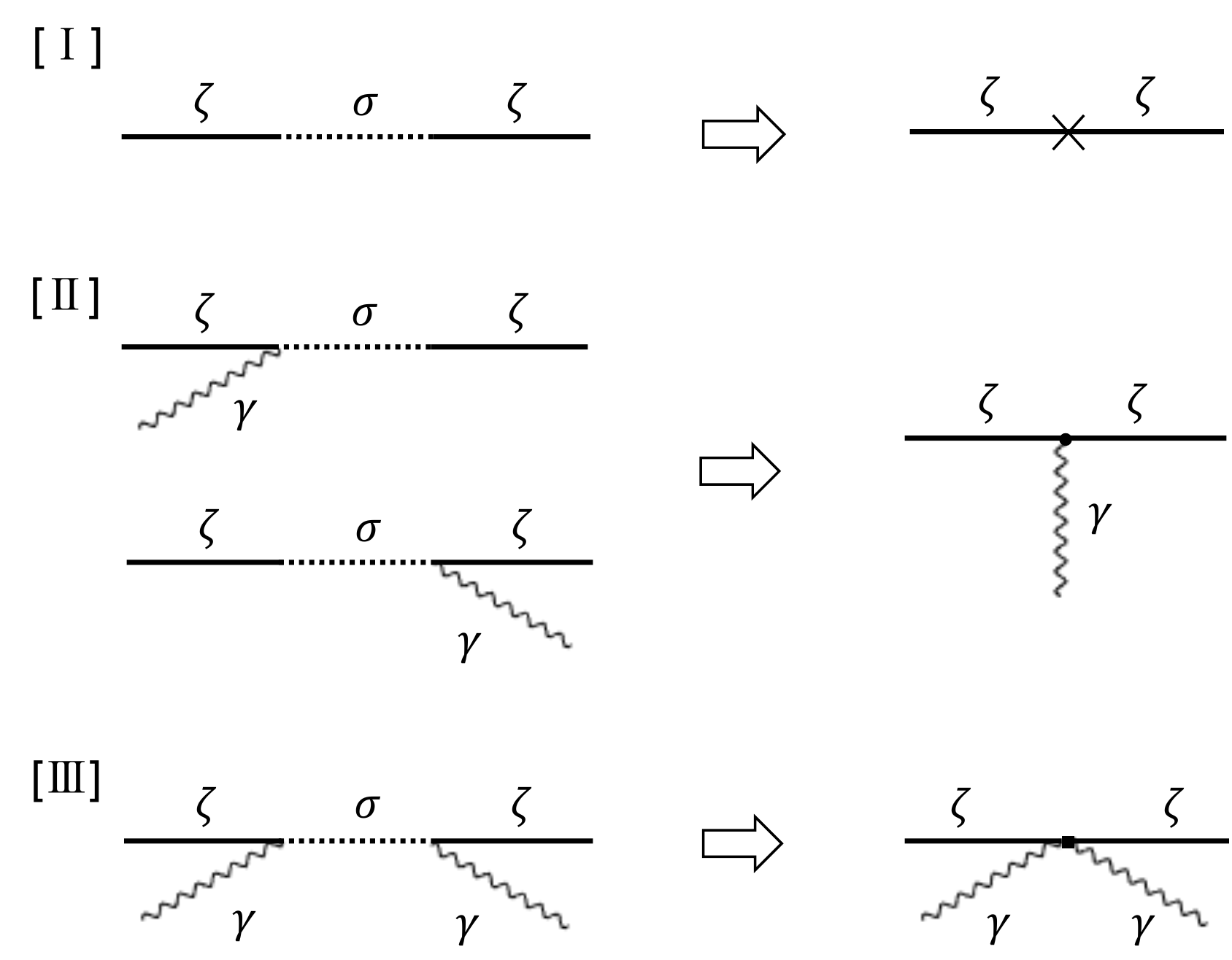}
  \caption{The simplified diagrams when $m\to\infty$}
  \label{effectivefigure}
\end{figure}
In these simplified diagrams, the interaction Hamiltonian $H_{I}(t)=-L_{I}(t)$ can be described as

\begin{eqnarray}
\textrm{[I]}\ H_{I}(t)&=&\left(-2\frac{c_1}{H}\right)^2 a^3(t)f(m)\int d^3 x\ \zetadot^2,\label{hamihami1}\\
\textrm{[I\hspace{-.1em}I]}\ H_{I}(t)&=&\left(-2\frac{c_1}{H}\right)\left(2\frac{c_4}{H}\right)a(t)f(m)\int d^3 x\ \zetadot\,\gaij\parti\partj\zetadot\ \times2,\label{hamihami2}\\
\textrm{[I\hspace{-.1em}I\hspace{-.1em}I]}\ H_{I}(t)&=&\left(2\frac{c_4}{H}\right)^2 a^{-1}(t)f(m)\int d^3 x\ \gaij\parti\partj\zetadot\,\gamma_{\alpha\beta}\partial_{\alpha}\partial_{\beta}\zetadot,\label{hamihami3}
\end{eqnarray}
using (\ref{quadzesig}) and (\ref{cubicgzsig}). Note that $f(m)$ is a function of the mass of $\sigma$, which is integrated out. In addition, be careful of the power exponent of $a(t)$.
The original exponent is three because of $\sqrt{-g}=a^3(t)$ in the action. However, if $\gamma$ once appears, the exponent is reduced by two because the $\gamma$ is produced from the inverse spatial metric $h^{ij}$, which has $a^{-2}(t)$. Also note that we are allowed to impose $\parti\partj$, which operated on $\sigma$ in (\ref{cubicgzsig}), on $\zetadot$ because in the soft-graviton limit, $\sigma$ and $\zetadot$ have the same momentum in the momentum space.
Finally, notice that there is a factor `$2$' in the last in (\ref{hamihami2}) because there are two diagrams originally as Figure \ref{effectivefigure} shows. \vspace{3mm}

Since the simplified diagrams have only one vertex, the computation is much easier than the original ones. Using the in-in formalism (\ref{ininformula}), the correlation function of $\mathcal{O}(t,\vx)$ ($=\zeta\zeta$, $\gamma\zeta\zeta$ and $\gamma\gamma\zeta\zeta$) is given by
\begin{equation}
\begin{split}
&\braket{0(t)|\mathcal{O}(t,\vx)|0(t)}\\
&\qquad=2\,\textrm{Im}\left[\int_{-\infty}^{t}dt'\ \braket{0|\mathcal{O}(t,\vx)H_{I}(t')|0}\right].
\end{split}\label{inineffective}
\end{equation}

Using the interaction Hamiltonian (\ref{hamihami1}),(\ref{hamihami2}) and (\ref{hamihami3}) into (\ref{inineffective}) respectively, and also using the two point functions shown in Table \ref{tipszg}, the computation can be performed as below:\vspace{3mm}

\leftline{\underline{{\bf [I] Computation of} ${\bm \zetzet}$}}
\begin{equation}
\begin{split}\nonumber
\langle\zeta(\vk_1)\zeta(\vk_2)\rangle&\quad=\quad(2\pi)^3 \delta^3 (\vk_1+\vk_2)\\
&\qquad\qquad\times f(m)\left(-2\frac{c_1}{H}\right)^2\left(\biarii\right)\left(\biari\right)\underline{\times\,2!}\\
&\qquad\qquad\quad\times2\,\textrm{Im}\left[\int_{-\infty}^{0}\frac{d\eta}{\eta^4 H^4}\eta^2 e^{ik_1\eta}\eta^2 e^{ik_2\eta}\right]
\end{split}
\end{equation}
\begin{equation}
\begin{split}
\qquad\qquad=\quad(2\pi)^3 \delta^3 (\vk_1+\vk_2)\,P_{\zeta}(k_2)\,\frac{c_1^2}{\epsilon H^4}\tilde{F}_1(m),
\end{split}\label{f1til}
\end{equation}
\begin{equation}\label{f1tildif}
\left(\tilde{F}_1 (m)\equiv-2H^2 f(m)\right)
\end{equation}\vspace{2mm}

\leftline{\underline{{\bf [I\hspace{-.1em}I] Computation of} ${\bm \gamzetzet}$ {\bf in the soft-graviton limit} ${\bm \vq\to0}$}}
\begin{equation}
\begin{split}\nonumber
&\langle\gamma^{s}(\vq)\zeta(\vk_1)\zeta(\vk_2)\rangle\quad=\quad(2\pi)^3\delta^3 (\vq+\vk_1+\vk_2)\ \epsilon_{ij}^{s}(\vq)(-i)^2 (k_2)_{i}(k_2)_{j}\\
&\qquad\qquad\qquad\qquad\qquad\qquad\times f(m)\left(-4\frac{c_1 c_4}{H^2}\right)\binashiq\left(\biarii\right)\left(\biari\right)\times2\,\underline{\times\,2!}\\
&\qquad\qquad\qquad\qquad\qquad\qquad\quad\times2\,\textrm{Im}\left[\int_{-\infty}^{0}\frac{d\eta}{\eta^2 H^2}\eta^2 e^{ik_1\eta}\eta^2 e^{ik_2\eta}\right]
\end{split}
\end{equation}
\begin{equation}
\begin{split}
\quad\qquad\qquad=\quad(2\pi)^3&\delta^3 (\vq+\vk_1+\vk_2)\ \epsilon_{ij}^{s}(\vq)\frac{(k_2)_{i}(k_2)_{j}}{(k_2)^2}\\
&\times P_{\gamma}(q)P_{\zeta}(k_2) \frac{c_1 c_4}{\epsilon H^2}\,\tilde{F}_2(m),
\end{split}\label{f2til}
\end{equation}
\begin{equation}\label{f2tildif}
\left(\tilde{F}_2 (m)\equiv2H^2 f(m)\right)
\end{equation}\vspace{2mm}

\leftline{\underline{{\bf [I\hspace{-.1em}I\hspace{-.1em}I] Computation of} ${\bm \gamgamzetzet}$ {\bf in the soft-graviton limit} ${\bm \vq_1,\vq_2\to0}$}}
\begin{equation}
\begin{split}\nonumber
\langle\gamma^{s_1}(\vq_1)\gamma^{s_2}(\vq_2)\zeta(\vk_1)\zeta(\vk_2)\rangle\quad=\quad&(2\pi)^3\delta^3 (\vq_1+\vq_2+\vk_1+\vk_2)\\
&\times\epsilon_{ij}^{s_1}(\vq_1)(-i)^2 (k_1)_{i}(k_1)_{j}\ \epsilon_{\alpha\beta}^{s_2}(\vq_2)(-i)^2 (k_2)_{\alpha}(k_2)_{\beta}\\
&\times f(m)\left(2\frac{c_4}{H}\right)^2\frac{H^2}{(q_1)^3}\frac{H^2}{(q_2)^3}\left(\biarii\right)\left(\biari\right)\\
&\quad\underline{\times\,2!\times2!}\\
&\qquad\times2\,\textrm{Im}\left[\int_{-\infty}^0 d\eta\ \eta^2 e^{ik_1\eta}\eta^2 e^{ik_2\eta}\right]
\end{split}
\end{equation}
\begin{equation}
\begin{split}
\qquad\qquad\qquad=\quad&(2\pi)^3\delta^3 (\vq_1+\vq_2+\vk_1+\vk_2)\\
&\quad\times\epsilon_{ij}^{s_1}(\vq_1)\frac{(k_2)_{i}(k_2)_{j}}{(k_2)^2}\epsilon_{\alpha\beta}^{s_2}(\vq_2)\frac{(k_2)_{\alpha}(k_2)_{\beta}}{(k_2)^2}\\
&\qquad\times P_{\gamma}(q_1)P_{\gamma}(q_2)P_{\zeta}(k_2)\frac{c_4^2}{\epsilon}\,\tilde{F}_3(m).
\end{split}\label{f3til}
\end{equation}
\begin{equation}\label{f3tildif}
\left(\tilde{F}_3 (m)\equiv-6H^2 f(m)\right)
\end{equation}\vspace{1mm}

Note that we use the integration by parts and the $i\epsilon$ prescription when we compute the integrals above (see Appendix \ref{app0}).
Also note that the underlined factors above are combinatorial factors for each diagram in Figure \ref{effectivefigure}.\vspace{3mm}

Comparing (\ref{f1til}), (\ref{f2til}) and (\ref{f3til}) with (\ref{f1ori}), (\ref{f2ori}) and (\ref{f3ori}), it follows that $\tilde{F}_1(m)$, $\tilde{F}_2(m)$ and $\tilde{F}_3(m)$ correspond to $F_1(m)$, $F_2(m)$ and $F_3(m)$ respectively in the $m\to\infty$ limit. Then, using (\ref{f1tildif}), (\ref{f2tildif}) and (\ref{f3tildif}),
\begin{eqnarray}
\tilde{G}_{12}(m)&\equiv&\frac{\tilde{F}_1(m)}{\tilde{F}_2(m)}\quad=\quad-1,\label{G12rela}\\
\tilde{G}_{23}(m)&\equiv&\frac{\tilde{F}_2(m)}{\tilde{F}_3(m)}\quad=\quad-\frac{1}{3}.\label{G23rela}
\end{eqnarray}
These results are consistent with the numerical analysis of the original diagrams shown in Table \ref{numericalresult}.\vspace{2mm}

The relations (\ref{G12rela}) and (\ref{G23rela}) can be useful in order to search for particles whose masses are much higher than $H\sim10^{14}$ GeV; although each function $F_1(m)$, $F_2(m)$ and $F_3(m)$ approaches zero as Figure \ref{fonem}, \ref{ftwom} and \ref{fthreem} show, the ratios among them take meaningful values.

\subsection{Genelarization}
In order to compute the correlation function including an arbitrary number of soft-gravitons, let us consider the action below instead of (\ref{eftaction3}):
\begin{equation}
\begin{split}
S_{I}=\int d^4 x\sqrt{-g}\,\delta g^{00}\biggl[C_0 \sigma&+C_1 g^{\mu\nu}\partial_{\mu}\partial_{\nu}\sigma\\
&+C_2g^{\mu\nu}g^{\rho\tau}\partial_{\mu}\partial_{\nu}\partial_{\rho}\partial_{\tau}\sigma+\cdots\biggl].
\end{split}
\end{equation}
Note that $C_0$ and $C_1$ correspond to $c_1$ and $c_4$ in the previous subsection respectively. If we consider just one graviton for each $g^{\mu\nu}$, it follows that the term including $C_{N}$ produces a coupling of $\zeta$, $\sigma$ and $N$ soft-gravitons. Therefore, using $\delta g^{00}=2\zetadot/H$ and $h^{ij}=-a^{-2}(t)\gaij$ to the first order, the interaction Hamiltonian $H_{I}(t)=-L_{I}(t)$ becomes
\begin{equation}
\begin{split}
H_{I}(t)=&-2\frac{C_0}{H}a^3(t)\int d^3 x\,\zetadot\sigma\\
&+2\frac{C_1}{H}a(t)\int d^3 x\,\zetadot\gaij\parti\partj\sigma\\
&-2\frac{C_2}{H}a^{-1}(t)\int d^3 x\,\zetadot\gaij\gamma_{\alpha\beta}\parti\partj\partial_{\alpha}\partial_{\beta}\sigma\\
&+\cdots\\
&+(-1)^{N+1}2\frac{C_{N}}{H}a^{3-2N}(t)\int d^3 x\,\zetadot\gaij\cdots\gamma_{\alpha\beta}\parti\partj\cdots\partial_{\alpha}\partial_{\beta}\sigma\\
&+\cdots.
\end{split}
\end{equation}

Using this Hamiltonian, we will compute two types of the correlation functions as the diagrams below:
\begin{figure}[H]
  \centering
  \includegraphics[width=9cm]{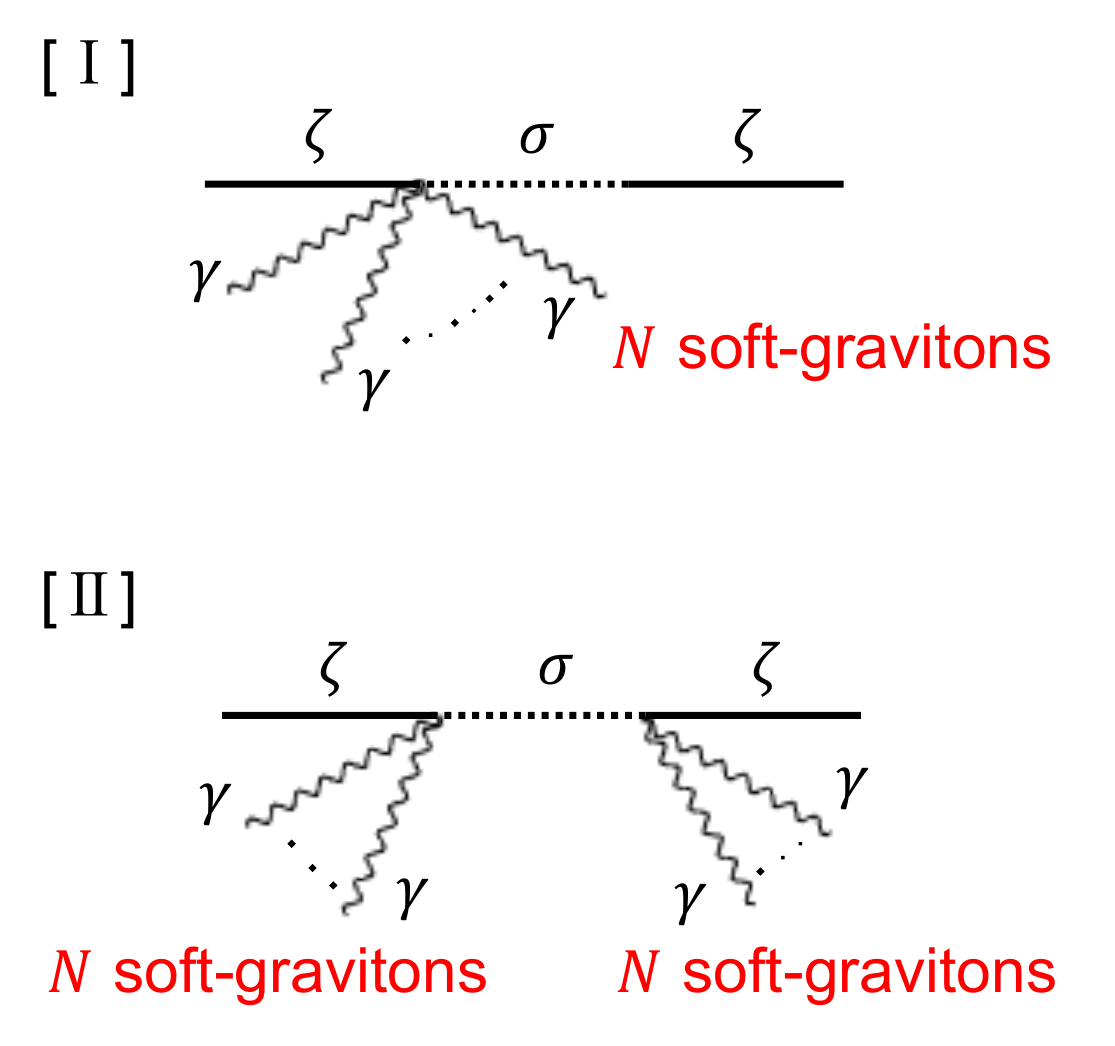}
  \caption{[I] `$(N,0)$ soft-gravitons' diagram\ [I\hspace{-.1em}I] `$(N,N)$ soft-gravitons' diagram}
  \label{General}
\end{figure}
Note that the diagrams [I] and [I\hspace{-.1em}I] are the generalization of [I\hspace{-.1em}I] and [I\hspace{-.1em}I\hspace{-.1em}I] in Figure \ref{diagrams} respectively.\vspace{3mm}

\subsubsection{Computations and results}
\leftline{\underline{\textbf{[I] Computation of the `$(N,0)$ soft-gravitons' diagram}}}\vspace{2mm}
The strategy for the computation is the same as (\ref{gzzmiddle}), so it follows that
\begin{equation}
\begin{split}
&\langle\gamma^{s_1}(\vq_1)\cdots\gamma^{s_{N}}(\vq_{N})\zeta(\vk_1)\zeta(\vk_2)\rangle\\
&=(2\pi)^3\delta^3 (\vq_1+\cdots +\vq_{N}+\vk_1+\vk_2)\\
&\qquad\times\epsilon_{ij}^{s_1}(\vq_1)\cdots\epsilon_{\alpha\beta}^{s_{N}}(\vq_{N})(-i)^2\cdots(-i)^2 (k_2)_{i}(k_2)_{j}\cdots(k_2)_{\alpha}(k_2)_{\beta}\\
&\qquad\times(-1)^{N}4\frac{C_0 C_{N}}{H^2}\frac{H^2}{q_1^3}\cdots\frac{H^2}{q_{N}^3}\left(\biarii\right)\left(\biari\right)\\
&\qquad\times\sigmatip\,\underline{\times\,2!\times N!}\\
&\times\Biggl[2\textrm{Re}\left[\int_{-\infty}^{0}\frac{d\eta_1}{\eta_1^4H^4}\int_{-\infty}^{0}\frac{d\eta_2}{(\eta_2 H)^{4-2N}}\eta_1^2e^{-ik_1\eta_1}\eta_2^2e^{ik_2\eta_2}(\eta_1\eta_2)^{\frac{3}{2}}\hankel(-k_2\eta_1)\hankelc(-k_2\eta_2)\right]\\
&\quad-2\textrm{Re}\left[\int_{-\infty}^{0}\frac{d\eta_1}{\eta_1^4H^4}\int_{-\infty}^{\eta_1}\frac{d\eta_2}{(\eta_2 H)^{4-2N}}\eta_1^2e^{ik_1\eta_1}\eta_2^2e^{ik_2\eta_2}(\eta_1\eta_2)^{\frac{3}{2}}\hankel(-k_2\eta_1)\hankelc(-k_2\eta_2)\right]\\
&\quad-2\textrm{Re}\left[\int_{-\infty}^{0}\frac{d\eta_1}{(\eta_1 H)^{4-2N}}\int_{-\infty}^{\eta_1}\frac{d\eta_2}{\eta_2^4H^4}\eta_1^2e^{ik_1\eta_1}\eta_2^2e^{ik_2\eta_2}(\eta_1\eta_2)^{\frac{3}{2}}\hankel(-k_2\eta_1)\hankelc(-k_2\eta_2)\right]\Biggl].
\end{split}\label{geneonemid}
\end{equation}
The underlined factor is a combinatorial factor for the diagram [I] in Figure \ref{General}.\vspace{2mm}

Then, setting $x\equiv-k_2 \eta_1$ and $y\equiv -k_2 \eta_2$,
\begin{equation}
\begin{split}
&\langle\gamma^{s_1}(\vq_1)\cdots\gamma^{s_{N}}(\vq_{N})\zeta(\vk_1)\zeta(\vk_2)\rangle\\
&=(2\pi)^3\delta^3 (\vq_1+\cdots +\vq_{N}+\vk_1+\vk_2)\\
&\qquad\times\epsilon_{ij}^{s_1}(\vq_1)\cdots\epsilon_{\alpha\beta}^{s_{N}}(\vq_{N})\frac{(k_2)_{i}(k_2)_{j}}{(k_2)^2}\cdots\frac{(k_2)_{\alpha}(k_2)_{\beta}}{(k_2)^2}\\
&\qquad\times P_{\gamma}(q_1)\cdots P_{\gamma}(q_{N})\ P_{\zeta}(k_2)\frac{\pi C_0 C_{N}}{\epsilon H^{4-2N}}N!\,e^{-\pi\textrm{Im}(\nu)}\\
&\qquad\times\Biggl[\textrm{Re}\left[\int_{0}^{\infty}dx\,x^{-\frac{1}{2}}e^{ix}\hankel(x)\int_{0}^{\infty}dy\,y^{2N-\frac{1}{2}}e^{-iy}\hankelc(y)\right]\\
&\qquad\quad-\textrm{Re}\left[\int_{0}^{\infty}dx\,x^{-\frac{1}{2}}e^{-ix}\hankel(x)\int_{x}^{\infty}dy\,y^{2N-\frac{1}{2}}e^{-iy}\hankelc(y)\right]\\
&\qquad\quad-\textrm{Re}\left[\int_{0}^{\infty}dx\,x^{2N-\frac{1}{2}}e^{-ix}\hankel(x)\int_{x}^{\infty}dy\,y^{-\frac{1}{2}}e^{-iy}\hankelc(y)\right]\Biggl].
\end{split}\label{geneoneres}
\end{equation}
We can compute these integrals just by using Table \ref{resulteasy} and \ref{resultdiff} (see Appendix \ref{app5}). Finally, we obtain
\begin{equation}
\begin{split}
\langle\gamma^{s_1}(\vq_1)&\cdots\gamma^{s_{N}}(\vq_{N})\zeta(\vk_1)\zeta(\vk_2)\rangle\\
&=(2\pi)^3\delta^3 (\vq_1+\cdots +\vq_{N}+\vk_1+\vk_2)\\
&\qquad\times\epsilon_{ij}^{s_1}(\vq_1)\cdots\epsilon_{\alpha\beta}^{s_{N}}(\vq_{N})\frac{(k_2)_{i}(k_2)_{j}}{(k_2)^2}\cdots\frac{(k_2)_{\alpha}(k_2)_{\beta}}{(k_2)^2}\\
&\qquad\times P_{\gamma}(q_1)\cdots P_{\gamma}(q_{N})\ P_{\zeta}(k_2)\frac{C_0 C_{N}}{\epsilon H^{4-2N}}\,R_{N}(m),
\end{split}\label{geneoneresres}
\end{equation}
where
\begin{equation}
\begin{split}
&R_{N}(m)\equiv\\
&\,2^{1-2N}(-1)^{N}N!\\
&\times\Biggl[\frac{\pi^2}{\cosh^2(\pi\mu)}\frac{1}{(2N)!}\left\{\left(2N-\frac{1}{2}\right)^2+\mu^2\right\}\cdots\left\{\left(\frac{3}{2}\right)^2+\mu^2\right\}\left\{\left(\frac{1}{2}\right)^2+\mu^2\right\}\\
&\quad-\frac{1}{\sinh(\pi\mu)}\,\textrm{Re}\Biggl[\,\sum_{n=0}^{\infty}(-1)^{n}(n+1)(n+2)\cdots(n+2N)\\
&\qquad\times\Biggl\{e^{\pi\mu}\frac{(1+n+2i\mu)(2+n+2i\mu)\cdots(2N+n+2i\mu)\times(N+\frac{1}{2}+n+i\mu)}{(\frac{1}{2}+n+i\mu)^2(\frac{3}{2}+n+i\mu)\cdots(2N-\frac{1}{2}+n+i\mu)(2N+\frac{1}{2}+n+i\mu)^2}\\
&\qquad -e^{-\pi\mu}\frac{(1+n-2i\mu)(2+n-2i\mu)\cdots(2N+n-2i\mu)\times(N+\frac{1}{2}+n-i\mu)}{(\frac{1}{2}+n-i\mu)^2(\frac{3}{2}+n-i\mu)\cdots(2N-\frac{1}{2}+n-i\mu)(2N+\frac{1}{2}+n-i\mu)^2}\Biggl\}\Biggl]\Biggl].
\end{split}\label{INmuzui}
\end{equation}
This result is consistent with (\ref{f2muzui}) if we set $N=1$.\vspace{4mm}

\leftline{\underline{\textbf{[I\hspace{-.1em}I] Computation of the `$(N,N)$ soft-gravitons' diagram}}}\vspace{2mm}
This computation is the same as (\ref{ggzzmiddle}):
\begin{equation}
\begin{split}
&\langle\gamma^{s_1}(\vq_1)\cdots\gamma^{s_{2N}}(\vq_{2N})\zeta(\vk_1)\zeta(\vk_2)\rangle\\
&=(2\pi)^3\delta^3 (\vq_1+\cdots +\vq_{2N}+\vk_1+\vk_2)\\
&\qquad\times\epsilon_{ij}^{s_1}(\vq_1)\cdots\epsilon_{\alpha\beta}^{s_{2N}}(\vq_{2N})(-i)^2\cdots(-i)^2 (k_2)_{i}(k_2)_{j}\cdots(k_2)_{\alpha}(k_2)_{\beta}\\
&\qquad\times4\frac{C_{N}^2}{H^2}\frac{H^2}{q_1^3}\cdots\frac{H^2}{q_{2N}^3}\left(\biarii\right)\left(\biari\right)\\
&\qquad\times\sigmatip\,\underline{\times\,2!\times (2N)!}\\
&\times\Biggl[\int_{-\infty}^{0}\frac{d\eta_1}{(\eta_1 H)^{4-2N}}\int_{-\infty}^{0}\frac{d\eta_2}{(\eta_2 H)^{4-2N}}\eta_1^2e^{-ik_1\eta_1}\eta_2^2e^{ik_2\eta_2}(\eta_1\eta_2)^{\frac{3}{2}}\hankel(-k_2\eta_1)\hankelc(-k_2\eta_2)\\
&-2\textrm{Re}\left[\int_{-\infty}^{0}\frac{d\eta_1}{(\eta_1 H)^{4-2N}}\int_{-\infty}^{\eta_1}\frac{d\eta_2}{(\eta_2 H)^{4-2N}}\eta_1^2e^{ik_1\eta_1}\eta_2^2e^{ik_2\eta_2}(\eta_1\eta_2)^{\frac{3}{2}}\hankel(-k_2\eta_1)\hankelc(-k_2\eta_2)\right]\Biggl].
\end{split}\label{genetwomid}
\end{equation}
The underlined factor is a combinatorial factor for the diagram [I\hspace{-.1em}I] in Figure \ref{General}.\vspace{2mm}

Then, setting $x\equiv-k_2 \eta_1$ and $y\equiv -k_2 \eta_2$,
\begin{equation}
\begin{split}
&\langle\gamma^{s_1}(\vq_1)\cdots\gamma^{s_{2N}}(\vq_{2N})\zeta(\vk_1)\zeta(\vk_2)\rangle\\
&=(2\pi)^3\delta^3 (\vq_1+\cdots +\vq_{2N}+\vk_1+\vk_2)\\
&\qquad\times\epsilon_{ij}^{s_1}(\vq_1)\cdots\epsilon_{\alpha\beta}^{s_{2N}}(\vq_{2N})\frac{(k_2)_{i}(k_2)_{j}}{(k_2)^2}\cdots\frac{(k_2)_{\alpha}(k_2)_{\beta}}{(k_2)^2}\\
&\qquad\times P_{\gamma}(q_1)\cdots P_{\gamma}(q_{2N})\ P_{\zeta}(k_2)\frac{\pi C_{N}^2}{\epsilon H^{4(1-N)}}(2N)!\,e^{-\pi\textrm{Im}(\nu)}\\
&\qquad\times\Biggl[\frac{1}{2}\left|\int_{0}^{\infty}dx\,x^{2N-\frac{1}{2}}e^{ix}\hankel(x)\right|^2\\
&\qquad\quad-\textrm{Re}\left[\int_{0}^{\infty}dx\,x^{2N-\frac{1}{2}}e^{-ix}\hankel(x)\int_{x}^{\infty}dy\,y^{2N-\frac{1}{2}}e^{-iy}\hankelc(y)\right]\Biggl].
\end{split}\label{genetwores}
\end{equation}
Finally, using Table \ref{resulteasy} and \ref{resultdiff} for the integrals (see Appendix \ref{app5}), we obtain
\begin{equation}
\begin{split}
\langle\gamma^{s_1}(\vq_1)&\cdots\gamma^{s_{2N}}(\vq_{2N})\zeta(\vk_1)\zeta(\vk_2)\rangle\\
&=(2\pi)^3\delta^3 (\vq_1+\cdots +\vq_{2N}+\vk_1+\vk_2)\\
&\qquad\times\epsilon_{ij}^{s_1}(\vq_1)\cdots\epsilon_{\alpha\beta}^{s_{2N}}(\vq_{2N})\frac{(k_2)_{i}(k_2)_{j}}{(k_2)^2}\cdots\frac{(k_2)_{\alpha}(k_2)_{\beta}}{(k_2)^2}\\
&\qquad\times P_{\gamma}(q_1)\cdots P_{\gamma}(q_{2N})\ P_{\zeta}(k_2)\frac{C_{N}^2}{\epsilon H^{4(1-N)}}\,S_{N}(m),
\end{split}\label{genetworesres}
\end{equation}
where
\begin{equation}
\begin{split}
&S_{N}(m)\equiv\\
&2^{-4N}\Biggl[\frac{\pi^2}{\cosh^2(\pi\mu)}\frac{1}{(2N)!}\left\{\left(2N-\frac{1}{2}\right)^2+\mu^2\right\}^2\cdots\left\{\left(\frac{3}{2}\right)^2+\mu^2\right\}^2\left\{\left(\frac{1}{2}\right)^2+\mu^2\right\}^2\\
&\quad-\frac{(2N)!}{\sinh(\pi\mu)}\,\textrm{Re}\Biggl[\,\sum_{n=0}^{\infty}(-1)^{n}(n+1)(n+2)\cdots(n+4N)\\
&\qquad\times\Biggl\{e^{\pi\mu}\frac{(1+n+2i\mu)(2+n+2i\mu)\cdots(4N+n+2i\mu)}{(\frac{1}{2}+n+i\mu)(\frac{3}{2}+n+i\mu)\cdots(2N+\frac{1}{2}+n+i\mu)^2\cdots(4N+\frac{1}{2}+n+i\mu)}\\
&\qquad -e^{-\pi\mu}\frac{(1+n-2i\mu)(2+n-2i\mu)\cdots(4N+n-2i\mu)}{(\frac{1}{2}+n-i\mu)(\frac{3}{2}+n-i\mu)\cdots(2N+\frac{1}{2}+n-i\mu)^2\cdots(4N+\frac{1}{2}+n-i\mu)}\Biggl\}\Biggl]\Biggl].
\end{split}\label{JNmuzui}
\end{equation}
This result is consistent with (\ref{f3muzui}) if we set $N=1$.

\subsubsection{Evaluation of $R_{N}(m)$ and $S_{N}(m)$}
The plots and the diagrams for $R_{N}(m)$ when $N=1,2,3$ and $4$ are shown as below:
\begin{figure}[H]
  \centering
  \includegraphics[width=11cm]{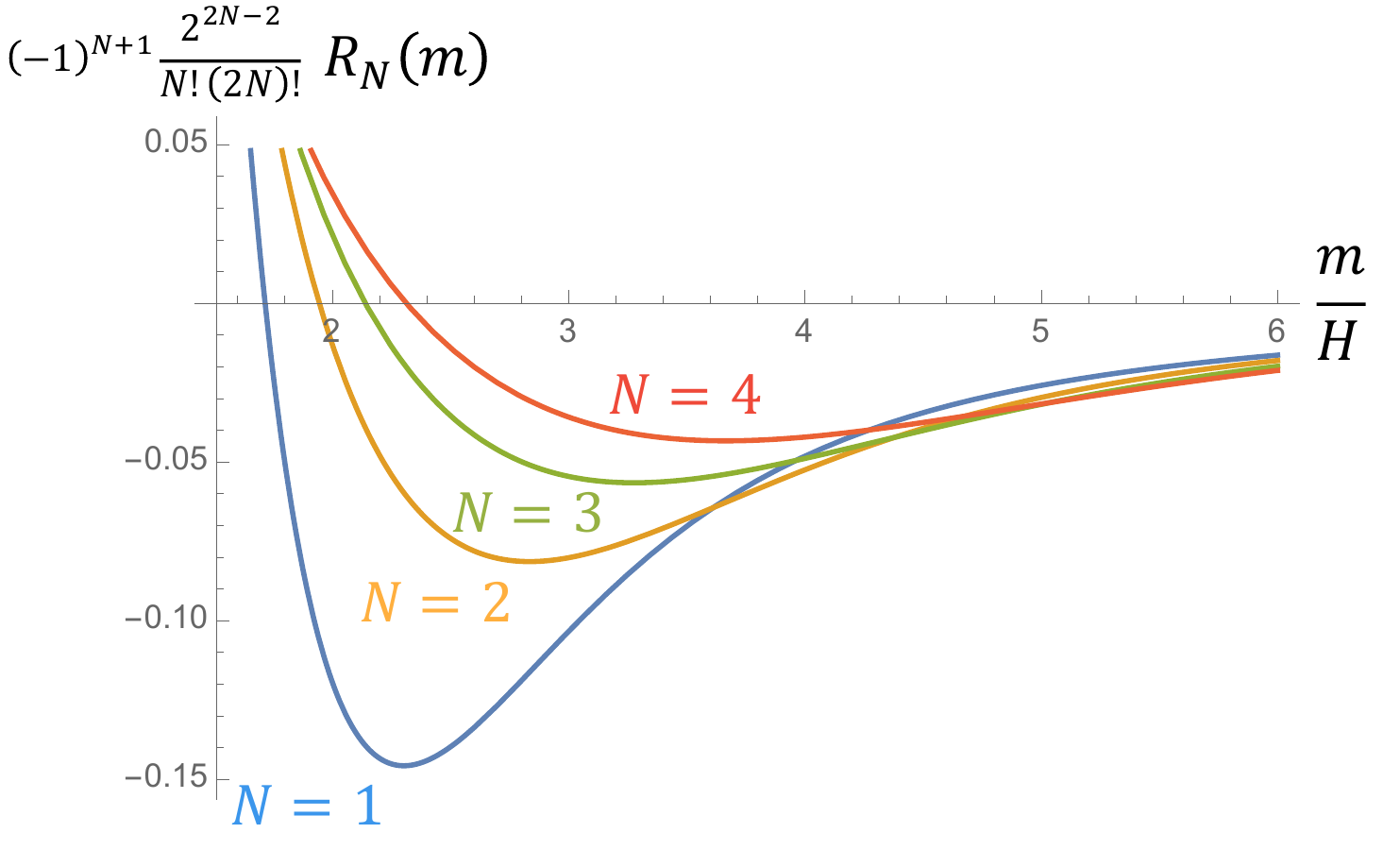}
  \caption{The plots for $(-1)^{N+1}\frac{2^{2N-2}}{N!(2N)!}R_{N}(m)$ when $N=1,2,3$ and $4$ in $m\geq1.5H$. The factor $(-1)^{N+1}\frac{2^{2N-2}}{N!(2N)!}$ is introduced for the functions to converge to the same value when $m\to\infty$; we will discuss this in the next subsection, and this factor is shown in (\ref{Itildef}).}
  \label{INfour}
\end{figure}
\begin{figure}[H]
  \centering
  \includegraphics[width=9cm]{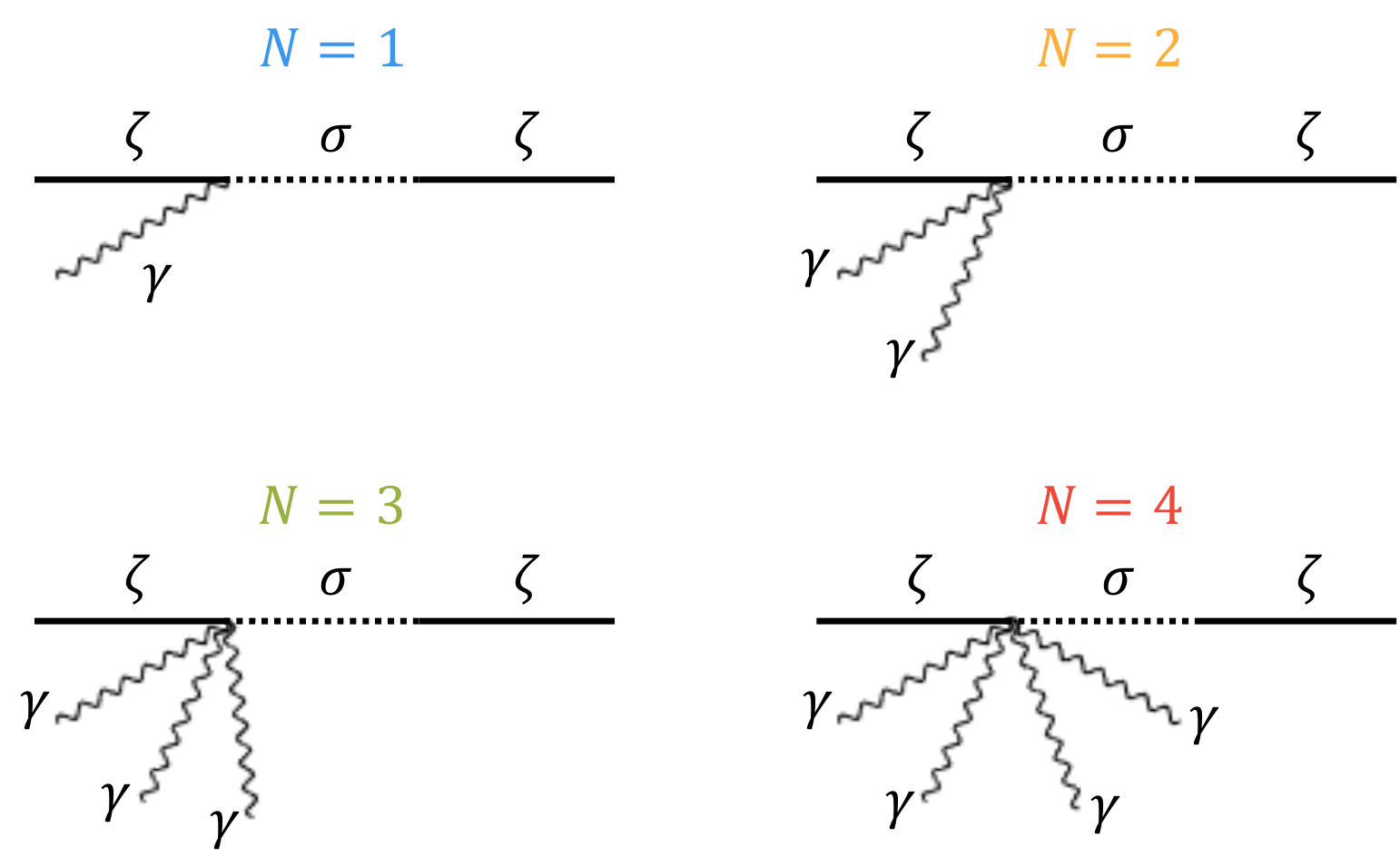}
  \caption{The diagrams for $R_{N}(m)$ when $N=1,2,3$ and $4$}
  \label{INdiagrams}
\end{figure}

As Figure \ref{INfour} shows, it follows that when the number of soft-gravitons is getting larger, the peak of the correlation function is shifted to larger mass of $\sigma$.\vspace{2mm}

Then, the plots and the diagrams for $S_{N}(m)$ when $N=1$ and $2$ are shown as below:
\begin{figure}[H]
  \centering
  \includegraphics[width=11cm]{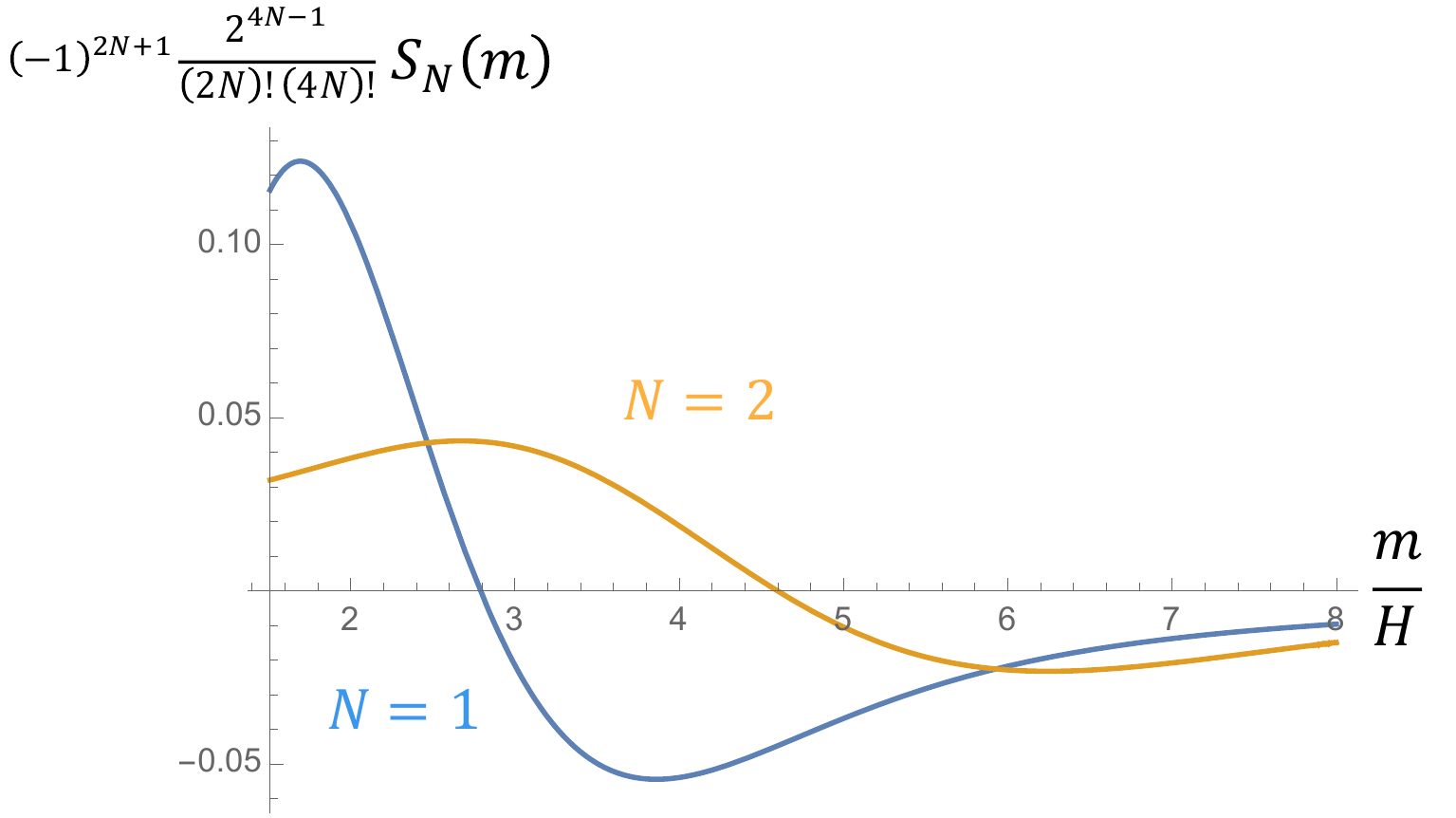}
  \caption{The plots for $(-1)^{2N+1}\frac{2^{4N-1}}{(2N)!(4N)!}S_{N}(m)$ when $N=1$ and $2$ in $m\geq1.5H$. The factor $(-1)^{2N+1}\frac{2^{4N-1}}{(2N)!(4N)!}$ is introduced for the same reason as Figure \ref{INfour}, and it is shown in (\ref{Jtildef}).}
  \label{JNtwo}
\end{figure}
\begin{figure}[H]
  \centering
  \includegraphics[width=9cm]{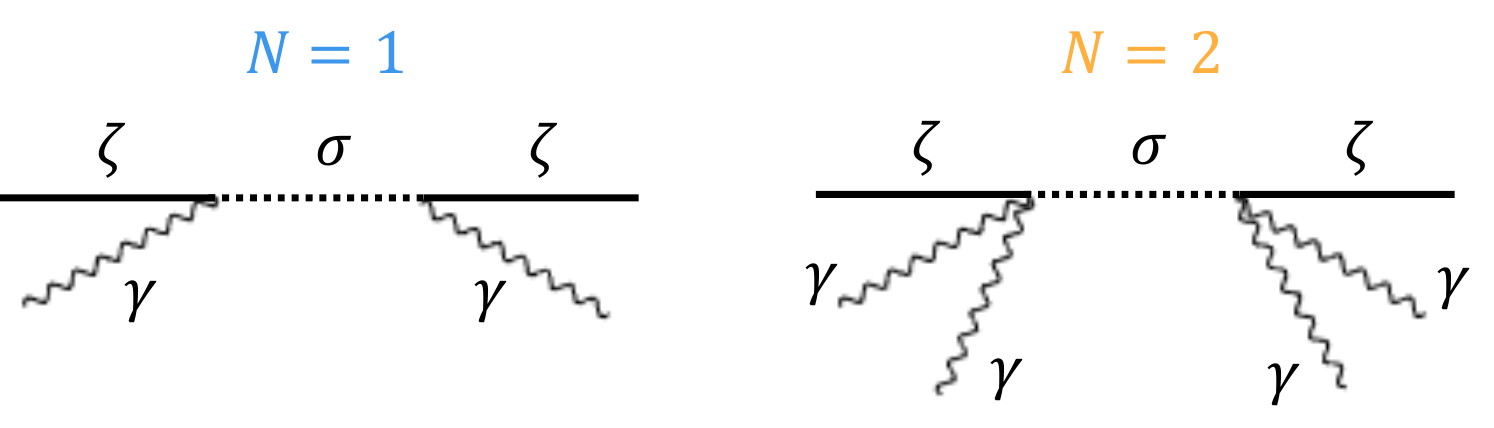}
  \caption{The diagrams for $S_{N}(m)$ when $N=1$ and $2$}
  \label{JNdiagrams}
\end{figure}

Also in this case, the peaks are shifted to larger mass of $\sigma$ when $N$ is getting larger.\vspace{2mm}

Finally, we perform the numerical analysis.
As (\ref{INmuzui}) and (\ref{JNmuzui}) show, the larger $N$ we set, the more complicated $R_{N}(m)$ and $S_{N}(m)$ become, so that the values of the functions tend to be unstable when $m$ is getting larger.
Therefore, we examine only $R_{1}(m)$, $R_2(m)$, $R_3(m)$ and $S_1(m)$, which have three soft-gravitons at most:\vspace{1mm}
\begin{table}[H]
  \centering
  \caption{The numerical analysis of $R_2(m)/R_1(m)$ and $R_2(m)/S_1(m)$}
  \begin{tabular}{|c||c|c|c|c|c|c|}\hline
  $\displaystyle m/H$&$\displaystyle 5.0$&$\displaystyle 10.0$&$\displaystyle 15.0$&$\displaystyle 20.0$&$\displaystyle 25.0$&$\displaystyle 30.0$\\ \hline\hline
  $\displaystyle R_2(m)/R_1(m)$&$\displaystyle -6.88167$&$\displaystyle -6.14334$&$\displaystyle -6.05725$&$\displaystyle -6.03114$&$\displaystyle -6.02061$&$\displaystyle -6.01494$\\ \hline
  $\displaystyle R_2(m)/S_1(m)$&$\displaystyle 1.61009$&$\displaystyle 1.90731$&$\displaystyle 1.96234$&$\displaystyle 1.97926$&$\displaystyle 1.98754$&$\displaystyle 1.99115$\\ \hline
  \end{tabular}\label{numericalresult2}
\end{table}

\begin{table}[H]
  \centering
  \caption{The numerical analysis of $R_3(m)/R_2(m)$}
  \begin{tabular}{|c||c|c|c|c|c|c|}\hline
  $\displaystyle m/H$&$\displaystyle 5.0$&$\displaystyle 7.0$&$\displaystyle 9.0$&$\displaystyle 11.0$&$\displaystyle 13.0$&$\displaystyle 15.0$\\ \hline\hline
  $\displaystyle R_3(m)/R_2(m)$&$\displaystyle -24.137$&$\displaystyle -24.242$&$\displaystyle -23.3081$&$\displaystyle -22.9589$&$\displaystyle -22.7956$&$\displaystyle -22.7469$\\ \hline
  \end{tabular}\label{numericalresult3}
\end{table}

From Table \ref{numericalresult2}, it clearly follows that $R_2(m)/R_1(m)\to-6$ and $R_2(m)/S_1(m)\to2$ when $m\to\infty$.
As for Table \ref{numericalresult3}, the next subsection tells the convergence value.

\subsubsection{$m\to\infty$ limit}
Let us examine the asymptotic behaviors of $R_{N}(m)$ and $S_{N}(m)$ when $m\to\infty$ in the same way as Section \ref{mtoinftysec}.\vspace{2mm}

If we assume that $\sigma$ can be integrated out when $m\to\infty$, the diagrams in Figure \ref{General} can be simplified as below:
\begin{figure}[H]
  \centering
  \includegraphics[width=11cm]{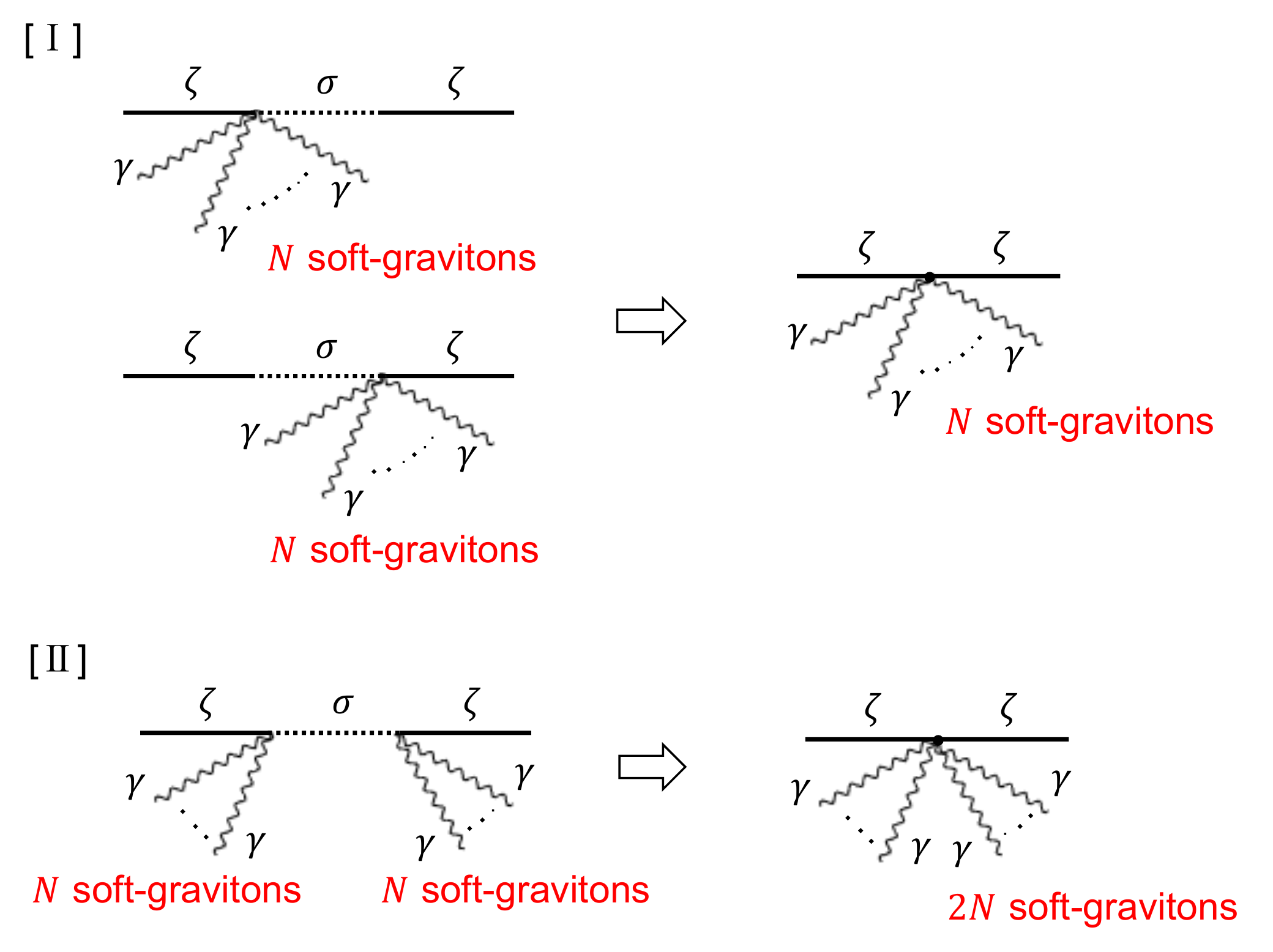}
  \caption{The simplified diagrams when $m\to\infty$}
  \label{simplifiedgene}
\end{figure}
Then, we can construct the interaction Hamiltonian in the same way as Section \ref{mtoinftysec}:
\begin{equation}\label{hamihamihami1}
\textrm{[I]}\ H_{I}(t)=(-1)^{N} 4\frac{C_0 C_{N}}{H^2} a^{3-2N}(t)f(m)\int d^3 x\ \zetadot\gaij\cdots\gamma_{\alpha\beta}\partial_{i}\partial_{j}\cdots\partial_{\alpha}\partial_{\beta}\zetadot\ \times2,
\end{equation}
\begin{equation}
\begin{split}
\textrm{[I\hspace{-.1em}I]}\ H_{I}(t)=4\frac{C_{N}^2}{H^2}a^{3-4N}(t)f(m)\int d^3 x\ \gaij&\cdots\gamma_{\alpha\beta}\partial_{i}\partial_{j}\cdots\partial_{\alpha}\partial_{\beta}\zetadot\\
&\times\gamma_{i'j'}\cdots\gamma_{\alpha'\beta'}\partial_{i'}\partial_{j'}\cdots\partial_{\alpha'}\partial_{\beta'}\zetadot.
\end{split}\label{hamihamihami2}
\end{equation}\vspace{2mm}
The strategy for the computation is the same as (\ref{f2til}) and (\ref{f3til}) (see Appendix \ref{app0} for the integrals below):\vspace{2mm}

\leftline{\underline{\textbf{[I] Computation of the `$(N,0)$ soft-gravitons' diagram {\sl (simplified)}}}}
\begin{equation}
\begin{split}\nonumber
\langle\gamma^{s_1}(\vq_1)\cdots&\gamma^{s_{N}}(\vq_{N})\zeta(\vk_1)\zeta(\vk_2)\rangle\\
&=(2\pi)^3\delta^3 (\vq_1+\cdots +\vq_{N}+\vk_1+\vk_2)\\
&\qquad\times\epsilon_{ij}^{s_1}(\vq_1)\cdots\epsilon_{\alpha\beta}^{s_{N}}(\vq_{N})(-i)^2\cdots(-i)^2 (k_2)_{i}(k_2)_{j}\cdots(k_2)_{\alpha}(k_2)_{\beta}\\
&\qquad\times f(m)(-1)^{N}4\frac{C_0 C_{N}}{H^2}\frac{H^2}{q_1^3}\cdots\frac{H^2}{q_{N}^3}\left(\biarii\right)\left(\biari\right)\times2\\
&\qquad\quad\underline{\times\,2!\times N!}\\
&\qquad\quad\times2\,\textrm{Im}\left[\int_{-\infty}^{0}\frac{d\eta}{(\eta H)^{4-2N}}\eta^2 e^{ik_1\eta}\eta^2 e^{ik_2\eta}\right]
\end{split}
\end{equation}\vspace{2mm}
\begin{equation}
\begin{split}
&=(2\pi)^3\delta^3 (\vq_1+\cdots +\vq_{N}+\vk_1+\vk_2)\\
&\qquad\times\epsilon_{ij}^{s_1}(\vq_1)\cdots\epsilon_{\alpha\beta}^{s_{N}}(\vq_{N})\frac{(k_2)_{i}(k_2)_{j}}{(k_2)^2}\cdots\frac{(k_2)_{\alpha}(k_2)_{\beta}}{(k_2)^2}\\
&\qquad\times P_{\gamma}(q_1)\cdots P_{\gamma}(q_{N})\ P_{\zeta}(k_2)\frac{C_0 C_{N}}{\epsilon H^{4-2N}}\,\tilde{R}_{N}(m),
\end{split}
\end{equation}
\begin{equation}\label{Itildef}
\left(\tilde{R}_{N}(m)\equiv H^2 f(m)\times(-1)^{N+1}\frac{N!(2N)!}{2^{2N-2}}\right)
\end{equation}\vspace{3mm}

\leftline{\underline{\textbf{[I\hspace{-.1em}I] Computation of the `$(N,N)$ soft-gravitons' diagram {\sl (simplified)}}}}
\begin{equation}
\begin{split}\nonumber
\langle\gamma^{s_1}(\vq_1)\cdots&\gamma^{s_{2N}}(\vq_{2N})\zeta(\vk_1)\zeta(\vk_2)\rangle\\
&=(2\pi)^3\delta^3 (\vq_1+\cdots +\vq_{2N}+\vk_1+\vk_2)\\
&\qquad\times\epsilon_{ij}^{s_1}(\vq_1)\cdots\epsilon_{\alpha\beta}^{s_{2N}}(\vq_{2N})(-i)^2\cdots(-i)^2 (k_2)_{i}(k_2)_{j}\cdots(k_2)_{\alpha}(k_2)_{\beta}\\
&\qquad\times f(m)\,4\frac{C_{N}^2}{H^2}\frac{H^2}{q_1^3}\cdots\frac{H^2}{q_{N}^3}\left(\biarii\right)\left(\biari\right)\,\underline{\times\,2!\times(2N)!}\\
&\qquad\times2\,\textrm{Im}\left[\int_{-\infty}^{0}\frac{d\eta}{(\eta H)^{4(1-N)}}\eta^2 e^{ik_1\eta}\eta^2 e^{ik_2\eta}\right]
\end{split}
\end{equation}\vspace{2mm}
\begin{equation}
\begin{split}
&=(2\pi)^3\delta^3 (\vq_1+\cdots +\vq_{2N}+\vk_1+\vk_2)\\
&\qquad\times\epsilon_{ij}^{s_1}(\vq_1)\cdots\epsilon_{\alpha\beta}^{s_{2N}}(\vq_{2N})\frac{(k_2)_{i}(k_2)_{j}}{(k_2)^2}\cdots\frac{(k_2)_{\alpha}(k_2)_{\beta}}{(k_2)^2}\\
&\qquad\times P_{\gamma}(q_1)\cdots P_{\gamma}(q_{2N})\ P_{\zeta}(k_2)\frac{C_{N}^2}{\epsilon H^{4(1-N)}}\,\tilde{S}_{N}(m).
\end{split}
\end{equation}
\begin{equation}\label{Jtildef}
\left(\tilde{S}_{N}(m)\equiv H^2 f(m)\times(-1)^{2N+1}\frac{(2N)!(4N)!}{2^{4N-1}}\right)
\end{equation}
The underlined factors above are combinatorial factors for each diagram in Figure \ref{simplifiedgene}.\vspace{3mm}

It follows that $\tilde{R}_{N}(m)$ and $\tilde{S}_{N}(m)$ correspond to $R_{N}(m)$ and $S_{N}(m)$ in the $m\to\infty$ limit respectively, by comparing these results with (\ref{geneoneresres}) and (\ref{genetworesres}).
Then, using (\ref{Itildef}) and (\ref{Jtildef}), we can find the relations:
\begin{eqnarray}
\frac{\tilde{R}_{N+1}(m)}{\tilde{R}_{N}(m)}&=&-\frac{(N+1)^2 (2N+1)}{2},\label{relationII}\\
\frac{\tilde{R}_{2N}(m)}{\tilde{S}_{N}(m)}&=&2.\label{relationIJ}
\end{eqnarray}
Note that the relation (\ref{relationIJ}) is trivial because this `2' is produced by the fact that there are two original diagrams in [I] in Figure \ref{simplifiedgene}.
On the other hand, the relation (\ref{relationII}) is crucial; it serves as a consistency relation, which relates $\langle\gamma^{s_1}\cdots\gamma^{s_{N+1}}\zeta\zeta\rangle$ to $\langle\gamma^{s_1}\cdots\gamma^{s_{N}}\zeta\zeta\rangle$ in the $m\to\infty$ limit in this model.
This relation can be useful when searching for particles whose masses are much higher than $10^{14}$ GeV.\vspace{2mm}

Finally, we can test the numerical analysis before.
When $N=1$, (\ref{relationII}) and (\ref{relationIJ}) become
\begin{equation}
\frac{\tilde{R}_{2}(m)}{\tilde{R}_{1}(m)}=-6,\quad\frac{\tilde{R}_{2}(m)}{\tilde{S}_{1}(m)}=2,
\end{equation}
which are consistent with Table \ref{numericalresult2}.
When $N=2$, (\ref{relationII}) becomes
\begin{equation}
\frac{\tilde{R}_{3}(m)}{\tilde{R}_{2}(m)}=-\frac{45}{2}.
\end{equation}
This is also consistent with Table \ref{numericalresult3}, although the numerical value is unstable when we set much larger $m$ by Mathematica 11.

\section{Summary}
In this paper, firstly we reviewed two theories:\vspace{2mm}

\leftline{\underline{\textbf{Maldacena's theory}}}\vspace{1mm}

In this theory, we add scalar and tensor fluctuations, $\zeta$ and $\gaij$, into the action of gravity and an inflaton field, and compute the three point functions of them using the in-in formalism.\vspace{2mm}

\leftline{\underline{\textbf{Effective Field Theory of Inflation}}}\vspace{1mm}

In this theory, the scalar fluctuation $\zeta$ is interpreted as a Nambu-Goldstone boson $\pi$, which is associated with a spontaneous breaking of time diffeomorphism invariance.
This theory can expand the Maldacena's theory because the broken symmetry allows us to consider more terms in a Lagrangian.\vspace{3mm}

Then, we applied the EFT method to introduce another scalar field $\sigma$ into the Maldacena's theory. Especially, we concentrated on $\zeta\sigma$ and $\gamma\zeta\sigma$ couplings, and computed some corrected correlation functions: $\zetzet$, $\gamzetzet$ and $\gamgamzetzet$ in the soft-graviton limit.
After that, we generalized the theory; we constructed couplings of $\zeta$, $\sigma$ and $N$ soft-gravitons, and computed $\langle\gamma^{s_1}\cdots\gamma^{s_{N}}\zeta\zeta\rangle$ for two generalized diagrams.
Then, by plotting it as a function of $m$ (the mass of $\sigma$) for several $N$'s, it followed that when the number of soft-gravitons is getting larger, the peak of the correlation function is shifted to larger mass of $\sigma$.
Finally, we derived the relation (\ref{relationII}), which relates $\langle\gamma^{s_1}\cdots\gamma^{s_{N+1}}\zeta\zeta\rangle$ to $\langle\gamma^{s_1}\cdots\gamma^{s_{N}}\zeta\zeta\rangle$ in the $m\to\infty$ limit, under the assumption that $\sigma$ is integrated out when $m\to\infty$.
Then we confirmed that the relation (\ref{relationII}) is consistent with the original results numerically.\vspace{2mm}

As mentioned in Introduction, the computational results of the correlation functions shown in (\ref{f1ori}) to (\ref{f3muzui}), or generally (\ref{geneoneresres}) with (\ref{INmuzui}) and (\ref{genetworesres}) with (\ref{JNmuzui}), may determine the mass of the unknown particle $\sigma$, if the future observation tells the values of the correlation functions.
The mass can be around $10^{14}$ GeV, which can be estimated as the energy scale during inflation and cannot be detected in terrestrial accelerators.
Such a way, which regards inflation as a particle detector, has been developed recently; considering higher spins\cite{Cosmo}\cite{Baumann}, the Standard Model background\cite{SMcosmo}, and so on.
Therefore, we hope that our results will give one of the hints to the future observation to seek for unknown particles.

\section*{Acknowledgement}
I would like to express my gratitude to Prof. Takahiro Kubota for guiding me to interesting topics, checking my computations, and encouraging me kindly.
I also thank Prof. Norihiro Iizuka and Prof. Tetsuya Onogi for giving me helpful comments at my presentations, Tetsuya Akutagawa, Tomoya Hosokawa and Yusuke Hosomi for many discussions, and all of the other members in the Particle Physics Theory Group in Osaka University for spending happy days.
Finally, I am grateful to my family for everyday support.

\appendix
\section{Computation of $\int_{-\infty}^{0}d\eta\,\eta^{2N}e^{2ik_2\eta}$}\label{app0}
Setting $x\equiv-k_2\eta$,
\begin{equation}\label{app01}
\int_{-\infty}^{0}d\eta\,\eta^{2N}e^{2ik_2\eta}\quad=\quad(k_2)^{-1-2N}\int_{0}^{\infty}dx\,x^{2N}e^{-2ix}.
\end{equation}
Then, using the integration by parts and the $i\epsilon$ prescription repetitively,
\begin{equation}
\begin{split}
\int_{0}^{\infty}dx\,x^{2N}e^{-2ix}\quad&=\quad\frac{N}{i}\int_{0}^{\infty}dx\,x^{2N-1}e^{-2ix}\\
&=\quad\frac{N}{i}\frac{N-\frac{1}{2}}{i}\int_{0}^{\infty}dx\,x^{2N-2}e^{-2ix}\\
&=\quad\cdots\\
&=\quad\frac{N}{i}\frac{N-\frac{1}{2}}{i}\cdots\frac{1}{i}\frac{\frac{1}{2}}{i}\times\frac{1}{2i}\\
&=\quad i(-1)^{N+1}\frac{(2N)!}{2^{2N+1}}.
\end{split}\label{app02}
\end{equation}
Finally, substituting (\ref{app02}) into (\ref{app01}), it follows that
\begin{equation}
\int_{-\infty}^{0}d\eta\,\eta^{2N}e^{2ik_2\eta}\quad=\quad(k_2)^{-1-2N}\,i(-1)^{N+1}\frac{(2N)!}{2^{2N+1}}.
\end{equation}

\section{Cancellation of $z^{-a_1}(\cdots)$ and $z^{-a_2}(\cdots)$}\label{app1}
The explicit form of $z^{-a}(\cdots)$ in (\ref{asymp1}) is
\begin{equation}
z^{-a}(-1)^{-a}2^{-2+2a}\frac{(1+2a+2l)\,\Gamma(\frac{1}{2}+a)}{\sqrt{\pi}\,\left(\frac{1}{2}+l\right)}.
\end{equation}
Substituting this into (\ref{indef1}) and using $a_1\equiv1/2-i\mu$, $a_2\equiv1/2+i\mu$ and $z\equiv2ix$,
\begin{equation*}
\begin{split}
x^{1+l-i\mu}&\frac{-2^{i\mu}\,\csch(\pi\mu)}{(1+l-i\mu)\,\Gamma(1-i\mu)}\\
&\times(2ix)^{-\frac{1}{2}+i\mu}(-1)^{-\frac{1}{2}+i\mu}2^{-1-2i\mu}\frac{(2+2l-2i\mu)\,\Gamma(1-i\mu)}{\sqrt{\pi}\,\left(\frac{1}{2}+l\right)}\\
+x^{1+l+i\mu}&\frac{2^{-i\mu}\,(1+\coth(\pi\mu))}{(1+l+i\mu)\,\Gamma(1+i\mu)}\\
&\times(2ix)^{-\frac{1}{2}-i\mu}(-1)^{-\frac{1}{2}-i\mu}2^{-1+2i\mu}\frac{(2+2l+2i\mu)\,\Gamma(1+i\mu)}{\sqrt{\pi}\,\left(\frac{1}{2}+l\right)}
\end{split}
\end{equation*}

\begin{equation}
=\quad\frac{x^{\frac{1}{2}+l}\left(-2i\right)^{-\frac{1}{2}}}{\sqrt{\pi}\,\left(\frac{1}{2}+l\right)}\left[-(-i)^{i\mu}\,\csch(\pi\mu)+(-i)^{-i\mu}\,\left(1+\coth(\pi\mu)\right)\right],
\end{equation}
which vanishes because 
\begin{equation}
\begin{split}
(-i)^{i\mu}\,&\csch(\pi\mu)\\
&=\ \frac{e^{\frac{\pi\mu}{2}}}{\sinh(\pi\mu)}\ =\ e^{-\frac{\pi\mu}{2}}\frac{\sinh(\pi\mu)+\cosh(\pi\mu)}{\sinh(\pi\mu)}\\
&\qquad\qquad\qquad=(-i)^{-i\mu}\,\left(1+\coth(\pi\mu)\right).
\end{split}
\end{equation}

\section{Calculations to derive Table \ref{resulteasy}}\label{app2}
Substituting the first term in (\ref{asymp1}) into (\ref{indef1}) and using $a_1\equiv1/2-i\mu$, $a_2\equiv1/2+i\mu$ and $z\equiv2ix$,
\begin{equation*}
\begin{split}
&x^{1+l-i\mu}\frac{-2^{i\mu}\csch(\pi\mu)}{(1+l-i\mu)\Gamma(1-i\mu)}\,(2ix)^{-1-l+i\mu}(-1)^{1-l+i\mu}2^{-1-2i\mu}\\
 &\quad\qquad\times\frac{(2+2l-2i\mu)}{\sqrt{\pi}}\frac{\Gamma(1-i\mu)\Gamma(-\frac{1}{2}-l)\Gamma(1+l-i\mu)}{\Gamma(-l-i\mu)}\\
&+x^{1+l+i\mu}\frac{2^{-i\mu}(1+\coth(\pi\mu))}{(1+l+i\mu)\Gamma(1+i\mu)}\,(2ix)^{-1-l-i\mu}(-1)^{1-l-i\mu}2^{-1+2i\mu}\\
 &\quad\qquad\times\frac{(2+2l+2i\mu)}{\sqrt{\pi}}\frac{\Gamma(1+i\mu)\Gamma(-\frac{1}{2}-l)\Gamma(1+l+i\mu)}{\Gamma(-l+i\mu)}
\end{split}
\end{equation*}
\begin{equation}
\begin{split}
=\quad&\frac{e^{\frac{\pi\mu}{2}}}{\sinh(\pi\mu)}\frac{(i/2)^{l}}{\sqrt{\pi}}\frac{i}{2}\\
&\times\frac{1}{\Gamma\left(\frac{3}{2}+l\right)}\,\Gamma\left(-\frac{1}{2}-l\right)\Gamma\left(\frac{3}{2}+l\right)\\
&\times\Gamma(1+l+i\mu)\Gamma(1+l-i\mu)\Biggl\{\frac{1}{\Gamma(1+l-i\mu)\Gamma(-l+i\mu)}-\frac{1}{\Gamma(1+l+i\mu)\Gamma(-l-i\mu)}\Biggl\}.
\end{split}\label{app21}
\end{equation}
Then, using the formula
\begin{equation}
\Gamma(z)\Gamma(1-z)=\frac{\pi}{\sin(\pi z)}
\end{equation}
for the second and the third lines in (\ref{app21}), it becomes
\begin{equation}
\begin{split}
&e^{\frac{\pi\mu}{2}}\frac{(i/2)^{l}}{\sqrt{\pi}}\frac{1}{\Gamma\left(\frac{3}{2}+l\right)}\frac{i}{2}\frac{1}{\sinh(\pi\mu)}\left(-\frac{\pi}{\cos(l\pi)}\right)\\
&\qquad\times\Gamma(1+l+i\mu)\Gamma(1+l-i\mu)\frac{1}{\pi}\Bigl\{\sin\left(\pi(-l+i\mu)\right)-\sin\left(\pi(-l-i\mu)\right)\Bigl\}.
\end{split}\label{app22}
\end{equation}
Finally, using the formula
\begin{equation}
\sin A-\sin B=2\cos\frac{A+B}{2}\sin\frac{A-B}{2}
\end{equation}
for the second line in (\ref{app22}), the result in Table \ref{resulteasy} follows (note that $\sin(i\pi\mu)=i\,\sinh(\pi\mu)$).

\section{Cancellation of the divergent terms in (\ref{tochu})}\label{app3}
Using (\ref{asymp1}) for the second term in (\ref{tochu2}) and substituting it into the integrals in (\ref{tochu}), the second and the third terms in (\ref{tochu}) become (note that $p\equiv1+l+m+2i\mu$ and $a_2\equiv1/2+i\mu$ for the second term while $p\equiv 1+l+m$ and $a_1\equiv1/2-i\mu$ for the third term)
\begin{equation*}
\begin{split}
&\frac{2^{-i\mu}\,\csch(\pi\mu)}{(1+l+i\mu)\Gamma(1+i\mu)}\frac{x^{2+l+m+n+2i\mu}}{1+m+n+i\mu}\\
&\quad\times(-2ix)^{-1-l-i\mu}(-1)^{1-l-i\mu}2^{-1+2i\mu}\frac{(2+2l+2i\mu)}{\sqrt{\pi}}\frac{\Gamma(1+i\mu)\Gamma(-\frac{1}{2}-l)\Gamma(1+l+i\mu)}{\Gamma(-l+i\mu)}\\
&-\frac{2^{i\mu}(1+\coth(\pi\mu))}{(1+l-i\mu)\Gamma(1-i\mu)}\frac{x^{2+l+m+n}}{1+m+n+i\mu}\\
&\quad\times(-2ix)^{-1-l+i\mu}(-1)^{1-l+i\mu}2^{-1-2i\mu}\frac{(2+2l-2i\mu)}{\sqrt{\pi}}\frac{\Gamma(1-i\mu)\Gamma(-\frac{1}{2}-l)\Gamma(1+l-i\mu)}{\Gamma(-l-i\mu)}
\end{split}
\end{equation*}
\begin{equation}
\begin{split}
=\quad&\frac{x^{1+m+n+i\mu}}{1+m+n+i\mu}\,\frac{e^{\frac{\pi\mu}{2}}}{\sinh(\pi\mu)}\frac{(-i/2)^{l}}{\sqrt{\pi}}\frac{-i}{2}\\
&\times\frac{1}{\Gamma\left(\frac{3}{2}+l\right)}\,\Gamma\left(-\frac{1}{2}-l\right)\Gamma\left(\frac{3}{2}+l\right)\\
&\times\Gamma(1+l+i\mu)\Gamma(1+l-i\mu)\Biggl\{\frac{1}{\Gamma(1+l-i\mu)\Gamma(-l+i\mu)}-\frac{1}{\Gamma(1+l+i\mu)\Gamma(-l-i\mu)}\Biggl\}.
\end{split}\label{app31}
\end{equation}
The second and the third lines above are the same as in (\ref{app21}), so following Appendix {\ref{app2}}, (\ref{app31}) becomes
\begin{equation}
-e^{\frac{\pi\mu}{2}}\frac{(-i/2)^{l}}{\sqrt\pi}\frac{\Gamma(1+l-i\mu)\Gamma(1+l+i\mu)}{\Gamma(l+\frac{3}{2})}\frac{x^{1+m+n+i\mu}}{1+m+n+i\mu},
\end{equation}
which eliminates the first term in (\ref{tochu}).

\section{Results of the integrals in (\ref{zzresult}), (\ref{gzzresult}) and (\ref{ggzzresult})}\label{app4}
All we have to do is just to substitute $-1/2$ or $3/2$ into $l$ and $m$ in the formulas in Table \ref{resulteasy} and \ref{resultdiff}. In order to derive the results below, we use the formulas:
\begin{eqnarray}
\Gamma(z+1)&=&z\Gamma(z),\\
\Gamma(z)\Gamma(1-z)&=&\frac{\pi}{\sin(\pi z)},\\
\Gamma\left(z+\frac{1}{2}\right)\Gamma\left(\frac{1}{2}-z\right)&=&\frac{\pi}{\cos(\pi z)}.
\end{eqnarray}

\subsection{Table {\ref{resulteasy}}}
\leftline{\underline{$l=-1/2$}}
\begin{equation}
\int_{0}^{\infty}dx\ x^{-\frac{1}{2}} e^{ix}\hankelm(x)=\frac{\sqrt{\pi}\,e^{\frac{\pi\mu}{2}}}{\cosh(\pi\mu)}(1-i)
\end{equation}

\leftline{\underline{$l=3/2$}}
\begin{equation}
\begin{split}
\int_{0}^{\infty}dx&\ x^{\frac{3}{2}} e^{ix}\hankelm(x)\\
&=\frac{\sqrt{\pi}\,e^{\frac{\pi\mu}{2}}}{\cosh(\pi\mu)}(-1+i)\times\frac{1}{8}\left(\frac{1}{4}+\mu^2\right)\left(\frac{9}{4}+\mu^2\right)
\end{split}
\end{equation}

\subsection{Table \ref{resultdiff}}
\leftline{\underline{$(l, m)=(-1/2,-1/2)$}}
\begin{equation}
\begin{split}
\int_{0}^{\infty}&dx\ x^{-\frac{1}{2}} e^{-ix}\hankelm(x)\int_{x}^{\infty}dy\ y^{-\frac{1}{2}} e^{-iy}\hankelmc(y)\\
&=\frac{1}{\pi\,\sinh(\pi\mu)}\sum_{n=0}^{\infty}(-1)^{n}\Biggl\{\frac{e^{2\pi\mu}}{(\frac{1}{2}+n+i\mu)^2}-\frac{1}{(\frac{1}{2}+n-i\mu)^2}\Biggl\}
\end{split}
\end{equation}

\leftline{\underline{$(l, m)=(3/2, -1/2)$}}
\begin{equation}
\begin{split}
\int_{0}^{\infty}&dx\ x^{-\frac{1}{2}} e^{-ix}\hankelm(x)\int_{x}^{\infty}dy\ y^{\frac{3}{2}} e^{-iy}\hankelmc(y)\\
&=-\frac{1}{4\pi\,\sinh(\pi\mu)}\,\sum_{n=0}^{\infty}(-1)^{n}(n+1)(n+2)\\
&\qquad\qquad\qquad\qquad\times\Biggl\{e^{2\pi\mu}\frac{(1+n+2i\mu)(2+n+2i\mu)}{(\frac{1}{2}+n+i\mu)^2(\frac{3}{2}+n+i\mu)(\frac{5}{2}+n+i\mu)}\\
&\qquad\qquad\qquad\qquad\qquad\qquad-\frac{(1+n-2i\mu)(2+n-2i\mu)}{(\frac{1}{2}+n-i\mu)^2(\frac{3}{2}+n-i\mu)(\frac{5}{2}+n-i\mu)}\Biggl\}
\end{split}
\end{equation}

\leftline{\underline{$(l, m)=(-1/2, \,3/2)$}}
\begin{equation}
\begin{split}
\int_{0}^{\infty}&dx\ x^{\frac{3}{2}} e^{-ix}\hankelm(x)\int_{x}^{\infty}dy\ y^{-\frac{1}{2}} e^{-iy}\hankelmc(y)\\
&=-\frac{1}{4\pi\,\sinh(\pi\mu)}\,\sum_{n=0}^{\infty}(-1)^{n}(n+1)(n+2)\\
&\qquad\qquad\qquad\qquad\times\Biggl\{e^{2\pi\mu}\frac{(1+n+2i\mu)(2+n+2i\mu)}{(\frac{1}{2}+n+i\mu)(\frac{3}{2}+n+i\mu)(\frac{5}{2}+n+i\mu)^2}\\
&\qquad\qquad\qquad\qquad\qquad\qquad-\frac{(1+n-2i\mu)(2+n-2i\mu)}{(\frac{1}{2}+n-i\mu)(\frac{3}{2}+n-i\mu)(\frac{5}{2}+n-i\mu)^2}\Biggl\}
\end{split}
\end{equation}

\leftline{\underline{$(l, m)=(3/2, \,3/2)$}}
\begin{equation}
\begin{split}
\int_{0}^{\infty}&dx\ x^{\frac{3}{2}} e^{-ix}\hankelm(x)\int_{x}^{\infty}dy\ y^{\frac{3}{2}} e^{-iy}\hankelmc(y)\\
&=\frac{1}{16\pi\,\sinh(\pi\mu)}\\
&\quad\quad\,\times\sum_{n=0}^{\infty}(-1)^{n}(n+1)(n+2)(n+3)(n+4)\\
&\qquad\times\Biggl\{e^{2\pi\mu}\frac{(1+n+2i\mu)(2+n+2i\mu)(3+n+2i\mu)(4+n+2i\mu)}{(\frac{1}{2}+n+i\mu)(\frac{3}{2}+n+i\mu)(\frac{5}{2}+n+i\mu)^2(\frac{7}{2}+n+i\mu)(\frac{9}{2}+n+i\mu)}\\
&\qquad\qquad-\frac{(1+n-2i\mu)(2+n-2i\mu)(3+n-2i\mu)(4+n-2i\mu)}{(\frac{1}{2}+n-i\mu)(\frac{3}{2}+n-i\mu)(\frac{5}{2}+n-i\mu)^2(\frac{7}{2}+n-i\mu)(\frac{9}{2}+n-i\mu)}\Biggl\}
\end{split}
\end{equation}

\section{Results of the integrals in (\ref{geneoneres}) and (\ref{genetwores})}\label{app5}
\subsection{Table \ref{resulteasy}}
\begin{equation}
\begin{split}
\int_{0}^{\infty}dx&\ x^{2N-\frac{1}{2}} e^{ix}\hankelm(x)\\
&=\frac{\sqrt{\pi}\,e^{\frac{\pi\mu}{2}}}{\cosh(\pi\mu)}\frac{2^{\frac{1}{2}-2N}(-1)^{N}}{(2N)!}e^{-i\frac{\pi}{4}}\\
&\quad\times\left\{\left(2N-\frac{1}{2}\right)^2+\mu^2\right\}\cdots\left\{\left(\frac{3}{2}\right)^2+\mu^2\right\}\left\{\left(\frac{1}{2}\right)^2+\mu^2\right\}
\end{split}
\end{equation}

\subsection{Table \ref{resultdiff}}
\begin{equation}
\begin{split}
\int_{0}^{\infty}&dx\ x^{-\frac{1}{2}} e^{-ix}\hankelm(x)\int_{x}^{\infty}dy\ y^{2N-\frac{1}{2}} e^{-iy}\hankelmc(y)\\
&=\frac{(2i)^{-2N}}{\pi\,\sinh(\pi\mu)}\,\sum_{n=0}^{\infty}(-1)^{n}(n+1)(n+2)\cdots(n+2N)\\
&\qquad\times\Biggl\{e^{2\pi\mu}\frac{(1+n+2i\mu)(2+n+2i\mu)\cdots(2N+n+2i\mu)}{(\frac{1}{2}+n+i\mu)^2(\frac{3}{2}+n+i\mu)\cdots(2N+\frac{1}{2}+n+i\mu)}\\
&\qquad\qquad-\frac{(1+n-2i\mu)(2+n-2i\mu)\cdots(2N+n-2i\mu)}{(\frac{1}{2}+n-i\mu)^2(\frac{3}{2}+n-i\mu)\cdots(2N+\frac{1}{2}+n-i\mu)}\Biggl\}
\end{split}
\end{equation}

\begin{equation}
\begin{split}
\int_{0}^{\infty}&dx\ x^{2N-\frac{1}{2}} e^{-ix}\hankelm(x)\int_{x}^{\infty}dy\ y^{-\frac{1}{2}} e^{-iy}\hankelmc(y)\\
&=\frac{(2i)^{-2N}}{\pi\,\sinh(\pi\mu)}\,\sum_{n=0}^{\infty}(-1)^{n}(n+1)(n+2)\cdots(n+2N)\\
&\qquad\times\Biggl\{e^{2\pi\mu}\frac{(1+n+2i\mu)(2+n+2i\mu)\cdots(2N+n+2i\mu)}{(\frac{1}{2}+n+i\mu)(\frac{3}{2}+n+i\mu)\cdots(2N+\frac{1}{2}+n+i\mu)^2}\\
&\qquad\qquad-\frac{(1+n-2i\mu)(2+n-2i\mu)\cdots(2N+n-2i\mu)}{(\frac{1}{2}+n-i\mu)(\frac{3}{2}+n-i\mu)\cdots(2N+\frac{1}{2}+n-i\mu)^2}\Biggl\}
\end{split}
\end{equation}

\begin{equation}
\begin{split}
&\int_{0}^{\infty}dx\ x^{2N-\frac{1}{2}} e^{-ix}\hankelm(x)\int_{x}^{\infty}dy\ y^{2N-\frac{1}{2}} e^{-iy}\hankelmc(y)\\
&=\frac{(2i)^{-4N}}{\pi\,\sinh(\pi\mu)}\,\sum_{n=0}^{\infty}(-1)^{n}(n+1)(n+2)\cdots(n+4N)\\
&\quad\times\Biggl\{e^{2\pi\mu}\frac{(1+n+2i\mu)(2+n+2i\mu)\cdots(4N+n+2i\mu)}{(\frac{1}{2}+n+i\mu)(\frac{3}{2}+n+i\mu)\cdots(2N+\frac{1}{2}+n+i\mu)^2\cdots(4N+\frac{1}{2}+n+i\mu)}\\
&\quad\qquad-\frac{(1+n-2i\mu)(2+n-2i\mu)\cdots(4N+n-2i\mu)}{(\frac{1}{2}+n-i\mu)(\frac{3}{2}+n-i\mu)\cdots(2N+\frac{1}{2}+n-i\mu)^2\cdots(4N+\frac{1}{2}+n-i\mu)}\Biggl\}
\end{split}
\end{equation}


\begin{thebibliography}{9}
\bibitem{Mal}J. Maldacena, ``Non-Gaussian features of primordial fluctuations in single field inflationary models," JHEP \textbf{0305}, 013 (2003), astro-ph/0210603.
\bibitem{Komatsu}N. Bartolo, E. Komatsu, S. Matarrese and A. Riotto, ``Non-Gaussianity from Inflation: Theory and Observations," Phys. Rep. \textbf{402}, 103-266 (2004), astro-ph/0406398. 
\bibitem{Effective}C. Cheung, P. Creminelli, A. L. Fitzpatrick, J. Kaplan and L. Senatore, ``The Effective Field Theory of Inflation," JHEP \textbf{03}, 014 (2008), hep-th/0709.0293.
\bibitem{Senatore}L. Senatore and M. Zaldarriaga, ``The Effective Field Theory of Multifield Inflation," JHEP \textbf{1204}, 024 (2012), hep-th/1009.2093.
\bibitem{Weinberg}S. Weinberg, ``Effective Field Theory for Inflation," Phys. Rev. D \textbf{77}, 123541 (2008), hep-th/0804.4291.
\bibitem{Quasi}X. Chen and Y. Wang, ``Quasi-Single-Field Inflation and Non-Gaussianities," JCAP \textbf{1004}, 027 (2010), hep-th/0911.3380.
\bibitem{Noumi}T. Noumi, M. Yamaguchi, and D. Yokoyama, ``EFT Approach to Quasi-Single-Field Inflation and Effects of Heavy Fields," JHEP \textbf{06}, 051 (2013), hep-th/1211.1624.
\bibitem{Cosmo}N. Arkani-Hamed and J. Maldacena, ``Cosmological Collider Physics,” hep-th/1503.08043.
\bibitem{Baumann}H. Lee, D. Baumann, G. L. Pimentel, ``Non-Gaussianity as a Particle Detector," JHEP \textbf{1612}, 040 (2016), hep-th/1607.03735.
\bibitem{SMcosmo}X. Chen, Y. Wang and Z. Xianyu, ``Standard Model Background of the Cosmological Collider," Phys. Rev. Lett. \textbf{118}, 261302 (2017), hep-th/1610.06597.
\bibitem{ADMpaper}R. L. Arnowitt, S. Deser and C. W. Misner, ``The Dynamics
of General Relativity," (1962), gr-qc/0405109.
\bibitem{Birrel}N. D. Birrell and P. C. W. Davies, ``Quantum Fields in Curved Space," Cambridge University Press, (1982).
\bibitem{ininpaper}J. S. Schwinger, ``Brownian Motion of a Quantum Oscillator," J. Math. Phys. \textbf{2}, 407 (1961); L. V. Keldysh, ``Diagram technique for nonequilibrium processes," Zh. Eksp. Teor. Fiz. \textbf{47}, 1515 (1964) [Sov. Phys. JETP \textbf{20}, 1018 (1965)].
\bibitem{TASI}D. Baumann, ``TASI Lectures on Inflation," hep-th/0907.5424.
\bibitem{Collins}H. Collins, ``Primordial non-Gaussianities from inflation," astro-ph/1101.1308.
\bibitem{Hinter}K. Hinterbichler, L. Hui and J. Khoury, ``An Infinite Set of Ward Identities for Adiabatic Modes in Cosmology," JCAP \textbf{1401}, 039 (2014), hep-th/1304.5527.
\bibitem{Ward2}L. Berezhiani and J. Khoury, ``Slavnov-Taylor Identities for Primordial Perturbations,” JCAP \textbf{1402}, 003 (2014), hep-th/1309.4461.
\bibitem{Higuchi}A. Higuchi, ``Forbidden Mass Range for Spin-2 Field Theory in De Sitter Spacetime,” Nucl. Phys.
\textbf{B282}, 397 (1987).
\bibitem{chenwang}X. Chen and Y. Wang, ``Quasi-Single-Field Inflation with Large Mass,” JCAP \textbf{1209}, 021 (2012), hep-th/1205.0160.
\end{thebibliography}
\end{document}